\documentclass[11pt,a4paper]{article} 
\pdfoutput=1
\usepackage{amsmath}
\usepackage{amssymb}
\usepackage{empheq}
\usepackage{bm}
\usepackage{color}
\usepackage{cite} 

\usepackage{array}
\usepackage{booktabs}
\usepackage{multirow}

\usepackage{hyperref}
\usepackage{graphicx}

\catcode `@=11 \@addtoreset{equation}{section} \catcode `@=12

\def\Tr{{\rm Tr}}

\begin{document}
\setlength{\oddsidemargin}{0cm}

\begin{titlepage}

\begin{center}
{\LARGE	
A Scaling Relation, $Z_m$-type Deconfinement Phases \\
and Imaginary Chemical Potentials \\
    in Finite Temperature Large-$N$ Gauge Theories }
\end{center}
\vspace{1.2cm}
\baselineskip 18pt 
\renewcommand{\thefootnote}{\fnsymbol{footnote}}

\begin{center}

Takehiro {\sc Azuma}$^{a}$\footnote{%
    E-mail address: azuma(at)mpg.setsunan.ac.jp
} and 
Takeshi {\sc Morita}$^{b,c}$\footnote{%
    E-mail address: morita.takeshi(at)shizuoka.ac.jp
}

\renewcommand{\thefootnote}{\arabic{footnote}}
\setcounter{footnote}{0}

\vspace{0.4cm}

{\it
    a. Institute for Fundamental Sciences, Setsunan University, 17-8 Ikeda Nakamachi, Neyagawa, Osaka, 572-8508, Japan  
    \vspace{0.2cm}
    \\

    b. Department of Physics,
    Shizuoka University \\
    836 Ohya, Suruga-ku, Shizuoka 422-8529, Japan 
    \vspace{0.2cm}
    \\
    c. Graduate School of Science and Technology, Shizuoka University\\
    836 Ohya, Suruga-ku, Shizuoka 422-8529, Japan
}

\end{center}


\vspace{1.5cm}

\begin{abstract}
We show that the effective potentials for the Polyakov loops in finite temperature SU$(N)$ gauge theories obey a certain scaling relation with respect to temperature in the large-$N$ limit. This scaling relation strongly constrains the possible terms in the Polyakov loop effective potentials. Moreover, by using the effective potentials in the presence of imaginary chemical potentials or imaginary angular velocities in several models, we find that phase transitions to $Z_m$-type deconfinement phases ($Z_m$ phase) occur, where the eigenvalues of the Polyakov loop are distributed $Z_m$ symmetrically. Physical quantities in the $Z_m$ phase obey the scaling properties of the effective potential. The models include Yang-Mills (YM) theories, the bosonic BFSS matrix model and ${\mathcal N}=4$ supersymmetric YM theory on $S^3$. Thus, the phase diagrams of large-$N$ gauge theories with imaginary chemical potentials are very rich and the stable $Z_m$ phase would be ubiquitous. Monte-Carlo calculations also support this. As a related topic, we discuss the phase diagrams of large-$N$ YM theories with real angular velocities in finite volume spaces.

\end{abstract}

\end{titlepage}

\tableofcontents

\section{Introduction and Summary}

Large-$N$ gauge theories \cite{tHooft:1973alw} play significant roles in theoretical physics. These theories are important not only in quantum chromodynamics (QCD) \cite{tHooft:1973alw, Makeenko:1979pb, Eguchi:1982nm, GONZALEZARROYO1983174, Lucini:2001ej}, which describes our real world, but also in quantum gravity, such as string theories \cite{Banks:1996vh, Ishibashi:1996xs, Berenstein:2002jq} and the gauge/gravity correspondence \cite{Maldacena:1997re, Witten:1998zw, Itzhaki:1998dd}.  For reviews, see Ref.~\cite{Aharony:1999ti, Lucini:2012gg}.

In this paper, we study SU$(N)$ gauge theories\footnote{The difference between SU$(N)$ and U$(N)$ is irrelevant in the large-$N$ limit, and our main results in this paper hold in both cases.} with adjoint matter fields at finite temperatures in the large-$N$ limit. One of the features of these theories is that they undergo the confinement/deconfinement transitions. (Some systems do not show such a phase transition, but these systems can be considered as cases with infinite or zero transition temperature\footnote{In some special theories such as the three-dimensional Chern-Simons theories \cite{Jain:2013py}, different phases appear. In this paper, we do not consider such special cases, but only systems in which the typical confinement and deconfinement phases appear.}.) The order parameters of this phase transition are the expectation values of the Polyakov loops \cite{Polyakov:1978vu}
\begin{align} 
    u_n:=\frac{1}{N} \Tr\left( {\rm P} e^{i \int_0^{n\beta} A_{\tau}  d \tau} \right), \quad n \in {\mathbf Z},
\end{align}
and $\langle u_1 \rangle=0 $ indicates the confinement phase and $\langle u_1 \rangle \neq 0 $ indicates the deconfinement phase, which breaks the $Z_N$ center symmetry of SU$(N)$, $u_n \to e^{2\pi n i/N} u_n$. Here $\beta$ is inverse temperature, $\tau$ is Euclid time and $A_\tau$ is the temporal component of the gauge field. In this paper, we take an unit $c=\hbar=k_B=1$. 

By applying the Landau-Ginzburg argument, the phase structure of large-$N$ gauge theories would be described by the effective potential for the Polyakov loops \cite{RevModPhys.53.43, PhysRevD.24.475, PhysRevD.25.2667, ALTES1994637}. In this paper, we report the following two new results on the Polyakov loop effective potentials and the phase structure of large-$N$ gauge theories:
\begin{itemize}

\item It is well known that Schwinger-Dyson equations strongly constrain dynamics of large-$N$ gauge theories \cite{Makeenko:1979pb, Eguchi:1982nm, Gocksch:1982en, GOCKSCH198381}. We develop these studies and show that the Polyakov loop effective potentials in general large-$N$ gauge theories obey a scaling relation with respect to temperature by using Schwinger-Dyson equations. It is expressed as Eq.~\eqref{F-scaling} or \eqref{F-scaling-fermion-odd}, depending on whether fermions are uncoupled or coupled. This scaling relation strongly constrains the possible terms in the effective potentials.

\item
Recently, a non-trivial result was reported in Refs.~\cite{Chen:2022smf, Chen:2024tkr} that the confining phases in the four-dimensional SU$(2)$ and SU(3) YM theories would be stabilized even at high temperatures by turning on the imaginary angular velocity, which is the imaginary chemical potential for the angular momentum. We extend this to various large-$N$ gauge theories, and find that there exists a $Z_m$-type deconfinement phase in which the eigenvalues of the Polyakov loop are distributed $Z_m$ symmetrically and the $Z_N$ center symmetry is broken to $Z_m$. This phase is called the ``$Z_m$ phase''. The $Z_m$ phase appears stably when the system has rotational symmetries or global U$(1)$ symmetries such as $R$-symmetries and their imaginary chemical potentials or imaginary angular velocities are introduced. Physical quantities in the $Z_m$ phase obey the scaling properties of the Polyakov loop effective action.
\end{itemize}

As concrete examples, we show these properties in several models: the gauged free matrix quantum mechanics \cite{Aharony:2003sx}, the bosonic BFSS matrix model \cite{Banks:1996vh}, the four-dimensional Yang-Mills (YM) theory on $S^3$ (small volume limit) \cite{Aharony:2003sx, Aharony:2005bq}, the four-dimensional ${\mathcal N}=4 $ supersymmetric YM (SYM) theory on $S^3$ (small volume limit)  \cite{Sundborg:1999ue, Aharony:2003sx} and the four-dimensional YM theory on $R^3$ (high temperature limit) \cite{RevModPhys.53.43}. In particular, we apply a Monte-Carlo calculation to the bosonic BFSS matrix model and observe transitions to $Z_m$ phases and the scaling properties. This confirms that the above two predictions are valid at the non-perturbative level. We also discuss the connection between the high temperature confinement phases in the SU$(2)$ and SU(3) YM theories \cite{Chen:2022smf, Chen:2024tkr} and our $Z_m$ phases in the large-$N$ case.

We briefly explain the importance of our results. First, the scaling relation strongly constrains the Polyakov loop effective potentials. Since the Polyakov loop effective potentials have been widely used in the studies of finite-temperature gauge theories (see Ref.~\cite{Fukushima:2017csk} for a recent review), the scaling relation would bring progress in these studies. Moreover, the discovery of stable $Z_m$ phases means that gauge theories have richer phase structures than previously thought. For example, even in the four-dimensional large-$N$ pure YM theory, we show that $Z_m$ phases ($m=$2, 3 and 4) can exist stably at high temperatures in addition to the usual deconfinement phase by introducing the imaginary angular velocity. This recalls the Roberge-Weiss transition \cite{ROBERGE1986734}, where the imaginary baryon chemical potential brings the interesting phase structure to QCD. Besides, by applying the gauge/gravity correspondence, we predict the existence of the gravity solutions corresponding to $Z_m$ phases, which have not been known so far.

 You may think that imaginary chemical potentials such as imaginary angular velocities are unphysical, but this is not true. In fact, the imaginary chemical potentials can be regarded as the temporal Aharonov–Bohm phases of the background gauge fields, which are introduced by gauging the corresponding symmetries. In particular, in quantum gravity, global symmetries are expected to be gauged, and then their imaginary chemical potentials would be related to some background gauge fields. The imaginary angular velocities are also realized as a non-trivial background metric \cite{Yamamoto:2013zwa, Chernodub:2020qah}.\\

As a byproduct of the investigation of imaginary angular velocities, we also discuss the dependence of the deconfinement transition temperatures on the real angular velocity in the four-dimensional large-$N$ YM theories on finite volume spaces ($S^3$  and $T^3$). Recently, the phase structures of YM theories in rotating systems \cite{Yamamoto:2013zwa, Jiang:2016wvv, Chernodub:2020qah, Fujimoto:2021xix, Braguta:2021jgn} have attracted much attention motivated by the study of neutron stars and collider experiments \cite{STAR:2017ckg}. However, there is no clear conclusion on how the deconfinement transition is affected by the angular velocity. There are two reasons why such an analysis is difficult. First, in order to realize a stable thermal equilibrium state by introducing an angular velocity, it is necessary to introduce an appropriate IR cut-off \cite{Chen:2022smf}, and the results may depend on it. Second, real angular velocities cause a sign problem generally, and Monte-Carlo computation is difficult.

We can avoid the first problem by considering the systems on finite volume spaces such as $S^3$ and $T^3$, where the definite IR cut-off is given. To avoid the second problem, we take the small volume limit of these systems to simplify the analysis. The phase diagram of the YM theory on a small $T^3$ has already been studied in Refs.~\cite{Morita:2010vi, Azuma:2023pzr}, and the transition temperature decreases as angular velocities increase. In this paper, we show that the same happens in the YM theory on a small $S^3$. These results suggest that the transition temperatures decrease with increasing angular velocities in finite volume systems. \\

Before proceeding to the main sections, we briefly review $Z_m$ phases in the previous studies on finite temperature large-$N$ gauge theories. The $Z_m$ phases have already been found in gauge theories coupled to the adjoint fermions that are periodic along the temporal circle \cite{Bedaque:2009md, Bringoltz:2009mi, Bringoltz:2009kb, Hollowood:2009sy, Bringoltz:2009fj}. (One motivation of these works is to investigate the center symmetry stabilization for the large-$N$ volume independence proposed by the authors of Ref.~\cite{Kovtun:2007py}.) These studies seem to be closely related to ours, since the periodic adjoint fermions in a thermal system can be realized by tuning an imaginary chemical potential $q\beta\mu_I = \pi$. (Recall that the imaginary chemical potential $\mu_I$ is also represented by the change of the charged field: $\Phi(\tau) \to \tilde{\Phi}(\tau)=e^{i q \mu_I \tau }  \Phi(\tau)  $, where $q$ is the charge, and it changes the periodicity by $e^{i q \beta \mu_I  } $.) Besides, $Z_m$ phases generally exist as unstable solutions in large-$N$ gauge theories even without introducing imaginary chemical potentials \cite{Azuma:2012uc}. Furthermore, their contributions to the spectral form factor have been discussed in Ref.~\cite{Chen:2022hbi}. In this way, although $Z_m$ phases are less well known than deconfinement phase, they are quite common in large-$N$ gauge theories. Our results on $Z_m$ phase strengthen this further. \\

This paper is organized as follows. In Sec.~\ref{sec-general-propreties}, we review some general properties of large-$N$ gauge theories at finite temperatures. We then argue that the Polyakov loop effective potentials obey a scaling relation. In Sec.~\ref{sec-free-MQM}, we study the gauged free matrix quantum mechanics and derive the Polyakov loop effective action explicitly. We will see that the effective action does indeed satisfy the scaling relation. Besides, we show that when imaginary angular velocities are introduced, phase transitions to $Z_m$ phases occur. In Sec.~\ref{sec-bBFSS} and Sec.~\ref{sec-YM}, we investigate the bosonic BFSS matrix model and four-dimensional Yang-Mills theory, respectively. Similarly to the free matrix model case, we will observe the scaling relation and $Z_m$ phases in these models too. Furthermore, we perform a Monte-Carlo calculation in the bosonic BFSS matrix model and confirm these properties numerically. Sec.~\ref{Sec_discussion} is devoted to discussions. In Appendix \ref{app-scaling-general}, some consequences of the scaling relation on observables are discussed. In particular, we will see that the values of the observables in $Z_m$ phases are related to those of conventional deconfinement phase. 
In Appendix \ref{subsubsec-MC_appendix}, we supplement the details of the numerical simulation presented in Sec.~\ref{subsubsec-MC_abridge}. 
We provide the details of the Fourier expansion regularization \cite{0706_1647,0707_4454} and depict the way the $Z_m$ phases emerge in the simulation.
In Appendix \ref{app-SYM}, we investigate the four-dimensional ${\mathcal N}=4 $ SYM theory on $S^3$ and observe $Z_m$ phases. In Appendix \ref{app-finite-N}, we study the four-dimensional SU($N$) YM theories at finite $N$ with imaginary angular velocities. We argue the connection between our results at large $N$ and the SU(2) and SU(3) cases studied in Ref.~\cite{Chen:2022smf}.

\section{Large-$N$ gauge theories at finite temperatures and a scaling relation}
\label{sec-general-propreties}

In this section, we review general properties of finite temperature large-$N$ gauge theories and introduce the effective potentials of the Polyakov loops. We then show that the effective potentials obey a scaling relation with respect to temperature. This scaling relation is useful not only for constraining the possible terms in the effective potentials but also for predicting the phase transition temperatures and physical quantities of $Z_m$ phases, which will be discussed in the next sections.

\subsection{Polyakov loop effective potential }
\label{subsec-Polyakov-loop-action}

Phases of SU$(N)$ gauge theories at temperature $T=1/\beta$ are characterized by the expectation values of the Polyakov loops $u_n$ winding around the Euclidean time circle $n$ times,
\begin{align} 
u_n=\frac{1}{N} \Tr\left( {\rm P} e^{i \int_0^{n\beta} A_{\tau}  d \tau} \right).
\label{un}
\end{align}
Here $\tau$ is the Euclidean time coordinate and $A_\tau$ is the temporal component of the SU$(N)$ gauge field.

In general, $ u_n $ depends on the spatial position.
However, in some situations, a saddle point solution that is independent of the position may dominate the path-integral, and analyses that ignore the position dependence are often used \cite{Fukushima:2017csk}.
Moreover, in matrix quantum mechanics \cite{Banks:1996vh, Berenstein:2002jq} and reduced models in 0+1 dimensions  \cite{Aharony:2003sx, AlvarezGaume:2005fv}, $ u_n $ is a constant mode from the beginning. Hereafter, we assume that $u_n $ has no position dependence with these situations in mind.

As a quantity related to $u_n $, the eigenvalue density of $A_\tau$ is also useful to characterize the phase.
We take the static diagonal gauge
\begin{align} 
	\beta A_\tau = {\rm diag}( \alpha_1 , \alpha_2,\cdots,  \alpha_N), \quad  \left( -\pi <  \alpha_k \le \pi, \quad \partial_\tau \alpha_k=0, \quad \sum_{k=1}^N \alpha_k =0 \quad \text{mod } 2\pi  \right).
	\label{gauge-diagonal}
\end{align}
Then, we define the eigenvalue density $\rho(\alpha)$ as
\begin{align}
	\rho(\alpha):=\frac{1}{N} \sum_{k=1}^N \delta(\alpha-\alpha_k), \qquad  (-\pi <  \alpha \le \pi).
	\label{rho}
\end{align}
By using $\rho(\alpha)$, the Polyakov loop $u_n$ can be written as 
\begin{align}
	u_n= \frac{1}{N} \sum_{k=1}^N e^{in \alpha_k} = \int_{-\pi}^{\pi} d\alpha \rho(\alpha) e^{in \alpha} .
	\label{polyakov_static_diag}
\end{align}
This equation tells us that $\{ u_n \}$ are the Fourier coefficients of $\rho(\alpha)$
\begin{align}
	\rho(\alpha)=\frac{1}{2\pi} \sum_{n \in {\mathbf Z}} u_n e^{-in \alpha}.
	\label{rho-un}
\end{align}
In the large-$N$ limit, $\{u_n \}$ behave as independent dynamical variables under the condition that $\rho(\alpha)$ is non-negative \cite{Aharony:2003sx}. \\

We now introduce the effective potential describing the thermodynamics of $\{u_n \}$. In finite temperature gauge theories, after taking the gauge \eqref{gauge-diagonal}, we may obtain the effective action for the Polyakov loops $u_n$ by integrating out the degrees of freedom other than $A_\tau$,
\begin{align} 
	Z=& \int {\mathcal D} A_\tau {\mathcal D} \cdots e^{-S} =   \int  \left(\prod_{n=1}^{\infty} d u_n d u_{-n}  \right)  e^{-S_{\rm eff} (\beta,\{ u_n \})}.
    \label{Z}
\end{align}
We call the effective action $S_{\rm eff} (\beta,\{ u_n \})$ the Polyakov loop effective action \cite{Fukushima:2017csk}.

We consider the possible terms of this effective action. For simplicity, we study the gauge theories only coupled to adjoint matter fields. Then the system has the $Z_N$ symmetry: $u_n \to e^{2\pi n i/N} u_n$, and $S_{\rm eff} (\beta,\{ u_n \})$ has to be invariant under this transformation.
Thus, the expansion of the action with respect to $u_n$ is restricted to the following form \cite{AlvarezGaume:2005fv},
\begin{align} 
S_{\rm eff} (\beta,\{ u_n \})=& \beta N^2 V(\beta,\{ u_n \}) ,
\label{S_eff-F}
\\ 
V(\beta,\{ u_n \}):=&\epsilon (\beta)+ \sum_{n=1}^\infty a_n(\beta)  u_n   u_{-n} + \sum_{k,l} b_{k,l}(\beta)  ( u_k u_l   u_{-k-l} + c.c. )  \nonumber \\
&+ \sum_{k,l,m} c_{k,l,m} (\beta) ( u_k u_l u_m   u_{-k-l-m} + c.c. ) +\cdots .
\label{F-expand}
\end{align}
Here $V(\beta,\{ u_n \})$ is the effective potential, and $\epsilon(\beta) $, $a_n(\beta) $, $b_{k,l}(\beta) $, $c_{k,l,m}(\beta) $, $
\cdots$ are the expansion coefficients, which are functions of the parameters of the system such as temperature, chemical potentials and the 't Hooft coupling, but only the inverse temperature dependence is given explicitly for later convenience.
In some special situations, these coefficients can be derived from the given Lagrangian, otherwise we have to tune them phenomenologically.
Assuming that the usual 't Hooft expansion holds, these coefficients are $O(1)$ quantities with respect to $N$. In the next subsection, we will see that these coefficients are further constrained by a scaling property of the Polyakov loop effective potential.

\subsection{A scaling relation of the Polyakov loop effective potential }
\label{subsec-Polyakov-loop-action-scaling}

We propose that the Polyakov loop effective potential \eqref{F-expand} in the large-$N$ limit obeys a scaling relation. This relation depends on whether fermions are coupled to the system or not.  Here we consider systems without fermions, and systems coupled to fermions are discussed in Sec.~\ref{app-scaling-derivation-fermion}. Then, our proposal for bosonic systems is that the effective potential \eqref{F-expand} satisfies the following scaling relation with respect to temperature,
\begin{align} 
	\left. V(  \beta,  \{ u_{n} \}) \right|_{u_n=0~(n  \notin m \mathbf{Z}),~ u_{mn} \to u_{n} } =
	  V( m \beta, \{ u_n \})  ,
			\label{F-scaling}
\end{align} 
where $m$ is an arbitrary positive integer\footnote{A more precise expression of the scaling relation \eqref{F-scaling} is as follows:
\begin{align} 
	\left. V(  \beta,\{ g_i \},  \{ u_{n} \}) \right|_{u_n=0~(n  \notin m \mathbf{Z}),~ u_{mn} \to u_{n} } =
	  V( m \beta,\{ g_i \}, \{ u_n \})  ,
\end{align}
where $\{ g_i \}$ are parameters other than temperature such as chemical potentials and coupling constants. These constants remain the same on the left and right-hand sides.}.
Here, the left-hand side is the effective potential \eqref{F-expand} with $u_n(\beta)$ set to zero if $n$ is not a multiple of $m$ and  $u_{mn}(\beta)$ set to $u_{n}(m\beta)$. 
(To emphasize the Polyakov loop at inverse temperature $\beta$, we write $u_n=u_n(\beta)$.) 
The right-hand side is the effective potential at the inverse temperature $m \beta$.
Roughly speaking, this scaling relation states that, if all $u_{n}(\beta)=0$ $(n  \notin m \mathbf{Z})$, the thermodynamics of $u_{mn}(\beta)$ at the inverse temperature $\beta$ is equivalent to the thermodynamics of $u_{n}(m\beta)$ at the inverse temperature $m\beta$.

Before providing the derivation of this scaling relation, which we will argue in the next subsection, here we show how the scaling relation strongly constrains the effective potential \eqref{F-expand}.
By applying the scaling relation \eqref{F-scaling} to the expansion of the effective potential \eqref{F-expand}, we obtain,
\begin{align} 
 &   \left. V(  \beta,  \{ u_{n} \}) \right|_{u_n=0~(n  \notin m \mathbf{Z}),~u_{mn} \to u_{n} } \nonumber \\
     &=\epsilon(\beta) + \sum_{n=1}^\infty a_{mn} (\beta) u_{n}   u_{-n}  + \sum_{k,l} b_{km,lm} (\beta) ( u_{k} u_{l}   u_{-k-l}  + c.c. )  \nonumber \\
    &= V( m \beta, \{ u_n \}) =\epsilon(m\beta) + \sum_{n=1}^\infty a_{n} (m\beta) u_{n}   u_{-n}  + \sum_{k,l} b_{k,l} (m\beta) ( u_{k} u_{l}   u_{-k-l}  + c.c. )  
    \end{align}
    Then, by comparing the coefficients of $u_n$, we obtain the following relations\footnote{A relation between the string tensions analogous to the scaling property of $a_n$ is obtained in Ref.~\cite{GOCKSCH198381}.}:
\begin{align} 
&\epsilon(\beta)=\epsilon(m\beta) \to \partial_\beta \epsilon (\beta)=0 ,
\label{e-scaling}
\\
&a_{mn}(\beta)=a_{n}(m\beta) \to  a_{m}(\beta)=a_{1}(m\beta) ,
\label{a-scaling}
\\
&b_{mk,ml}(\beta)=b_{k,l}(m\beta) \to b_{m,m}(\beta)=b_{1,1}(m\beta),~ b_{m,2m}(\beta)=b_{1,2}(m\beta), \cdots.
\end{align}
The second and third equations are obtained by taking $n=1$ in the left equations. It is easy to obtain similar relations for the higher order expansion coefficients of the Polyakov loop effective potential \eqref{F-expand}. 

In this way, the scaling relation \eqref{F-scaling} strongly constrains the possible terms in the effective potential. In particular, the expansion \eqref{F-expand} now becomes
\begin{align} 
    V(\beta,\{ u_n \})=&\epsilon + \sum_{n=1}^\infty a_1(n\beta)  |u_n|^2  + O(u_n^3) ,
    \label{F-expand-scaling}
\end{align}
where $\epsilon$ does not depend on $\beta$.
Thus, when $|u_n|$ are small, the thermodynamics of the system is controlled just by a single function $a_1(\beta)$. For example, if the system is in a confinement phase where $u_n \simeq 0$, this effective potential would be particularly useful, as we will demonstrate soon.

\subsection{Derivation of the scaling relation}
\label{app-scaling-derivation}

We show the derivation of the scaling relation for the Polyakov loop effective potential introduced in the previous subsection. The scaling relation changes depending on the presence of fermions. We consider bosonic systems in Sec.~\ref{app-scaling-derivation-boson} and systems coupled to fermions in Sec.~\ref{app-scaling-derivation-fermion}. 

\subsubsection{Derivation of the scaling relation \eqref{F-scaling} in bosonic systems}
\label{app-scaling-derivation-boson}

We derive the scaling relation \eqref{F-scaling} in bosonic systems by using Schwinger-Dyson equations (loop equations), which provide non-perturbative relations between loop operators. Schwinger-Dyson equations are particularly useful in large-$N$ gauge theories, since the large-$N$ factorization drastically simplifies them \cite{Makeenko:1979pb}, and in fact, it has been shown in Refs.~\cite{Eguchi:1982nm, GOCKSCH198381, Gocksch:1982en} that dynamics of large-$N$ gauge theories are strongly constrained. The scaling relation, which we will derive here, is an extension of these results.

We consider $d$-dimensional U($N$) gauge theories ($d \ge 1$), which may be coupled to bosonic adjoint matter fields. (Here, we take U($N$) rather than SU($N$), since Schwinger-Dyson equations are simpler but their difference is irrelevant at large $N$.) When gauge theories are at finite temperatures, there exist not only the usual closed loop operators (Wilson loops) but also loop operators winding around the Euclidean temporal circle. For example, the following loop operator exists
\begin{align} 
    W(C_n,\beta):= 
    \frac{1}{N}
        \Tr {\rm P} \exp \left( i \int_{C_n} A_\mu dx^\mu \right) ,
        \label{Wilson-loop}
\end{align}
where $C_n$ is a contour winding around the temporal circle $n$ times. 
(Here we have restored the position dependence of $A_\tau$ and the diagonal gauge \eqref{gauge-diagonal} has not been taken either.) 
This loop operator is closed through the periodicity of the field $A_\mu(\tau, \vec{x})=A_\mu(\tau+n\beta, \vec{x})$. Thus, in the coordinate space $- \infty < \tau < \infty $, the contour $C_n$ can be regarded as an ``open path" connecting the point $(\tau, \vec{x})$ to the point $(\tau+n\beta, \vec{x})$ \cite{Gocksch:1982en}. We call $W(C_n,\beta)$ for $n \neq 0$ an open loop operator.

\begin{figure}
    \begin{center}
            \includegraphics[scale=1]{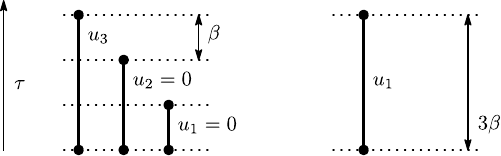}
    \caption{
    Polyakov loops at inverse temperature $\beta$ (left panel) and $3\beta$ (right panel). 
    They are open loop operators in the coordinate space $- \infty < \tau < \infty $.
    If $u_{n}(\beta)=0$ ($n  \notin 3 \mathbf{Z}$), only the same Polyakov loops appear in both the inverse temperature $\beta$ and $3\beta$. Thus, the Polyakov loops $u_{3n}(\beta)$ and $u_{ n}(3\beta)$ should obey the same Schwinger-Dyson equations.
    }
    \label{fig-SD}
    \end{center}
\end{figure}

The Polyakov loops $u_n$ defined in Eq.~\eqref{un} are also a kind of open loop operators. See Fig.~\ref{fig-SD}. Also,  there are loop operators with several local operators inserted in the path, 
\begin{align} 
    \frac{1}{N}
    \Tr \left[ {\rm P} e^{ i \int_{C_n} A_\mu dx^\mu} F_{\mu\nu}(x_1)   \right],~
    \Tr \left[ {\rm P} e^{ i \int_{C_n} A_\mu dx^\mu} \Phi(x_1)  \right],~
        \frac{1}{N}
        \Tr \left[ {\rm P} e^{ i \int_{C_n} A_\mu dx^\mu} \Phi(x_1) \Phi(x_2)  \right],~ \cdots,
    \end{align}
    where $x_1$ and $x_2$ are the points on the path $C_n$, and 
    $F_{\mu\nu}$ is the field strength and $\Phi$ is an adjoint matter field.
Hereafter, we symbolically denote by $W(C_n,\beta)$ the loop operator, including these operators.

The Schwinger-Dyson equations for the loop operators $W(C_n,\beta)$  describe the joining and/or splitting of the loops (See, for example, chapter 12 of Ref.~\cite{Makeenko:2002uj}). If there are no fermions, the Schwinger-Dyson equation with respect to the loop operator $W(C_n,\beta)$ is schematically written as \cite{Gocksch:1982en}
\begin{align} 
  \sum_{C'_n} g' \left\langle  W(C'_n,\beta) \right\rangle &= \sum_k \sum_{C''_{n-k},~C'''_k} \left\langle W(C''_{n-k},\beta) W(C'''_{k},\beta) \right\rangle \nonumber \\
  &= \sum_k \sum_{C''_{n-k},~C'''_k} \left\langle W(C''_{n-k},\beta) \right\rangle \left\langle W(C'''_{k},\beta) \right\rangle + O(1/N^2),
  \label{SD-equation}
\end{align}
where the large-$N$ factorization has been used in the second line. (The explicit expression for the Schwinger-Dyson equation of pure Yang-Mills theory is given in Eq.~(11) of Ref.~\cite{Gocksch:1982en}.)
The left-hand side represents the local operator insertions to the loop operator $W(C_n,\beta)$ through the Lagrangian density, where $\{C'_n \}$ symbolically denotes the operator insertions to $W(C_n,\beta)$ and $\{g' \}$ are some constants depending on the parameters of the Lagrangian such as the masses of the matters, coupling constants and chemical potentials but not temperature. (Recall that the Lagrangian density does not depend on temperature.) The right-hand side represents the splitting of the loop operator $W(C_n,\beta)$ into $W(C''_{n-k},\beta)$ and $W(C'''_k,\beta)$.
The total winding number is $n$ on both the left and right-hand sides, and it is conserved\footnote{\label{ftnt-fundamental}If fundamental matters are coupled, there are genuine open loop operators connecting the fundamental and anti-fundamental matters, and the winding number is not conserved. Thus, the discussion in this section is not applicable.
Similarly, if there is a matter field that breaks the $Z_N$ center symmetry, we cannot apply our discussion.
}. 
There are other types of Schwinger-Dyson equations for multiple loop operators and they also have similar structures \cite{GOCKSCH198381}.

The key feature of the Schwinger-Dyson equation \eqref{SD-equation} at large $N$ is that the temperature dependence arises only through the non-zero expectation values of the open loop operators $W(C_k,\beta)$ ($k \neq 0$) \cite{Gocksch:1982en}.
Thus, for example, if we formally set all the expectation values of $W(C_k,\beta)$ to zero except $k=0$, the set of the Schwinger-Dyson equations is equivalent to that of zero temperature, where only $W(C_0,\beta)$ exists\footnote{\label{ftnt-zero-temperature} We can show that $\langle W(C_n,\beta) \rangle=0$ $(n \neq 0)$ can always be a consistent solution of the Schwinger-Dyson equations as follows. At zero temperature, the periodicity of the temporal Euclidean circle can be ignored, and the stable solution of the the Schwinger-Dyson equations is given by $\langle W(C_0,\beta=\infty ) \rangle$ and $\langle W(C_n,\beta=\infty ) \rangle =0$ ($n \neq 0$), where $\langle W(C_0,\beta=\infty ) \rangle$ take certain values. Since the Schwinger-Dyson equation \eqref{SD-equation} does not depend on $\beta$, this can be a solution at any temperature. Therefore,  $\langle W(C_0,\beta ) \rangle=\langle W(C_0,\beta=\infty ) \rangle$ and $\langle W(C_n,\beta ) \rangle =0$ ($n \neq 0$) is a consistent solution. This solution would be stable when the system is confined, while it becomes unstable above the transition temperature. Note that the stability of the solution is not important for the derivation of the scaling relation.}.
 This is due to the large-$N$ factorization, since  $\langle W(C_k,\beta) \rangle =0$ does not mean $\langle W(C_k,\beta) W(C_n,\beta) \rangle =0$ at finite $N$.

Now what happens if we set all the expectation values of the open loop operators to zero except $ \langle W(C_{mk},\beta) \rangle$ ($k \in  \mathbf{Z}$) for a positive integer $m$ in the Schwinger-Dyson equation \eqref{SD-equation}?
In this case, the equation becomes
\begin{align} 
    \sum_{C'_{mn}} g' \left\langle  W(C'_{mn},\beta) \right\rangle &= \sum_k \sum_{C''_{m(n-k)},~C'''_{mk}} \left\langle W(C''_{m(n-k)},\beta) \right\rangle \left\langle W(C'''_{mk},\beta) \right\rangle + O(1/N^2).
    \label{SD-equation-m}
  \end{align}
This is the Schwinger-Dyson equation at inverse temperature $m \beta$ by identifying $ \langle W(C_{mk},\beta) \rangle = \langle W(C_{k},m\beta) \rangle$\footnote{
Similarly to the argument in footnote \ref{ftnt-zero-temperature}, for a positive integer $m$, $ \langle W(C_n,\beta) \rangle=0$ $(n \notin m  \mathbf{Z} )$ and $\langle W(C_{mk},\beta) \rangle=\langle W(C_k, m \beta ) \rangle$ $(k \in \mathbf{Z} )$ is a consistent solution of the Schwinger-Dyson equations.
}. See Fig.~\ref{fig-SD} for the $m=3$ case.

Recall that the Schwinger-Dyson equations provide the correlations between the loop operators. On the other hand, the Polyakov loop effective potential \eqref{F-expand} also provides the correlations of the Polyakov loops $u_n$ after the integration of other fields. Therefore, the above properties of the Schwinger-Dyson equations should be reproduced by the Polyakov loop effective potential \eqref{F-expand}. Hence, the Polyakov loop effective potential satisfies the scaling relation
\begin{align} 
	\left. V(  \beta,  \{ u_{n} \}) \right|_{u_n=0~(n  \notin m \mathbf{Z}),~ u_{mn} \to u_{n} } =
	  V( m \beta, \{ u_n \})  ,
			\label{F-scaling-app}
\end{align} 
 which ensures that, if we set $u_n(\beta)=0$ ($n \notin m \mathbf{Z}$), the correlations of the Polyakov loops $u_{mn}(\beta)$ are equivalent to those of $u_{n}(m\beta)$. This is the derivation of the scaling relation of the effective action \eqref{F-scaling}.

\subsubsection{Scaling relation in systems coupled to fermions}
\label{app-scaling-derivation-fermion}

If adjoint fermions $\Psi$ are coupled, the scaling relation \eqref{F-scaling}(= \eqref{F-scaling-app}) are modified as follows.
Suppose that the fermions $\Psi$ are inserted to the loop operator $ W(C_{n},\beta) $. Since the fermions satisfy the anti-periodic boundary condition $\Psi(\tau+\beta)=-\Psi(\tau)$, the Schwinger-Dyson equation for $ W(C_{n},\beta) $ derived from the variation of the fermion $\Psi+ \delta \Psi $ is schematically written as 
\begin{align} 
    \sum_{C'_n} g' \left\langle  W(C'_n,\beta) \right\rangle =& \sum_k \sum_{C''_{n-k},~C'''_k}  (-1)^k  \left\langle W(C''_{n-k},\beta) \right\rangle \left\langle W(C'''_{k},\beta) \right\rangle   + O(1/N^2).
    \label{SD-equation-fermions}
\end{align}
Note that additional sign factors coming from the ordering of $\delta \Psi$ may appear in this equation but they are not important in our arguments, and we have omitted them.
Here, the sign factor $(-1)^k$ appears in the right-hand side, since the splitting into the loop $C'''_k$ occurs through the relation $\Psi(\tau+ k\beta)=(-1)^k \Psi(\tau)$. We will see shortly that this sign factor requires a modification of the scaling relation \eqref{F-scaling-app}, depending on whether $m$ is odd or even\footnote{If the fermions are periodic $\Psi(\tau+ \beta)=\Psi(\tau)$, this sign factor does not appear and the scaling relation is identical to the bosonic one. Such fermions appear when imaginary chemical potentials are appropriately tuned. }.

Let us consider the odd $m=2l+1$ case. If we set all the expectation values of the open loop operators to zero except $ \langle W(C_{mk},\beta) \rangle$ ($k \in  \mathbf{Z}$, $m=2l+1$, $l \in \mathbf{N}$ ), the Schwinger-Dyson equation for $ \langle W(C_{mk},\beta) \rangle$ becomes 
\begin{align} 
    \sum_{C'_{mn}} g' \left\langle  W(C'_{mn},\beta) \right\rangle =& \sum_k \sum_{C''_{m(n-k)},~C'''_{mk}}  (-1)^{k}  \left\langle W(C''_{m(n-k)},\beta) \right\rangle \left\langle W(C'''_{mk},\beta) \right\rangle   + O(1/N^2).
    \label{SD-equation-fermions-odd}
\end{align}
This equation is equivalent to that for $ \langle W(C_{k}, (2l+1)\beta) \rangle$ at the inverse temperature $(2l+1)\beta$, since the sign factor has become $(-1)^{ (2l+1)k}=(-1)^{k}$.
This property requires that the Polyakov loop effective potential satisfies the following relation
\begin{align} 
    &	\left. V(  \beta,  \{ u_{n} \}) \right|_{u_n=0~(n  \notin (2l+1) \mathbf{Z}),~ u_{(2l+1)n} \to u_{n} } =
          V( (2l+1) \beta, \{ u_n \})   .     \label{F-scaling-fermion-odd}
\end{align} 
This is similar to the scaling relation \eqref{F-scaling-app} in the bosonic systems.

For even $m=2l$, if we set all the expectation values of the open loop operators to zero except $ \langle W(C_{mk},\beta) \rangle$ ($k \in  \mathbf{Z}$, $m=2l$, $l \in \mathbf{N}$ ), the Schwinger-Dyson equation \eqref{SD-equation-fermions} becomes
\begin{align} 
    \sum_{C'_{mn}} d' \left\langle  W(C'_{mn},\beta) \right\rangle =& \sum_k \sum_{C''_{m(n-k)},~C'''_{mk}}   \left\langle W(C''_{m(n-k)},\beta) \right\rangle \left\langle W(C'''_{mk},\beta) \right\rangle   + O(1/N^2),
\end{align}
where the sign factor disappears, since $(-1)^{ (2l)k}=1$. This is not the Schwinger-Dyson equation for $ \langle W(C_{k}, 2l\beta) \rangle$ at the inverse temperature $2l \beta$ but that in the same system with periodic fermions $\Psi(\tau+ 2l\beta)=\Psi(\tau)$. Thus, the partition function in this case is not a thermal one but the Witten index $\Tr \left( (-1)^{F}e^{-2l\beta H} \right)$ \cite{Witten:1982df}. Then, the Polyakov loop effective potential has to satisfy 
\begin{align} 
    &	\left. V(  \beta,  \{ u_{n} \}) \right|_{u_n=0~(n  \notin 2l \mathbf{Z}),~ u_{2ln} \to u_{n} } =
    V_P( 2l \beta, \{ u_n \}),      
    \label{F-scaling-fermion-even}
\end{align} 
where $V_P$ is the effective potential for the system with the periodic fermions. In this way, the scaling relation \eqref{F-scaling-app} has been modified.

In Sec.~\ref{subsec-Polyakov-loop-action-scaling}, we have argued that the scaling property of the effective action in the bosonic system constrains the expansion coefficients in Eq.~\eqref{F-expand} as shown in Eq.~\eqref{a-scaling}. Similarly, in the system coupled to the fermions, the scaling properties \eqref{F-scaling-fermion-odd} and \eqref{F-scaling-fermion-even} require the relations:
\begin{align} 
 \partial_\beta \epsilon =0, \quad
   a_{2n+1}(\beta)=a_{1}((2n+1)\beta), \quad  a_{2n}(\beta)=a_{2}(n\beta), \quad \cdots.
   \label{a-scaling-fermion}
\end{align}
Note that $a_2(\beta) = a_{P1}(2\beta) $, where $a_{P1}$ is the coefficient of $|u_1|^2$ in the effective potential $V_P$.  
Thus, the expansion \eqref{F-expand} of the effective potential for small $u_n$ becomes
\begin{align} 
    V(\beta,\{ u_n \})=&\epsilon + \sum_{n=0}^\infty a_{1}((2n+1)\beta)  |u_{2n+1}|^2
    + \sum_{n=1}^\infty a_{2}(n\beta)  |u_{2n}|^2
    + O(u_n^3) .
    \label{F-expand-scaling-fermion}
\end{align}
Hence, the thermodynamics of the system for small $u_n$ is controlled by the two functions $a_1(\beta)$ and $a_2(\beta)$.

\subsection{Confinement/Deconfinement transition}
\label{subsec-phase-general}

\begin{figure}
    \begin{center}
        \begin{tabular}{ccc}
            \begin{minipage}{0.33\hsize}
                \begin{center}
                    \includegraphics[scale=0.5]{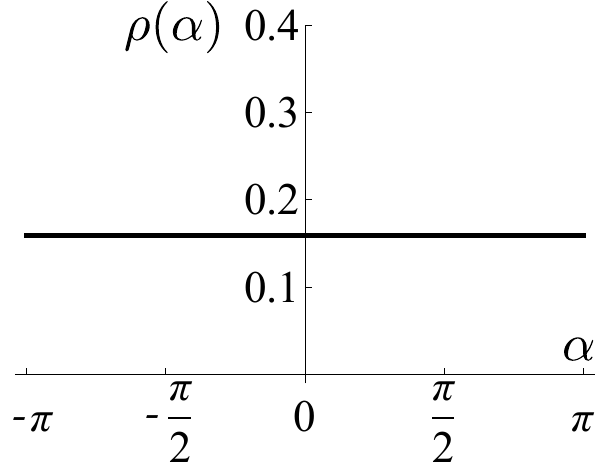}\\
                    uniform distribution
                \end{center}
            \end{minipage}
            \begin{minipage}{0.33\hsize}
                \begin{center}
                    \includegraphics[scale=0.5]{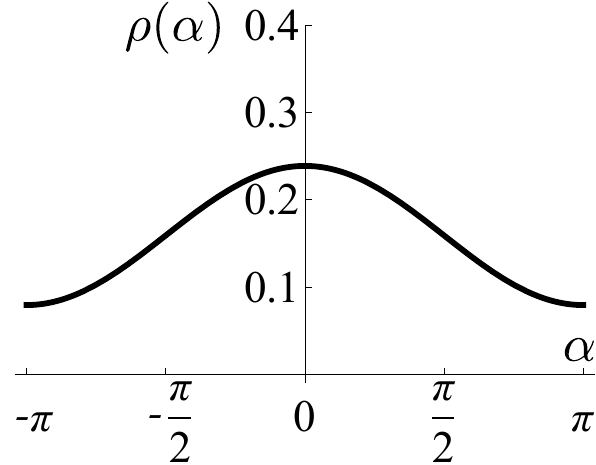}\\
                    non-uniform distribution
                \end{center}
            \end{minipage}
            \begin{minipage}{0.33\hsize}
                \begin{center}
                    \includegraphics[scale=0.5]{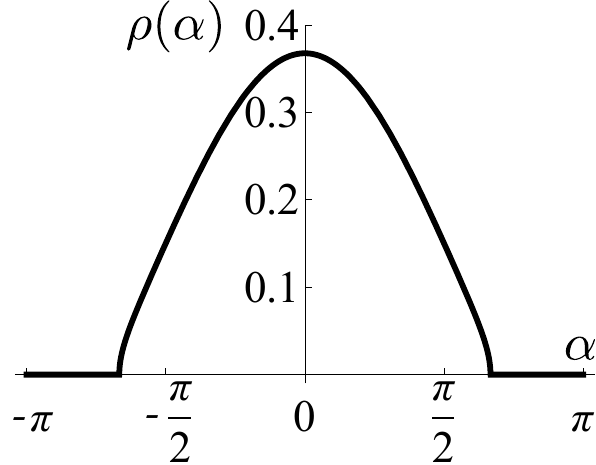}\\
                    gapped distribution
                \end{center}
            \end{minipage}
        \end{tabular}
        \caption{
            Three typical eigenvalue distributions $\rho(\alpha)$ of $A_\tau$ \eqref{rho}.
            In usual gauge theories, the uniform distribution (the left panel) is favored at low temperatures, while the gapped distribution (the right panel) is favored at high temperatures.
            Depending on the models, the non-uniform distribution (the center panel) appear at middle temperatures near the Hagedorn temperature $T_H$.
        }
        \label{fig-rho}
    \end{center}
\end{figure}

Using the effective potential \eqref{F-expand-scaling}, we discuss some general properties of the phase structure of gauge theories. For simplicity, we consider the cases where fermion does not exist. Suppose that the coefficient $ a_1 (\beta) $ of $|u_1|^2$ is positive near $T=0$. Then, through Eq.~\eqref{a-scaling}, the coefficients $a_n (\beta) =a_1(n\beta)$ of all $|u_n|^2$ are also positive there. Thus, $u_n=0$ ($\forall n$) is a stable solution, and the eigenvalue density of $A_\tau$ is obtained from Eq.~\eqref{rho-un} as
\begin{align} 
\rho(\alpha)=\frac{1}{2\pi}.
\end{align}
Therefore the eigenvalues distribute uniformly as plotted in Fig.~\ref{fig-rho} (the left panel), and the $Z_N$ center symmetry is preserved.

Since $u_1=0$, this is a confinement phase. In this case, the free energy $F$ becomes\footnote{In large-$N$ theories, the  saddle-point approximation would work, and the effective potential \eqref{F-expand} of a saddle-point solution can be regarded as the pseudo free energy.}
\begin{align} 
  F/N^2:=\frac{1}{\beta N^2} \log Z =  V(\beta,\{ u_n =0\})=\epsilon,
\end{align}
which is a constant independent of temperature. This property is called large-$N$ volume independence in confinement phase \cite{Eguchi:1982nm, Gocksch:1982en}. Since the free energy is independent of temperature, the entropy is also zero (more precisely it is $O(1)$).

Suppose that, as temperature increases from zero, $ a_1(\beta)$ reaches zero at $T=T_H$,
\begin{align} 
a_1(1/T_H)=0  .
\end{align}
Then the configuration $u_1=0$ becomes unstable beyond $T=T_H$. $T_H$ is called the Hagedorn temperature.
Since $a_1>0$ for $0 < T < T_H$, $a_n$ ($n \ge 2$) have to be positive at $T=T_H$ from Eq.~\eqref{a-scaling}. Thus, it might naively seem that, at $T \simeq T_H$ (but $T>T_H $), the configuration $u_1=1$ and $u_n=0$ ($n \ge 2$) may appear to reduce the effective potential \eqref{F-expand-scaling}. (Note that $|u_n|\le 1$ is always satisfied from Eq.\eqref{polyakov_static_diag}.) However, this configuration clearly does not satisfy the condition $\rho(\alpha)\ge 0 $ from Eq.~\eqref{rho-un}. Therefore, $T>T_H$ generally requires a detailed analysis as done in Refs.~\cite{Aharony:2003sx, Liu:2004vy, AlvarezGaume:2005fv, Okuyama:2017pil}, and we skip it here. It is known that, when $T$ is sufficiently higher than $T_H$, the eigenvalues typically collapse as shown in the right panel of Fig.~\ref{fig-rho}. In this phase, $u_1\neq 0$ and the $Z_N$ center symmetry is broken. Thus, it can be considered as a deconfinement phase. There, the entropy is $O(N^2)$, since the free energy $F$ would have a nontrivial temperature dependence. Hence, through the phase transition, the entropy suddenly increases from $O(1)$ to $O(N^2)$.

Also, an intermediate phase between these two phases (the center of Fig.~\ref{fig-rho}) may appear near $T=T_H$. It is easy to show that $u_1 \neq 0$ in this phase, and it is also a kind of the deconfinement phase, but unlike the high temperature phase (the right panel of Fig.~\ref{fig-rho}), there is no gap in the eigenvalue distribution. To distinguish these two deconfinement phases, the intermediate phase is called the non-uniform phase and the high temperature one is called the gapped phase. The transition between these two phases is known as the Gross-Witten-Wadia transition \cite{Gross:1980he, Wadia:2012fr}.

\subsubsection{$u_m$ in the confinement phase}
\label{app-scaling-no-fermion-um-conf}

The scaling relation also constrains observables. As an example, we evaluate  $|u_n|$ in the confinement phase.
In the confinement phase, the Polyakov loops $|u_n|$ are all zero in the large-$N$ limit. However, we can evaluate the leading $1/N$ correction. In the effective potential \eqref{F-expand-scaling}, the higher order terms with respect to $u_n$ are negligible in the confinement phase, and the quadratic term $a_n|u_n|^2$ would dominate. Since this is approximately a gaussian integral in the path integral \eqref{Z} \cite{Aharony:2003sx}, we obtain \cite{Azuma:2014cfa}
\begin{align} 
	\langle |u_n| \rangle =  \frac{1}{2N} \sqrt{\frac{\pi}{\beta a_n}} \left( 1+ O\left(\frac{1}{N^2} \right) \right) .
	\label{u_n-conf}
\end{align}
This expression would be reliable if the system is sufficiently below the transition point $a_n=0$, and all $u_n$ are small.
In bosonic systems, by applying the scaling relation for $a_n$ \eqref{a-scaling} to Eq.~\eqref{u_n-conf}, we obtain
\begin{align} 
	\langle | u_n (\beta) | \rangle= \sqrt{n} \langle | u_1 (n \beta) | \rangle   , \quad (\text{confinement phase in bosonic system}) .
	\label{u_n-conf-2}
\end{align}
For the systems coupled to fermions, since the scaling relation is modified as per Eq.~\eqref{a-scaling-fermion}, this becomes 
\begin{align} 
    \langle | u_{2l+1} (\beta) | \rangle=& \sqrt{2l+1} \langle | u_1 ((2l+1) \beta) | \rangle,\qquad 
    \langle | u_{2l} (\beta) | \rangle= \sqrt{l} \langle | u_2 (l \beta) | \rangle.
    \label{u_n-conf-fermion} 
\end{align}
Therefore, the scaling relation predicts correlations between different $u_n$.
Note that, in $d$-dimensional ($d\ge 2$) quantum field theories, $a_n$ \eqref{F-expand} is proportional to the volume of the space. 
Thus, when the volume is infinity, $|u_n|$ becomes zero by Eq.~\eqref{u_n-conf}. However,  $|u_n|^2$ times the volume would be finite and Eqs.~\eqref{u_n-conf-2} and \eqref{u_n-conf-fermion}  provide non-trivial predictions.

We should emphasize that these relations should be satisfied in any large-$N$ gauge theories with adjoint matters. Indeed, we will see that the relation \eqref{u_n-conf-2} holds in the bosonic BFSS model, which would be a strongly coupled system, by using a Monte-Carlo calculation in Sec.~\ref{confinement_MC}. 
In this way, the scaling relation is useful to predict observables.

\section{$Z_m$ phases in the gauged free matrix quantum mechanics}
\label{sec-free-MQM}

We have studied the usual confinement/deconfinement transitions in large-$N$ gauge theories by using the effective action \eqref{F-expand-scaling} in the previous section. From now, we discuss that, once imaginary chemical potentials are introduced, $ a_1(\beta)$ behaves unusually and the above discussions are not applicable. Then, $Z_m$-type deconfinement phases ($Z_m$ phases) can appear, which are different from the typical phases in large-$N$ gauge theories. Furthermore, the scaling relation of the effective potential plays an important role in determining the phase transition temperatures and physical quantities of the $Z_m$ phases. In the following sections, we will demonstrate this by studying several models.\\

First, we consider the $D$-dimensional SU($N$) gauged free matrix quantum mechanics \cite{Aharony:2003sx},
\begin{align}
    \label{action-Free-MQM}
    S=         & 
    \int_0^{\beta} \hspace{-2mm} d\tau  
    \Tr 
    \left\{ 
 \sum_{I=1}^D  \frac{1}{2}
    \left(
    D_\tau X^I \right)^2+
    \frac{1}{2} M^2 X^{I2} 	\right\}.
\end{align}
Here, $X^I$ ($I=1,\cdots,D$) is a traceless $N \times N$ hermitian matrix and $M$ is the mass, and $ D_\tau:= \partial_\tau -i[A_\tau, ]$ is the covariant derivative of SU$(N)$.
This is a toy model for $N$ (toy) D0-branes in $D$-dimensional space, where each of the diagonal components of $X^I$ represents the position of the D0-brane. Although this model is very simple, it is known to undergo the large-$N$ confinement/deconfinement transition \cite{Aharony:2003sx} and is useful to benchmark our proposals about the scaling properties and $Z_m$ phases.

To see $Z_m$ phases, we will introduce the imaginary angular velocities (the chemical potentials for the angular momenta, which is also related to $R$-charge chemical potentials in supersymmetric gauge theories \cite{Hawking:1999dp}). For this purpose, it is convenient to define $Z^I:= \left( X^I+i X^{\tilde{D}+I} \right)/\sqrt{2} $, ($I=1,\cdots, \tilde{D}$, $2\tilde{D} \le D$). Then, the action \eqref{action-Free-MQM} is invariant under the rotation on the $(I,\tilde{D}+I)$ plane: $Z^I \to e^{i \theta} Z^I$, ($\theta$= const.), and the angular momenta are conserved. Thus, we can introduce the (real) chemical potentials (angular velocities) $\Omega$ to the Euclidean action \eqref{action-Free-MQM}, and it becomes\cite{Morita:2010vi, Azuma:2023pzr}
\begin{align}
    \label{action-Free-MQM-omega}
    S=         & 
    \int_0^{\beta} \hspace{-2mm} dt  
    \Tr 
    \left\{ 
    \sum_{I=1}^{\tilde{D}}
    \left(
    D_t -\Omega \right) Z^{ I \dagger}
    \left(
    D_t +\Omega \right) Z^I  + M^2 Z^{ I \dagger}	Z^I 
    + \sum_{I=2\tilde{D}+1}^D \frac{1}{2}
    \left(
    D_t X^I \right)^2+
    \frac{1}{2} M^2 X^{I2} 	\right\}.
\end{align}
For simplicity, we have introduced a common angular velocity $\Omega$ for the $\tilde{D}$ planes\footnote{As can be seen from the form of the action \eqref{action-Free-MQM-omega}, the action is not real for a real $\Omega$. Therefore, systems with finite angular velocities generally have a sign problem in Monte-Carlo calculations. We can avoid this by taking $\Omega$ pure imaginary.}.

In this model we can easily perform the path integrals for $X^I$ and $Z^I$ exactly, and obtain the Polyakov loop effective potential \eqref{F-expand} at large $N$ \cite{Morita:2010vi, Azuma:2023pzr},
\begin{align}
    \label{effective-action-free-MQM}
    V(\beta, \{ u_n \})= &  \epsilon + \sum_{n=1}^{\infty} a_n |u_n|^2 ,
\end{align}
where
\begin{align}
    \epsilon= & \frac{D M}{2},
    \label{e-free-MQM} \\
    a_n= & \frac{1}{n\beta} \left[
        1-\tilde{D}  e^{- n\beta M }(e^{n\beta \Omega}+e^{-n\beta \Omega})- (D-2\tilde{D})  e^{- n\beta M }
    \right].
    \label{an-free-MQM-Re}
\end{align}
This result is for $ |{\rm Re} (\Omega) |<M$, and the path integral diverges when $ |{\rm Re} (\Omega) | \ge M$. This divergence indicates that the system is unstable at $ |{\rm Re} (\Omega) | \ge M$. 

Let us confirm that the obtained effective potential \eqref{effective-action-free-MQM} is consistent with the scaling relation \eqref{F-expand-scaling}. In Eq.~\eqref{e-free-MQM}, $\epsilon$, which is just the zero-point energies of $X^I$ and $Z^I$, is independent of temperature. Also, Eq.~\eqref{an-free-MQM-Re} shows that, in $a_n$, $n$ and $\beta$ appear only in the pair $\beta n$, and the relation \eqref{a-scaling} is realized. Therefore, the scaling relation \eqref{F-expand-scaling} is satisfied in this model.

\subsection{Phase structure of the free matrix quantum mechanics at $\Omega=0$}
\label{subsec-free-MQM-zero}

We discuss the phase structure of the effective potential \eqref{effective-action-free-MQM}. For simplicity, we first study the $\Omega=0$ case. As discussed in section \ref{subsec-phase-general}, the phase structure is determined by $a_1$, and, at $\Omega=0$, it becomes 
\begin{align}
    a_1(T)= & \frac{1}{\beta} \left(
        1-D  e^{-\beta M} 
    \right).
    \label{a1-free-MQM-0}
\end{align}
Since $ a_1$ is positive near $T=0$, the system is confined at low temperatures.

As temperature increases, $a_1$ reaches zero and the system is deconfined. The transition temperature can be calculated by solving $a_1(T)=0$, and we obtain
\begin{align} 
  T_H:=\frac{M}{\log D}.
  \label{TH-free-MQM}
\end{align}
This is the Hagedorn temperature of the model \eqref{action-Free-MQM}.
Beyond $T=T_H$, $a_1(T)$ monotonically decreases, and the deconfinement phase is always stable.

\begin{figure}
    \begin{center}
        \begin{tabular}{cc}
            \begin{minipage}{0.5\hsize}
                \begin{center}
                    \includegraphics[scale=0.7]{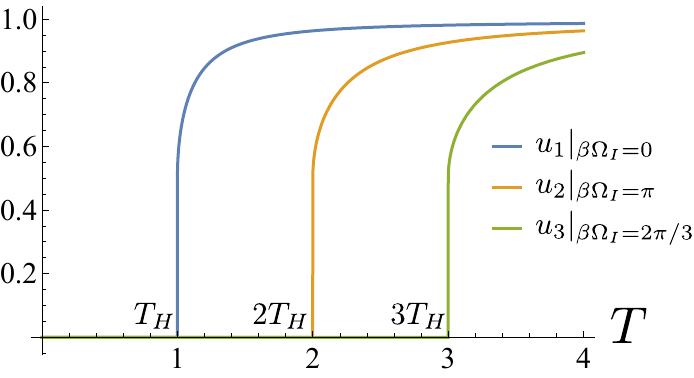}\\
                \end{center}
            \end{minipage}
            \begin{minipage}{0.5\hsize}
                \begin{center}
                    \includegraphics[scale=0.7]{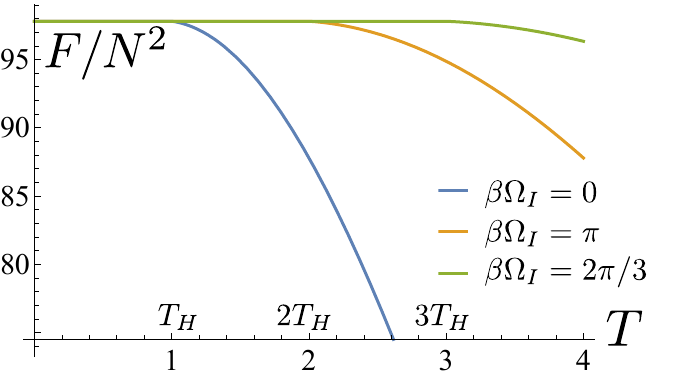}\\
                \end{center}
            \end{minipage}
        \end{tabular}
        \caption{$u_n(T)$ and $F(T)$ in the free matrix quantum mechanics \eqref{action-Free-MQM-omega} with an imaginary chemical potential $\Omega_I$. The three curves show the results at three different imaginary chemical potentials:  $\beta \Omega_I =0$, $2\pi/3$ and $\pi$. There, transitions from the confinement phase to the $Z_1$ phase ($\beta \Omega_I =0$), the $Z_2$ phase ($\beta \Omega_I =\pi$) and the $Z_3$ phase ($\beta \Omega_I =2\pi/3$) occur at $T=T_H$, $2T_H$ and $3T_H$, respectively. $u_n$ satisfy the relation $u_1(T)|_{Z_1} =u_2(T/2)|_{Z_2}= u_3(T/3)|_{Z_3}$. Also, the free energies in the three phases are related by the relation $F(T)|_{Z_1} =F(T/2)|_{Z_2}= F(T/3)|_{Z_3}$. These are the consequence of the scaling relaiton \eqref{F-scaling}. Note that these plots are obtained from Eqs.~\eqref{un-Z1-free-MQM} and \eqref{F-Z1-free-MQM}, and they are not accurate away from the phase transition points because of the approximation \eqref{large-D}.  (To improve the approximation \eqref{large-D}, $D=50$ is used in these plots. We have taken $M=\log D$ so that $T_H=1$ in Eq.~\eqref{TH-free-MQM}).
        }
        \label{fig-free-MQM-F}
    \end{center}
\end{figure}

To understand the properties of the phases, we evaluate the free energy $F$ and the Polyakov loops $u_n$.
In the region $T<T_H$, $a_n>0$ and, as discussed in Sec.~\ref{subsec-phase-general}, we obtain
\begin{align} 
  F/N^2=\epsilon, \quad u_n=O(1/N).
\end{align}
These results are plotted in Fig.~\ref{fig-free-MQM-F}.

On the other hand, the analysis at $T>T_H$ is generally difficult \cite{Aharony:2003sx}. 
However, we can use an approximation near $T=T_H$ \cite{Azuma:2023pzr}.
By substituting $T=T_H+\Delta T$ to $a_n=a_1(T/n)$, we obtain
\begin{align} 
  a_n = \frac{1}{n \beta} \left(1- \left( \frac{1}{D} \right)^{n-1}  D^{\frac{n \Delta T}{T_H}  }  \right) .
\end{align}
Thus,  if we regard $D$ as a large number, it approximately becomes
\begin{align} 
    a_1 = \frac{1}{ \beta} \left(1-   D^{\frac{ \Delta T}{T_H}  }  \right), \qquad
    a_n \simeq \frac{1}{n \beta} \qquad (n \ge 2),
    \label{large-D}  
\end{align}
for a small $\Delta T$. This approximation is valid up to $\Delta T \sim T_H/2$, where the correction to $a_2$ becomes $O(1)$. With this approximation, we can perform the path integral of the effective potential \eqref{effective-action-free-MQM} and we obtain the solution at $T>T_H$ ($a_1<0$) \cite{Rossi:1996hs, Liu:2004vy, AlvarezGaume:2005fv, Okuyama:2017pil, Azuma:2023pzr}.
\begin{align} 
&u_1=u_{-1}=   \frac{1+w}{2}, \qquad u_{ n} =u_{-n} = \frac{w^2}{n-1} P^{(1,2)}_{n-2}(2w-1) ,\quad (n \ge 2) , \nonumber \\
&w:= \sqrt{\frac{-a_1}{1-a_1}},
\label{un-Z1-free-MQM}
\end{align}
where $P^{(1,2)}_{n-2}(z)$ denotes the Jacobi polynomial. Since $u_1 \neq 0$, this solution is a deconfinement phase. The eigenvalue density \eqref{polyakov_static_diag} is given by  
\begin{align} 
  \rho (\alpha)= \frac{g_0}{\pi} \cos \left(\frac{\alpha}{2}\right) {\rm Re}  \sqrt{\frac{1}{g_0}-\sin^2\frac{\alpha}{2} } , \quad g_0:=\frac{1}{1-w}, 
  \label{density-free-MQM}
\end{align}
and it is gapped as plotted in the right panel of Fig.~\ref{fig-rho}. Here, the gauge in which the peak of the eigenvalue distribution is at $\alpha=0$ has been taken in Eq.~\eqref{un-Z1-free-MQM}. The temperature dependence of $u_1$ is plotted in the left panel of Fig.~\ref{fig-free-MQM-F}. Note that the non-uniform phase does not appear in this model.

The free energy of this solution is also calculated through the saddle point approximation \cite{Liu:2004vy, AlvarezGaume:2005fv, Okuyama:2017pil, Azuma:2023pzr}
\begin{align} 
  F/N^2= \epsilon - \frac{1 }{2\beta} \left[\frac{w}{1-w} + \log (1-w)   \right].
  \label{F-Z1-free-MQM}
\end{align}
This result is plotted in Fig.~\ref{fig-free-MQM-F} (the right panel). Since the derivative of the free energy at $T=T_H$ is discontinuous, it is a first order transition \cite{Aharony:2003sx}.

\subsection{Phase structure of the free matrix quantum mechanics with real angular velocities}
\label{subsec-free-MQM-real}

Before studying the phase structure with imaginary angular velocities, we briefly show the results for real angular velocities $\Omega$. For simplicity, we take $D=2$ and $\tilde{D}=1$. Then, the phase structure is determined by $a_1$, which now becomes
\begin{align}
    a_1(T,\Omega)= & \frac{1}{\beta} \left[
        1-2  e^{-M\beta} \cosh(\beta \Omega) 
    \right].
    \label{a1-free-MQM}
\end{align}
Again, $ a_1 >0$  near $T=0$ and $\Omega=0$, and the transition occurs on the curve $a_1(T,\Omega)=0$. This curve is plotted in Fig.~\ref{fig-free-MQM} (the left panel), and the transition temperature decreases with angular velocity. The system is destabilized when $|\Omega| \ge M$.

\subsection{Phase structure of the free matrix quantum mechanics with imaginary angular velocities}
\label{subsec-free-MQM-Im}

We set $\Omega=i \Omega_I$ ($\Omega_I \in {\mathcal R}$) in the effective potential \eqref{effective-action-free-MQM} and study the effect of the imaginary angular velocity. As in the previous subsection, we take $D=2$ and $\tilde{D}=1$ for simplicity. In this case, $a_n$ becomes
\begin{align}
    a_n(T,\Omega_I)= & \frac{1}{n\beta} \left[
        1-2  e^{-n\beta M} \cos( n\beta \Omega_I) 
    \right].
    \label{an-free-MQM}
\end{align}
This equation shows that the system has a periodicity $\Omega_I =\Omega_I+2\pi /\beta$ and symmetry $\Omega_I \to - \Omega_I$. This periodicity is due to the fact that the angular velocity is introduced in the partition function in the form $Z=\Tr\left[ \exp(-\beta(H-i\Omega_I J))\right]$ and the angular momentum $J$ is an integer. Because of this periodicity, it is convenient to vary $\beta \Omega_I$ when investigating the phase diagrams.

Now we examine the phase structure of this system. Near $T=0$, since $a_n >0 $, the confinement phase $u_n=0$ is stable. Then, the phase transition to the deconfinement phase is expected to occur at $a_1(T,\Omega_I)=0$. However, as is clear from Eq.~\eqref{an-free-MQM}, $a_1(T,\Omega_I)$ is always positive in the region $ \pi/2 < \beta \Omega_I < 3\pi/2$ where $\cos( \beta \Omega_I) <0 $. Therefore, $u_1=0$ is stable in this region even at high temperatures.

\begin{figure}
    \begin{center}
        \begin{tabular}{cc}
            \begin{minipage}{0.5\hsize}
                \begin{center}
                    \includegraphics[scale=0.6]{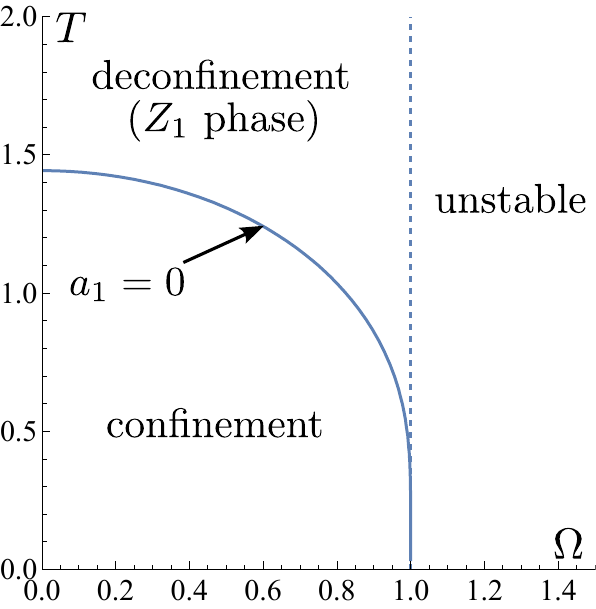}\\
                    $T$ vs. $\Omega$
                \end{center}
            \end{minipage}
            \begin{minipage}{0.5\hsize}
                \begin{center}
                    \includegraphics[scale=0.6]{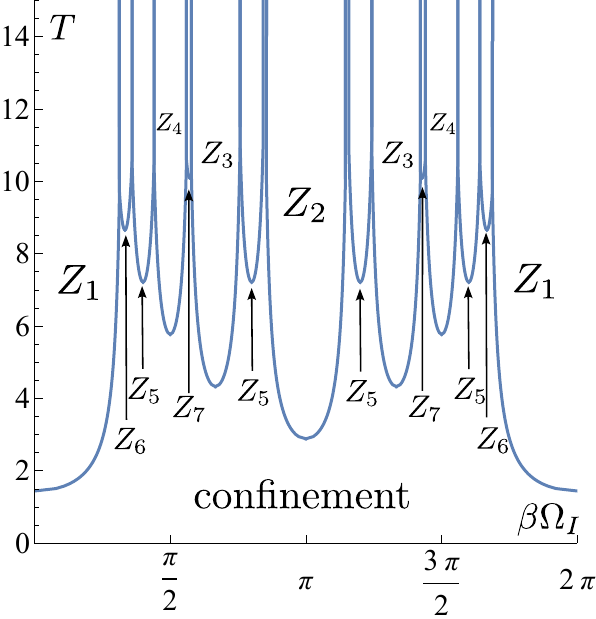}\\
                    $T$ vs. $\beta \Omega_I$
                \end{center}
            \end{minipage}
        \end{tabular}
        \caption{
Phase diagrams of the free matrix quantum mechanics \eqref{action-Free-MQM-omega} ($M=1$, $D=2$ and $\tilde{D}=1$) with a real angular velocity $\Omega$ (the left panel) and with an imaginary angular velocity $\Omega_I$ (the right panel). For the real angular velocity, only the conventional confinement phase and the deconfinement phase ($Z_1$ phase) appear, and the transition occurs on the curve $a_1=0$ \eqref{a1-free-MQM}. Also, the system is destabilized when $|\Omega| \ge M=1$. On the other hand, for the imaginary angular velocity, the $Z_m$ phases ($m=1,2,3,4,5,6$ and 7) appear and the transitions to these phases from the confinement phase occur on the curves $a_m=0$ \eqref{an-free-MQM}. Also, we speculate that the transitions between the $Z_m$ phase and the $Z_n$ phase occur on the curves $a_m=a_n$. 
        }
        \label{fig-free-MQM}
    \end{center}
\end{figure}

However, this is not the end of the story\footnote{In Ref.~\cite{Azuma:2023pzr}, the phase diagrams of the bosonic BFSS matrix model with imaginary chemical potentials were studied. There, only the curve $a_1=0$ was investigated, and $Z_m$ phases ($m \ge 2$) were not observed.}. In the region $ 3\pi/4 < \beta \Omega_I < 5\pi/4 $, while $a_1$ is positive, $a_2$ can be negative. Thus, $u_1= 0$ and $u_2 \neq 0$ are stable there. 

We investigate the situation $u_1= 0$ and $u_2 \neq 0$ in detail. For the sake of concreteness, we take $\beta \Omega_I =\pi$. In this case, $a_n$ becomes
\begin{align}
    a_n(T,\Omega_I=\pi/\beta)= & \frac{1}{n\beta} \left[
        1- (-1)^n 2  e^{-nM\beta}  
    \right].
    \label{an-free-MQM-pi}
\end{align}
Thus, $a_{2n+1} >0$ at any temperature for $\forall n$, and $u_{2n+1}=0$ is always stable. This means that the condition $u_{2n+1}=0$ imposed by hand on the left-hand side of the scaling relation \eqref{F-scaling} at $m=2$ is thermodynamically realized. Then, by substituting $u_{2n+1}=0$ and $m=2$ into to Eq.~\eqref{F-scaling}, the effective potential satisfies the following relation,
\begin{align} 
	\left. V(  \beta,  \{ u_{n} \},\Omega_I=\pi/\beta) \right|_{u_{2n+1}=0,~ u_{2n} \to u_{n}  } =
	  V( 2 \beta, \{ u_n \}, \Omega_I=\pi/\beta),
      \label{f-scaling-Z2}
\end{align} 
where the $\Omega_I$ dependence has been explicitly written. The effective potential on the right-hand side is at the inverse temperature $2\beta$, and it has a periodicity $\Omega_I=\Omega_I+2\pi/(2\beta)$.
By using this periodicity, $\Omega_I$ on the right-hand side can be eliminated,
\begin{align} 
	\left. V(  \beta,  \{ u_{n} \},\Omega_I=\pi/\beta) \right|_{u_{2n+1}=0,~ u_{2n} \to u_{n}  }=
	  V( 2 \beta, \{ u_n \}, \Omega_I=0)  .
      \label{F-Z2-free-MQM}
\end{align} 
Since the effective potential governs the thermodynamic properties of the system, this relation states that the thermodynamics for $(T,\beta\Omega_I)=(T,\pi)$ is equivalent to that for $(T,\beta\Omega_I)=(T/2,0)$.
Here the Polyakov loops are related by
\begin{align} 
    u_{2n}(T,\beta\Omega_I=\pi)=  u_{n}(T/2,\beta\Omega_I=0), \quad     u_{2n+1}(T,\beta\Omega_I=\pi)=  0.
\label{un-Z2-free-MQM}
\end{align}

\begin{figure}
    \begin{center}
        \begin{tabular}{ccc}
            \begin{minipage}{0.33\hsize}
                \begin{center}
                    \includegraphics[scale=0.5]{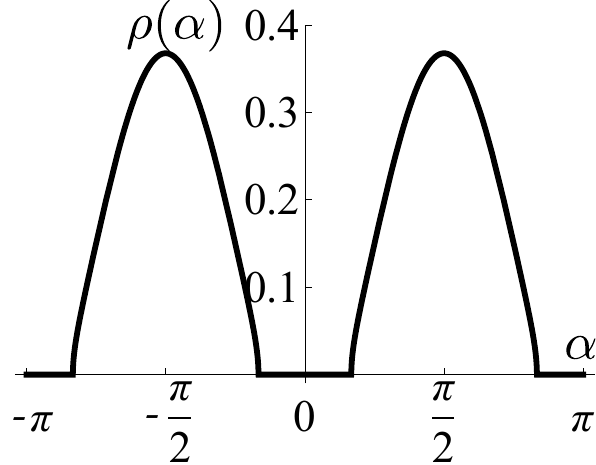}\\
                    $Z_2$ phase \\
                \end{center}
            \end{minipage}
            \begin{minipage}{0.33\hsize}
                \begin{center}
                    \includegraphics[scale=0.5]{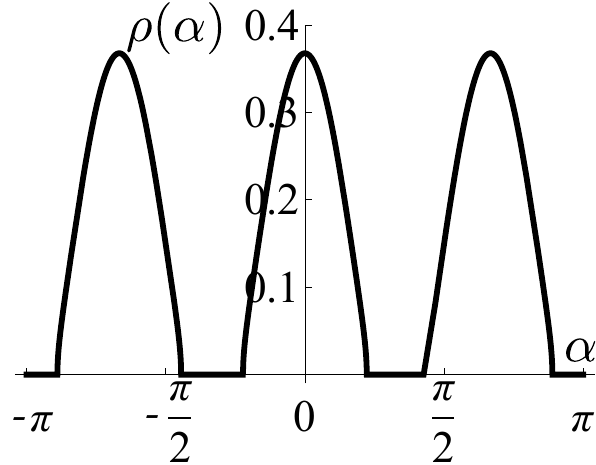}\\
                    $Z_3$ phase \\
                \end{center}
            \end{minipage}
            \begin{minipage}{0.33\hsize}
                \begin{center}
                    \includegraphics[scale=0.4]{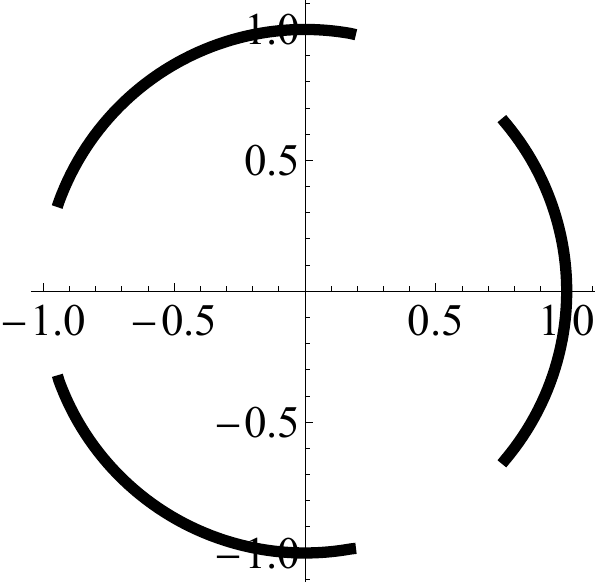}\\
                    $\{ e^{ i \alpha_k } \}$ of $Z_3$ phase \\
                \end{center}
            \end{minipage}
        \end{tabular}
        \caption{
            The eigenvalue density functions $\rho(\alpha)$ of the $Z_2$ phase (the left panel) and the $Z_3$ phase (the center panel) in the free matrix quantum mechanics \eqref{action-Free-MQM-omega}.
            The right panel is the scatter plot of $\{ e^{ i \alpha_k } \}$ ($k=1,\cdots,N$) for the $Z_3$ phase, and three cuts exist symmetrically.
         }
        \label{fig-rho-Z2}
    \end{center}
\end{figure}

The thermodynamics of the system for $\beta \Omega_I=0$ has already been investigated in Sec.~\ref{subsec-free-MQM-zero}, and the results there can be applied to the $\beta \Omega_I=\pi$ case.
For example, the free energy $F$ at $\beta \Omega_I=\pi$ is simply given by Eq.~\eqref{F-Z1-free-MQM} by replacing temperature $T \to T/2$. This result is plotted in Fig.~\ref{fig-free-MQM-F}, and the phase transition for $\beta \Omega_I=\pi$ occurs at $T=2T_H$.  Furthermore, the eigenvalue density $\rho(\alpha)$ of $A_\tau$ ($T>2T_H$) can also be obtained as shown in Fig.~\ref{fig-rho-Z2} (the left panel). (See Eq.~\eqref{rho-Zm} for the derivation.) It is two copies of the distribution \eqref{density-free-MQM} at $T/2$ with an appropriate rescaling. Since this configuration is $Z_2$ symmetric, we call this phase the $Z_2$ phase. This phase can be considered as a kind of deconfinement phase because the eigenvalue distribution is not uniform, and the $Z_N$ center symmetry is broken to $Z_2$. Hereafter, we call the usual deconfinement phase \eqref{un-Z1-free-MQM} for $\beta \omega_I=0$ ``the $Z_1$ phase" to distinguish it from the $Z_2$ symmetric deconfinement phase.\\

As we have seen, the thermodynamic properties at temperature $T$ for $\beta \Omega_I=\pi$ are equivalent to those at temperature $T/2$ for $\beta \Omega_I=0$ owing to the scaling relation \eqref{F-scaling}. To reach this conclusion, we have used only the stability of $u_{2n+1}=0$ and the periodicity $\beta \Omega_I = \beta \Omega_I+2\pi$. Therefore, this conclusion always holds for systems where these properties are satisfied, regardless of the details of the models. This is a strong consequence of the scaling relation \eqref{F-scaling}.\\

We have studied the $\beta \Omega_I =\pi$ case, and now we move to general $\Omega_I$. Again, the transition from the confinement phase to the $Z_2$ phase would occur on the curve $a_2=0$. These are plotted in Fig.~\ref{fig-free-MQM} (the right panel).

Similarly, the transition from the confinement phase to the phase with the $Z_m$ symmetric eigenvalue distribution (let us call it the $Z_m$ phase) would occur on the curve $a_m=0$ ($m=3,4,5,6$ and 7). All these curves are related through the scaling relation \eqref{a-scaling} as plotted in Fig.~\ref{fig-free-MQM}. Also, the eigenvalue density of the $Z_3$ phase is shown in Fig.~\ref{fig-rho-Z2} (the middle and right panels). In these phases, the $Z_N$ center symmetry is broken to $Z_m$.

Beyond the transition curves $a_m=0$, evaluating the phase diagram is more difficult because we cannot use the approximation \eqref{large-D} in general, and it is not easy to find a stable solution. We speculate that the $Z_m$ phase with the smallest $a_m$ is favored. This is because, in the $Z_m$ symmetric eigenvalue distribution, $|u_m|$ takes the largest value among $\{ |u_n| \}$ and this configuration would minimize the effective potential when $a_m$ is the smallest. If so, the transition between the $Z_m$ phase and the $Z_n$ phase occurs on the curve $a_m=a_n(<0)$. With this speculation, the phase diagram beyond the curves $a_m=0$ is plotted in Fig.~\ref{fig-free-MQM}.\\

We have seen that when $\beta \Omega_I=\pi$, the quantities in the $Z_2$ phase can be derived from those in the $Z_1$ phase for $\beta \Omega_I=0$ by using the relation \eqref{f-scaling-Z2}. The same is possible for other $Z_m$ phases.
From the scaling relation \eqref{F-scaling} and the periodicity of $\Omega_I$, the following relation is satisfied:
\begin{align} 
	\left. V(  \beta,  \{ u_{n} \}, \beta \Omega_I=2\pi k/m) \right|_{u_n=0~(n  \notin m \mathbf{Z}),~ u_{mn} \to u_{n} } =
	  V( m \beta, \{ u_n \}, \beta \Omega_I=0), \quad  k \in \mathbf{Z}.
      \label{F-Zm-free-MQM}
\end{align} 
Thus, if $u_n=0$ ($n  \notin m \mathbf{Z}$) is stable at $\beta \Omega_I=2\pi k/m$, this relation is thermodynamically favored, and the quantities at  $(T,\beta\Omega_I)=(T,2 k \pi/m)$ is equivalent to those at $(T/m, 0)$. Then, the phase transition between the confinement phase and the $Z_m$ phase occurs at $(T,\beta\Omega_I)=(mT_H,2 k \pi/m)$.
Again, we have only used the stability of $u_n=0$ ($n  \notin m \mathbf{Z}$) and the periodicity of $\Omega_I$, and this relation will be satisfied independently of the details of the models. In Appendix \ref{app-scaling-general}, we provide a generalization of this relation for various quantities at general $\Omega_I$.

\begin{figure}
    \begin{center}
        \begin{tabular}{cc}
            \begin{minipage}{0.5\hsize}
                \begin{center}
                    \includegraphics[scale=0.6]{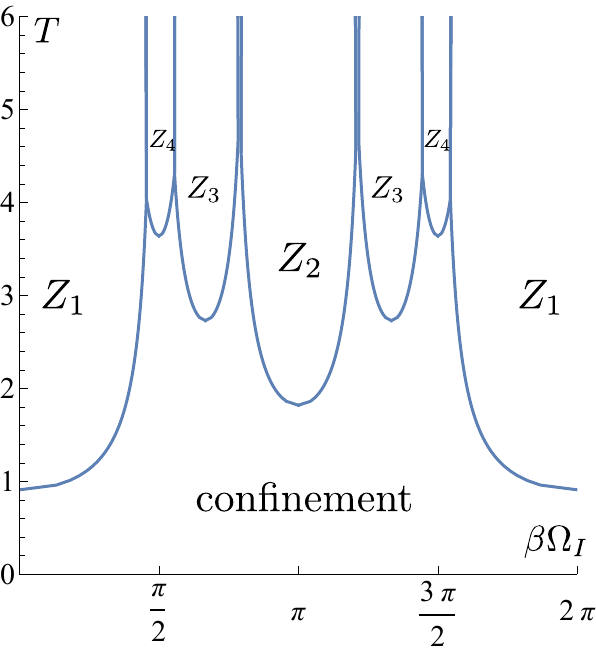}\\
                    $(D,\tilde{D})=(3,1)$ 
                \end{center}
            \end{minipage}
            \begin{minipage}{0.5\hsize}
                \begin{center}
                    \includegraphics[scale=0.6]{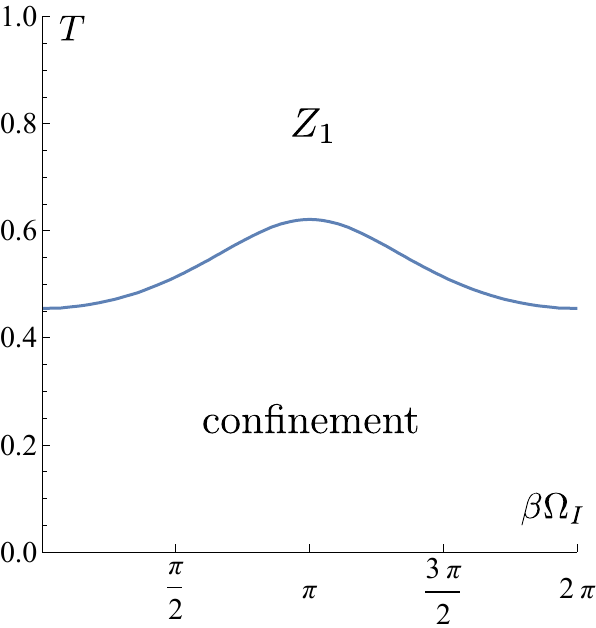}\\
                    $(D,\tilde{D})=(9,1)$ 
                \end{center}
            \end{minipage}
        \end{tabular}
        \caption{
Phase diagrams of the free matrix quantum mechanics \eqref{action-Free-MQM-omega} with imaginary chemical potentials $\Omega_I$ in $(D,\tilde{D})=(3,1)$ and $(9,1)$. We have taken $M=1$. The transitions from the confinement phase to the $Z_m$ phases occur on the curves $a_m=0$ \eqref{an-free-MQM-Re}. Similarly to the $(D,\tilde{D})=(2,1)$ case in Fig.~\ref{fig-free-MQM}, the phase diagrams beyond the transition curves $a_m=0$ are speculative. As the ratio $\tilde{D}/D$ decreases, the number of the $Z_m$ phases tends to decrease.
        }
        \label{fig-phase-free-MQM-d9d6}
    \end{center}
\end{figure}

So far we have considered the $(D,\tilde{D})=(2,1)$ case, and it is easily extended to general $(D,\tilde{D})$. The phase diagrams of $(D,\tilde{D})=(3,1)$ and $(D,\tilde{D})=(9,1)$ are shown in Fig.~\ref{fig-phase-free-MQM-d9d6}. 
In the $(D,\tilde{D})=(3,1)$ case, the $Z_m$ phases ($m=1,2,3$ and 4) appear. In the $(D,\tilde{D})=(9,1)$ case, only phase transitions from the confinement phase to the $Z_1$ phase occur. Generally, if $\tilde{D}$ is small compared to $D$ such that $a_1<a_m$ $(m \ge 2)$, only $Z_1$ phase would appear.\\

In this section, we have shown that the $Z_m$ phases appear by introducing imaginary angular velocities in the free matrix quantum mechanics \eqref{action-Free-MQM-omega}. Also, the physical quantities in the $Z_m$ phase obey the scaling relation \eqref{F-scaling}. As discussed below, $Z_m$ phases also appear in various large-$N$ gauge theories.

\section{$Z_m$ phases in the bosonic BFSS matrix model}
\label{sec-bBFSS}

It is worth asking whether the $Z_m$ phases and the scaling properties confirmed in the free matrix quantum mechanics appear in systems with interactions. For this purpose, we consider the $D+1$ dimensional SU($N$) pure Yang-Mills theory on a $D$-dimensional torus $T^D$,
\begin{align} 
    S=\frac{1}{g'^2}  \int_0^\beta dt \int_{T^D}d^Dx \Tr\left( F_{\mu\nu}^2 \right).
\label{action-YM-TD}
\end{align}
By taking a small volume limit on $T^D$, the theory would be reduced to the matrix quantum mechanics,
\begin{align}
    \label{action-BFSS}
    S
    =
    \int_0^{\beta} \hspace{-2mm} dt  
    \Tr 
    \Biggl\{ 
    \sum_{I=1}^D
    \frac{1}{2}
    \left(
    D_t X^I \right)^2
    -
    \sum_{I,J=1}^D \frac{g^2}{4} [X^I,X^J]^2
    \Biggr\},
\end{align}
where $X^I$ is the Kaluza-Klein zero mode of the spatial component of the gauge field $A^I$ in the original theory \eqref{action-YM-TD}. This model is called the bosonic BFSS matrix model \cite{Banks:1996vh}, which is a kind of the large-$N$ reduced model \cite{Eguchi:1982nm}. This model has rotational symmetries similar to the free matrix quantum mechanics \eqref{action-Free-MQM}\footnote{In the original $D+1$ dimensional theory \eqref{action-YM-TD}, the rotational symmetries are broken by the torus compactification. However, once the Kaluza-Klein reduction is done, the model \eqref{action-BFSS} recovers the rotational symmetries.}, and we can introduce the angular velocities as
\begin{align}
    \label{action-BFSS-J}
    S
    = & 
    \int_0^{\beta} \hspace{-2mm} dt  
    \Tr 
    \Biggl\{ 
    \sum_{I=1}^{\tilde{D}}
    \left(
    D_t -\Omega \right) Z^{I \dagger }
    \left(
    D_t +\Omega \right) Z^I 	+
    \sum_{I=2\tilde{D}+1}^D
    \frac{1}{2}
    \left(
    D_t X^I \right)^2
    -
    \sum_{I,J=1}^D \frac{g^2}{4} [X^I,X^J]^2
    \Biggr\}.
\end{align}
Here we have taken the common angular velocity $\Omega$ on the $\tilde{D}$ planes as we did in the free matrix quantum mechanics \eqref{action-Free-MQM-omega}.

Unlike the free matrix quantum mechanics, the Polyakov loop effective potential cannot be derived exactly due to the quartic interactions of $X^I$. However, it is known that this model can be analyzed approximately by using the large-$D$ expansion \cite{Hotta:1998en, Mandal:2009vz, Morita:2010vi} and the minimum sensitivity method \cite{Kabat:1999hp, Morita:2020liy, Azuma:2023pzr}. In this study, we use the minimum sensitivity method \cite{Stevenson:1981vj}. We will also solve the model by using Monte-Carlo calculations.

\subsection{Results from the minimum sensitivity method}

By using the minimum sensitivity method, the Polyakov loop effective potential at the two-loop order in the bosonic BFSS matrix model with angular velocities \eqref{action-BFSS-J} has been computed in Ref.~\cite{Azuma:2023pzr} (see Eqs.~(A.19)-(A.20) in Ref.~\cite{Azuma:2023pzr}). When $|u_n| \simeq 0$, the effective potential is given by
\begin{align}
    V(\{ u_n \},m_X,m_Z)= &  \epsilon (m_X,m_Z)+ \sum_{n=1}^\infty  a_n(\beta,\mu, m_X,m_Z) |u_n|^2 
    + \textrm{O} (u_n u_m u_{-n-m} )  .
    \label{effective-action-app}
\end{align}
Here
\begin{align}
    \epsilon (m_X,m_Z):= & \frac{(D-2\tilde{D}) m_X}{4}+\frac{\tilde{D} m_Z}{2} \nonumber \\
                   & +	\frac{\lambda}{8}
    \left[
        \frac{2\tilde{D}(2 \tilde{D}-1)}{m_Z^2}
        +\frac{4 \tilde{D}(D-2 \tilde{D})}{m_X m_Z}
        +\frac{(D-2 \tilde{D})(D-2 \tilde{D}-1)}{m_X^2}
        \right],
    \label{f0}
\end{align}
\begin{align}
    &n\beta a_n(\beta,\Omega, m_X,m_Z) := \nonumber \\
     & 1
    - \tilde{D} \left(
    1+\frac{n\beta m_Z}{2}
    \right) z^{n} (q^n+q^{-n})
    - (D-2\tilde{D}) \left(
    1+\frac{n\beta m_X}{2}
    \right) x^{n}  \nonumber \\
    &+ \frac{n\beta\lambda}{4m_Z^2}
    \left[ 2 \tilde{D} (\tilde{D}+1) z^{2n} + \tilde{D}(\tilde{D}-2) z^{2n}(q^{-2n }+q^{2n }) 
    +	2 \tilde{D} (2 \tilde{D}-1) z^{n}(q^{-n }+q^{n }) 
    \right]
    \nonumber                                                                             \\
         & + \frac{n\beta \lambda}{2m_X m_Z} \tilde{D}(D-2 \tilde{D})
    \left[ 2 x^n + (1+x^n) z^{n}(q^{-n }+q^{n }) 
    \right]
    \nonumber                                                                             \\
         & +\frac{n\beta \lambda}{4m_X^2}(D-2 \tilde{D})(D-2 \tilde{D}-1)  (x^{2n}+2x^{n})
    \label{An-bBFSS},
\end{align}
where $x := e^{-\beta m_X}$, $z := e^{-\beta m_Z}$, $q := e^{-\beta \Omega}$ and $\lambda:=g^2N$. Here $m_X$ and $m_Z$ are trial masses for $X^I$ and $Z^I$, respectively, and they are determined by the conditions:
\begin{align} 
    \frac{\partial V}{\partial m_X} =\frac{\partial V}{\partial m_Z}=0  .
    \label{minimum-F}
\end{align}
Note that this effective potential is consistent with the scaling properties \eqref{F-expand-scaling}, since $ \epsilon (m_X,m_Z)$ does not depend on temperature and, in $a_n$, $\beta$ and $n$ are only included in the form $n\beta$\footnote{From the analysis in Appendix A of Ref.~\cite{Azuma:2023pzr}, 
we can confirm that the coefficients of the third order of $u_n$ satisfy the scaling relation \eqref{F-scaling} as well.}.

Since the phase diagram with real angular velocities has been studied in Ref.~\cite{Azuma:2023pzr}, we now introduce the imaginary angular velocity as $\Omega=i \Omega_I $ and analyze only this case. First, we examine the behavior of the system at low temperatures. Since $a_1 >0$ at $T = 0$, $|u_n|=0$ is stable at sufficiently low temperatures and the system is in a confinement phase. When $|u_n|=0$, the conditions \eqref{minimum-F} become $ \partial \epsilon /\partial m_X =\partial \epsilon /\partial m_Z=0 $, and the solution is given by 
\begin{align}
    m_X=m_Z=m_0:=(D-1)^{1/3}\lambda^{1/3}.
    \label{m-2-loop}
\end{align}
These are the values of the trial masses $m_X$ and $m_Z$ in the confinement phase.

\begin{figure}
    \begin{center}
        \begin{tabular}{cc}
            \begin{minipage}{0.5\hsize}
                \begin{center}
                    \includegraphics[scale=0.6]{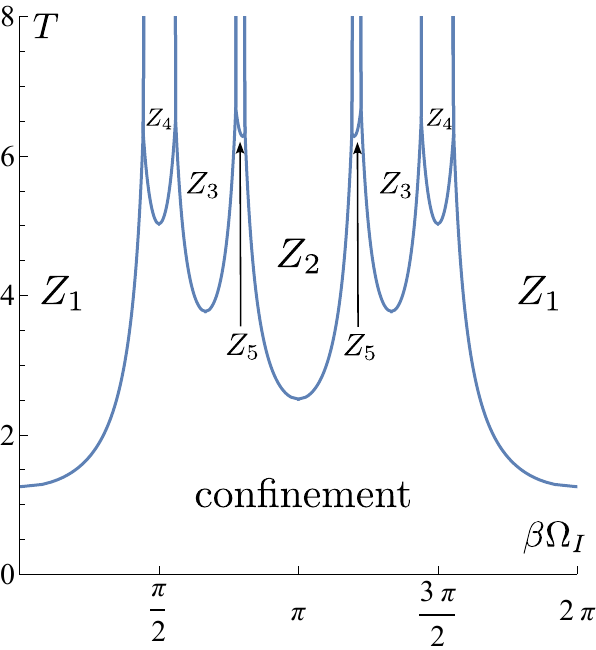}\\
                    $(D,\tilde{D})=(3,1)$
                \end{center}
            \end{minipage}
            \begin{minipage}{0.5\hsize}
                \begin{center}
                    \includegraphics[scale=0.6]{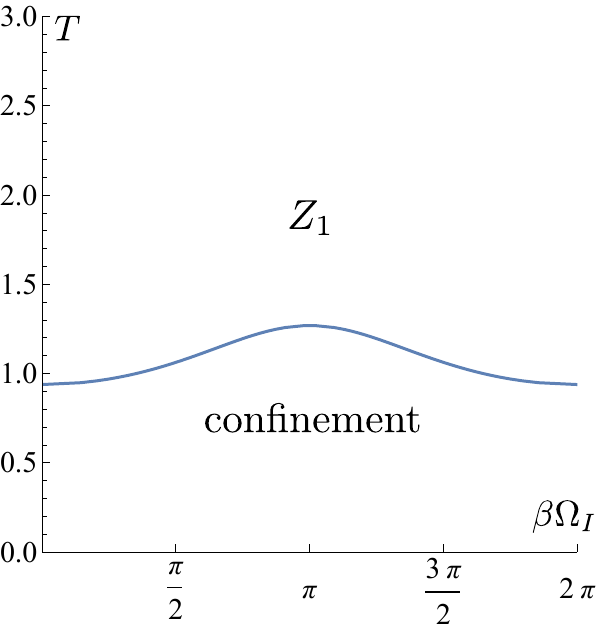}\\
                    $(D,\tilde{D})=(9,1)$
                \end{center}
            \end{minipage}
        \end{tabular}
        \caption{
Phase diagrams of the bosonic BFSS matrix model \eqref{action-BFSS-J} with imaginary chemical potentials $\Omega_I$ for $(D,\tilde{D})=(3,1)$ and $(D,\tilde{D})=(9,1)$. $\lambda=1$ has been taken. The transitions from the confinement phase to the $Z_m$ phases would occur on the curves $a_m=0$ given by Eq.~\eqref{critical-point}. The phases beyond the transition curves $a_m=0$ are highly speculative.}
        \label{fig-phases-bBFSS}
    \end{center}
\end{figure}

By using these obtained masses, $a_n$ in the confinement phase is determined as $a_n=a_n(\beta,\Omega_I, m_0,m_0) $ in Eq.~\eqref{An-bBFSS}. Then, similarly to the free matrix quantum mechanics case, the phase transitions from the confinement phase to $Z_m$ phases occur when\footnote{The details of the confinement/deconfinement phase transition of this model at $\Omega_I=0$ have been investigated in Refs.~\cite{Azuma:2014cfa, Azuma:2023pzr, Bergner:2019rca, Morita:2020liy}. It is a first order transition for smaller $D$ and would become second order for large $D$, ($D \ge 26$ at least \cite{Bergner:2019rca}). When it is a second order transition, the non-uniform phase sketched in Fig.~\ref{fig-rho} (the middle panel) appears as an intermediate stable phase \cite{Aharony:2003sx, AlvarezGaume:2005fv}. Then the scaling relation \eqref{F-scaling} predicts that the $Z_m$ type non-uniform phases also appear near the transition curve $a_m=0$. } 
\begin{align}
    a_m(\beta,\Omega_I, m_0,m_0) =0.
 \label{critical-point}
\end{align}
These equations determine the phase diagram at low temperatures, and the results for $(D,\tilde{D})=(3,1)$ and $(D,\tilde{D})=(9,1)$ are plotted in Fig.~\ref{fig-phases-bBFSS}. These are qualitatively similar to the phase diagrams of the free matrix quantum mechanics, and whether transitions to $Z_m$ phases occur depends on $D$ and $\tilde{D}$.

Exploring the phase diagrams beyond the transition lines $a_m=0$ is more difficult. In the case of the free matrix model, we speculated that the $Z_m$ phase with the smallest $a_m$ is favored. However, we cannot use it in the bosonic BFSS matrix model, since $a_m$ depends on the trial masses $m_X$ and $m_Z$ and they are not constants in the $Z_m$ phases. We also have to take care of higher order terms for $u_n$ in the effective potential \eqref{effective-action-app}, which we can ignore in the confinement phase. In Fig.~\ref{fig-phases-bBFSS}, we determine the phase boundaries between the $Z_m$ phase and the $Z_n$ phase by ignoring these difficulties and just solving $a_m=a_n$, where we set $m_X$ and $m_Z$ to the values of the confinement phase \eqref{m-2-loop}. This prescription is not accurate but it may not be qualitatively so bad.

%

\subsection{Results of the numerical simulation}
\label{subsubsec-MC_abridge}

In this section, we present the numerical simulation of the bosonic BFSS matrix model \eqref{action-BFSS-J}. We focus on the $(D,\tilde{D})=(3,1)$  case and take the matrix size $N$ to be $30$. Also, we focus on the pure-imaginary $\Omega$ case, where we recall $\Omega =i \Omega_I$ ($\Omega_I$ is real). By using the scaling relaiton \eqref{F-scaling}, we can predict several relations between the expectation values of observables in $Z_m$ phases as discussed in Appendix \ref{app-scaling-general}, and they will be tested in the numerical calculation. The details of our numerical analysis via the Fourier expansion regularization are presented in Appendix \ref{subsubsec-MC_appendix}.

\subsubsection{$Z_m$ phases and scaling relation} 
\label{scale_MC}

In order to investigate the phase structures, we set $\beta \Omega_I=0$, $2\pi/3$, $\pi$ and $4\pi/3$, and evaluate the observables by changing temperature. Here, we adopt the Fourier expansion regularization \eqref{fourier_tr_boson}, with the parameters \eqref{parameters_MC}. 

The scaling property \eqref{F-Zm-free-MQM} and the stability of $u_n$ in the minimum sensitivity analysis predict that, if the transition to a $Z_1$ phase from a confinement phase occurs at $(T,\beta \Omega_I)=(T_H,0)$, then the transition to the $Z_2$ phase occurs at $(T,\beta \Omega_I)=(2T_H,\pi)$, and the transition to the $Z_3$ phase occurs at $(T,\beta \Omega_I)=(3T_H,2\pi/3)$ and $(3T_H,4\pi/3)$.

\begin{figure} [htbp]
    \centering
    \includegraphics[width=0.49\textwidth]{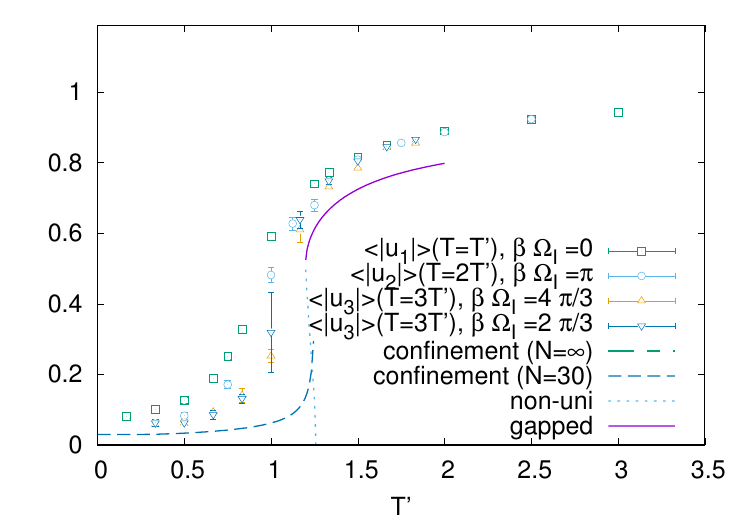}
    \includegraphics[width=0.49\textwidth]{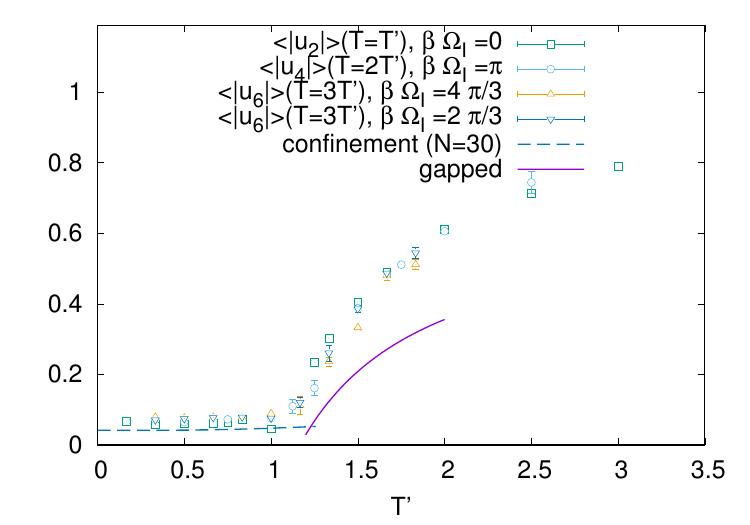}
    \includegraphics[width=0.49\textwidth]{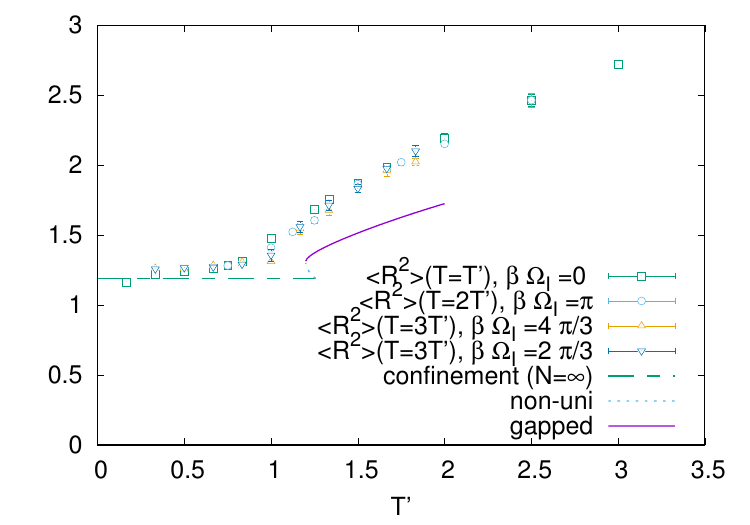}
    \includegraphics[width=0.49\textwidth]{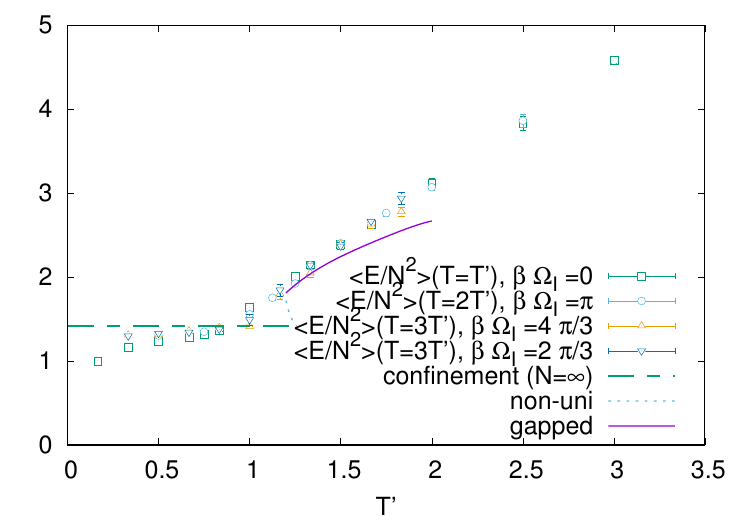}
    \caption{
    Monte-Carlo results of the bosonic BFSS matrix model \eqref{action-BFSS-J} with an imaginary angular velocity $\Omega_I$ for $(D,\tilde{D})=(3,1)$ and $N=30$.    
$\langle |u_k| \rangle$ (top), $\langle R^2 \rangle$ (bottom, left) and $\langle \frac{E}{N^2} \rangle$ (bottom, right) at $\beta \Omega_I=0$, $\beta \Omega_I=\pi$ and $\beta \Omega_I = \frac{2\pi}{3}$, $\frac{4\pi}{3}$ are plotted for the temperature $T=T'$, $T=2T'$ and $T=3T'$, respectively. The scaling relation \eqref{F-scaling} predicts that the $T'$ dependence of these observables is coincident at large $N$, if transitions to the $Z_m$ phases occur. Our numerical results show that they are indeed coincident, which confirm our conjecture. The discrepancies of $u_n$ for $T<T_H  \simeq 1.1$ can also be explained as we discuss in Sec.~\ref{confinement_MC}. The lines denote the result of the minimum sensitivity method at $\beta \Omega_I=0$ for the temperature $T=T'$. The dashed lines represent those of the confinement phase at $N=30$ or $N = \infty$. The dotted and solid lines represent those of the non-uniform and gapped phase at $N=\infty$, respectively.}\label{scaling_behavior}
\end{figure}

In Fig.~\ref{scaling_behavior}, the observables at $\beta \Omega_I=0$, $\beta \Omega_I = \pi$ and $\beta \Omega_I = \frac{2\pi}{3}, \frac{4\pi}{3}$ are plotted for the temperature $T=T'$, $T=2T'$ and $T=3T'$, respectively. If our prediction is correct, the transition temperatures $T'$ at each $\beta \Omega_I$ are coincident. Indeed Fig.~\ref{scaling_behavior} shows that all the transitions occur near $T'=1.1$. In addition, the values of the Polyakov loops $u_n$ (the top two panels in Fig.~\ref{scaling_behavior}) indicate that the eigenvalue distributions are $Z_m$ symmetric, and the $Z_m$ phases are realized in our Monte-Carlo simulation\footnote{In our numerical simulations, we evaluate $|u_n|$ instead of taking a gauge such that all $u_n$ are real. We do not observe any issue when we compare these two quantities, as we explain in Footnote 6 of Ref.~\cite{Azuma:2023pzr}.}. Also the obtained transition temperature $T_H \simeq 1.1$ is consistent with the existent numerical results at $\beta \Omega_I=0$ \cite{0704_3183,Kawahara:2007fn, Azuma:2014cfa}. Furthermore, the $T'$ dependence of all the observables with the different $\beta \Omega_I$ are almost coincident, and it is consistent with the scaling properties of the observables shown in Eqs.~\eqref{E-Z_m}, \eqref{u_n-Z_m} and \eqref{R-Z_m} in Appendix \ref{app-scaling-no-fermion}. These results verify the scaling relation and the existence of the stable $Z_m$ phases in the bosonic BFSS matrix model.

The numerical results are also compared with those of the minimum sensitivity method (two-loop) at $\beta \Omega_I=0$ calculated in the previous subsection and Sec.~6.2 of Ref.~\cite{Azuma:2023pzr}. Although they match qualitatively, the quantitative disparity is large particularly in the deconfinement phase $T \ge T_H$. This is because, in the minimum sensitivity method, we ignored the $O(u_n^3)$ terms in the effective action \eqref{effective-action-app} and used the approximation \eqref{large-D} for $T \ge T_H$, which is not reliable at $D=3$ \cite{Azuma:2023pzr}. Besides, the discrepancy of the transition temperature $T_H$ is about 15\% (see Table III in Ref.~\cite{Morita:2020liy}, which shows that the minimum sensitivity works better for larger $D$). 
Since the minimum sensitivity method is a kind of mean field approximation \cite{Stevenson:1981vj}, this discrepancy may indicate that the bosonic BFSS matrix model at $D=3$ is strongly coupled. Thus, our numerical results provide evidence that our scaling relation works properly even in strongly coupled systems.

\subsubsection{Polyakov loops $ \langle | u_n | \rangle$ in the confinement phase} 
\label{confinement_MC}

We next study the behaviors of $u_n$ in the confinement phase ($T <T_H \simeq 1.1$).
As we have discussed in Sec.~\ref{app-scaling-no-fermion-um-conf}, although $ \langle | u_n |\rangle =0$ in the confinement phase at large $N$, they receive the $1/N$ corrections and the scaling relation \eqref{F-scaling} predicts the relation \eqref{u_n-conf-2}. This relation is rewritten as
\begin{align} 
    \frac{1}{\sqrt{n}}	\langle  | u_n (nT,\Omega_I) | \rangle=  \langle | u_1 (T,\Omega_I) | \rangle , \quad (\text{confinement phase}) .
	\label{u_n-conf-MC}
\end{align}
Indeed,  $ | u_n |$ in our numerical results (the top two panels in Fig.~\ref{scaling_behavior}) show  $ \langle | u_n | \rangle \neq 0$. 
We evaluate $u_n$ in detail numerically, at $\beta \Omega_I=0$ for simplicity, and test the scaling relation \eqref{u_n-conf-MC}.

\begin{figure} [htbp]
    \centering
    \includegraphics[width=0.49\textwidth]{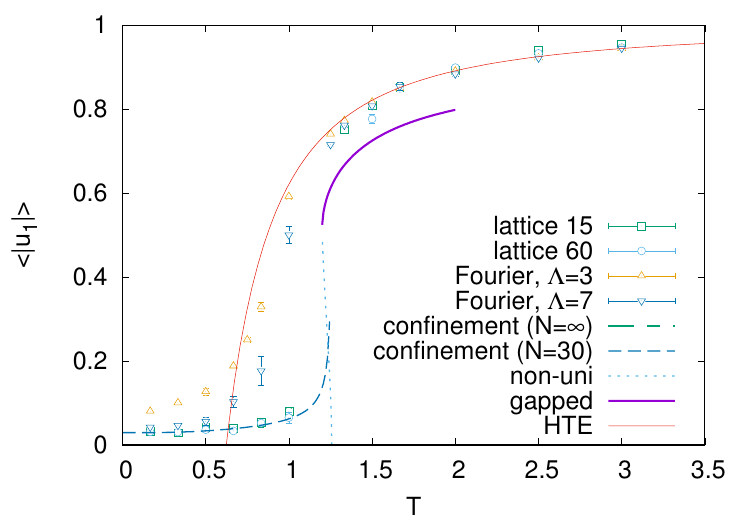}
    \includegraphics[width=0.49\textwidth]{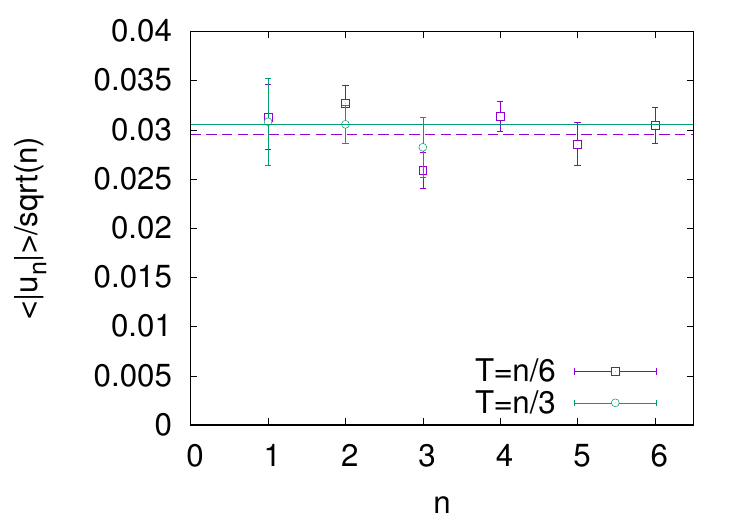}
    \caption{(Left) We compare $\langle |u_1| \rangle$ obtained by the lattice regularization with 15 and 60 lattice sites with that of the Fourier expansion regularization with $\Lambda=3$ and 7. In both cases, we take $D=3$, $N=30$, $\Omega_I=0$. The solid line ``HTE" denotes $u_1$ obtained by the high temperature expansion \cite{0710_2188} at $D=3$, $N = \infty$, $\Omega_I=0$. 
    The other lines are the predictions from the minimum sensitivity. The lattice regularization agrees with the minimum sensitivity in the confinement phase $(T \lesssim 1.0) $.
    (Right) $\frac{\langle |u_n| \rangle}{\sqrt{n}}$ is plotted for the temperature $T=\frac{n}{6}$ ($n=1,2,3,4,5,6$) and $T=\frac{n}{3}$ ($n=1,2,3$) for $D=3$, $N=30$, $\Omega=0$. They take almost the same value, which verify the scaling relation \eqref{u_n-conf-MC}. The dashed and solid lines represent $u_1$ obtained by the minimum sensitivity method at $T=\frac{1}{6}$ and $T=\frac{1}{3}$ as given by Eq.~\eqref{u1_MS}.}\label{scaling_behavior_cofinement}
\end{figure}

However, in the confinement phase $(T \lesssim 1.1) $, the discrepancy of $|u_1|$ between the minimum sensitivity and the numerical simulation is larger than those for the other observables (the top left panel in Fig.~\ref{scaling_behavior}). We guess that this is attributed to the fluctuations in the Fourier expansion regularization \eqref{fourier_tr_boson} at low temperature. To illustrate this, we compare $\langle |u_1| \rangle$ between the Fourier expansion regularization at $\Lambda=3$, 7 and the lattice regularization with 15 and 60 lattice sites, whose details are presented in Sec.~5.1 of Ref.~\cite{Azuma:2023pzr}. In Fig.~\ref{scaling_behavior_cofinement} (Left) the result at $D=3$, $N=30$, $\Omega_I=0$ is presented. Although the results of the Fourier expansion regularization and the lattice regularization agree in the deconfinement phase $(T \gtrsim 1.4) $, they do not agree in the confinement phase\footnote{Although Fig.~\ref{scaling_behavior_cofinement} (Left) shows that increasing $\Lambda$ improves the result of $\langle |u_1| \rangle$ at low temperature, this still obscures the behavior below the critical point $T_H \simeq 1.1$. Also, with the Fourier expansion regularization at $D=3$, $N=30$, $\Omega_I=0$, $\Lambda=7$,  $\langle |u_1| \rangle$ is $0.042(5)$ and $0.046(6)$ at $T=\frac{1}{6}$ and $T=\frac{1}{3}$, respectively, which is less close to the result of the minimum sensitivity method \eqref{u1_MS} than that of the lattice regularization presented in Fig.~\ref{scaling_behavior_cofinement} (Right). On the other hand, our Fourier expansion analysis except $|u_1|$ for $T \lesssim 1.4$ would be reliable, since $|u_1|$ agrees with the lattice regularization for  $T \gtrsim 1.4 $ and other quantities ($u_2$, $R^2$, $E$) agree with those of the minimum sensitivity $T \lesssim 1.0$ as shown in Fig.~\ref{scaling_behavior}, which imply the fluctuation of $|u_1|$ is less relevant in these quantities. } $(T \lesssim 1.0) $, while those of the lattice regularization agree with the minimum sensitivity nicely. Thus, the fluctuation for $|u_1|$ in the Fourier expansion regularization is larger in this region, and it would cause the discrepancy. 

Thus, in order to test the relation \eqref{u_n-conf-MC}, we apply the lattice regularization, instead of the Fourier expansion regularization. Here, we focus on the $D=3, N=30$ and $\Omega_I=0$ case with 60 lattice sites. In Fig.~\ref{scaling_behavior_cofinement} (Right), we plot $\frac{\langle |u_n |\rangle}{\sqrt{n}}$ for the temperature $T=\frac{n}{6}$ ($n=1,2,3,4,5,6$) and $T=\frac{n}{3}$ ($n=1,2,3$), which are all $T<1.1$. This nicely confirms the property \eqref{u_n-conf-MC}, verifying the scaling relation \eqref{F-scaling} in the bosonic BFSS matrix model \eqref{action-BFSS}.

Also, they are nicely compared with the value of $u_1$:
\begin{eqnarray}
 |u_1| = \left\{ \begin{array}{ll} 0.029564 & (T=\frac{1}{6}) \\ 0.0305833 & (T=\frac{1}{3}) \end{array} \right. , \label{u1_MS}
\end{eqnarray}
which is obtained by Eq.~\eqref{u_n-conf} and $a_1(\beta,0,m_0,m_0)$ of the minimum sensitivity \eqref{An-bBFSS}. It may indicate that the minimum sensitivity analysis is better at lower temperatures in the confinement phase. Recall that the confinement phase is controlled by just a single function $a_1(T)$ in the Polyakov loop effective action \eqref{F-expand-scaling}, while higher order terms of $u_n$ contribute in the deconfinement phase. This may explain that the minimum sensitivity works better at low temperatures.

\section{$Z_m$ phases in four-dimensional pure Yang-Mills theories}
\label{sec-YM}

In the previous two sections, we studied the models of matrix quantum mechanics. In this section, we will further study two models related to the four-dimensional pure Yang-Mills theories, which is also related to the bosonic BFSS matrix model \eqref{action-BFSS} at $D=3$. One is the YM theory on a small $S^3$ and the other is the YM theory on $R^3$ at high temperatures. In both cases, we can use free field approximations and investigate the models analytically. We will see that $Z_m$ phases appear in these models as well.

\subsection{Pure Yang-Mills theory on $S^3$}
\label{subsec-S3}

We analyze the finite temperature SU$(N)$ Yang-Mills theory on $S^3$,
\begin{align} 
    S=\frac{1}{g^2}  \int_0^\beta dt \int_{S^3}d^3x \Tr\left( F_{\mu\nu}^2 \right).
\end{align}
On $S^3$, the gauge field except $A_\tau$ possess masses proportional to the inverse of the radius of the sphere. Thus, the perturbative expansion works for a small sphere, and we study the leading order (free limit) of this perturbative calculation \cite{Aharony:2003sx, Aharony:2005bq}. 

There are two commuting angular momenta on $S^3$, and we introduce two angular velocities $\Omega_1$ and $\Omega_2$ correspondingly \cite{Hawking:1999dp}. The Polyakov loop effective potential with these angular velocities can be read from the result of ${\mathcal N}=4$ SYM theory in Ref.~\cite{Murata:2008bg}, and it becomes
\begin{align} 
V(\beta,\{ u_n \})=&   \frac{11}{120}+
    \sum_{n=1}^\infty a_n   |u_n|^2, \quad
    a_n:= \frac{1}{n \beta} \left( 1-z_V(x^n) \right) ,
    \label{effective-action-free-YM}
\end{align}
where $x:=e^{-\beta}$ and we have taken the radius of the sphere to be $1$.
Although $x$ was used as a symbol for a similar quantity in the previous section, we do not think this is confusing. 
The first term is the the Casimir energy of the gauge fields on $S^3$ \cite{Balasubramanian:1999re, Aharony:2003sx}. $z_V(x)$ is called the single-particle partition function for the spatial gauge fields, which is defined as  \cite{Murata:2008bg}
\begin{align} 
    z_V(x)=&\frac{x^2(1+x^2-x^{1+\Omega_1}-x^{1-\Omega_1}-x^{1+\Omega_2}-x^{1-\Omega_2}+x^{\Omega_1+\Omega_2}+x^{-\Omega_1-\Omega_2})}{(1-x^{1+\Omega_1})(1-x^{1+\Omega_2})(1-x^{1-\Omega_1})(1-x^{1-\Omega_2})}+ \left(\Omega_2 \to - \Omega_2 \right) .
    \label{z_V}
\end{align} 
Note that this result is valid only when $|\Omega_1| <1$ and $|\Omega_2| <1$. Otherwise the partition function diverges, which indicates the instability of the system.

The effective potential \eqref{effective-action-free-YM} is consistent with the scaling relation \eqref{F-expand-scaling}, since $n$ and $\beta$ appear only in the combination $n \beta $ in $a_n$. Besides, the effective action has a periodicity $\Omega_i \to \Omega_i + 2\pi i/\beta $.\\

Using this effective potential \eqref{effective-action-free-YM}, we analyze the phase structure. 
Similarly to the free matrix quantum mechanics case, this effective potential has only the quadratic terms for $u_n$, which is a feature of free theories. Thus, we can apply the analysis in the free matrix quantum mechanics.

First, for simplicity, we take $\Omega_1=\Omega_2=0$. Then $a_1$ becomes
\begin{align} 
\beta a_1  =  1-\frac{2x^2(3-x)}{(1-x)^3}.
    \end{align} 
This is positive at $T=0$ and reaches zero at  $T=T_H \simeq 0.76$, where the confinement/deconfinement transition occurs. For $T>T_H$, $a_1$ is always negative and only the standard deconfinement phase appears.

\subsubsection{Phase diagram of the Yang-Mills theory on $S^3$ with a real angular velocity}
\label{subsubsec-S3-Re}

\begin{figure}
    \begin{center}
        \begin{tabular}{cc}
            \begin{minipage}{0.5\hsize}
                \begin{center}
                    \includegraphics[scale=0.6]{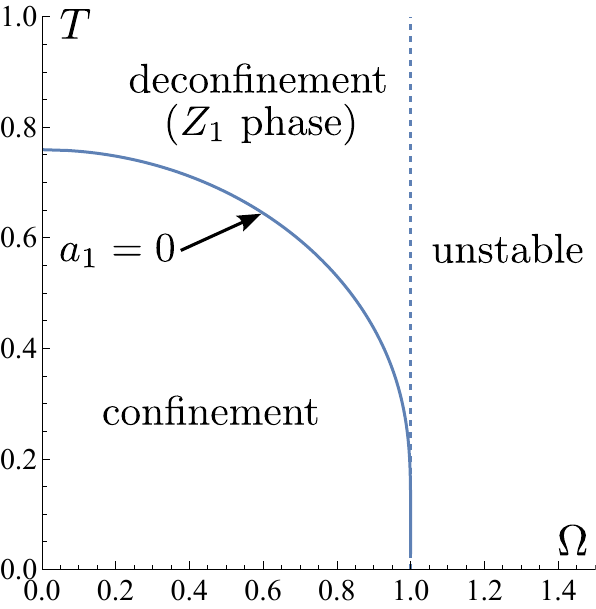}\\
                    $T$ vs. $\Omega$
                \end{center}
            \end{minipage}
            \begin{minipage}{0.5\hsize}
                \begin{center}
                    \includegraphics[scale=0.6]{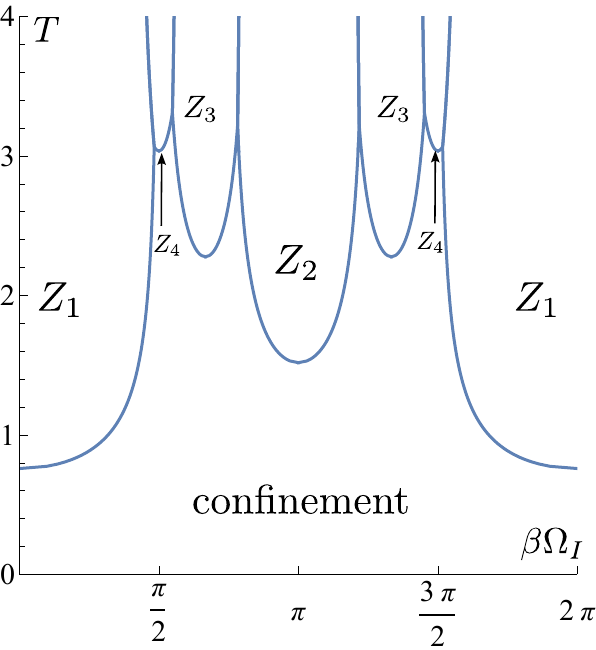}\\
                    $T$ vs. $\beta \Omega_I$
                \end{center}
            \end{minipage}
        \end{tabular}
        \caption{
Phase diagrams of the four-dimensional Yang-Mills theory on $S^3$ \eqref{effective-action-free-YM} with a real angular velocity $\Omega$ (the left panel) and with an imaginary angular velocity $\Omega_I$  (the right panel). For the real angular velocity, the phase diagram is very similar to the free matrix quantum mechanics case. For the imaginary angular velocity, the $Z_m$ phases ($m=1,2,3$ and 4) appear and the transitions to these phases from the confinement phase occur on the curves $a_m=0$ \eqref{an-YM-S3-Im}. Also, the transitions between the $Z_m$ phase and the $Z_n$ phase would occur on the curves $a_m=a_n$.
        }
        \label{fig-S3}
    \end{center}
\end{figure}

Taking $\Omega_1=\Omega >0$ and $\Omega_2=0$ in the effective potential \eqref{effective-action-free-YM}, we study the phase diagram of the system with a real angular velocity. (Since the results do not change qualitatively even if both $\Omega_1$ and $\Omega_2$ are set to non-zero, we omit showing the results for $\Omega_2 \neq 0$ here.) Then $a_1$ becomes
\begin{align} 
    \beta a_1= 1-\frac{2x^2(1-x+2 \cosh \beta \Omega)}{(1-x)(1+x^2-2 x \cosh \beta \Omega )}.
    \end{align} 
Similarly to the free matrix quantum mechanics case, only the confinement phase and the $Z_1$ phase appears and the transition occurs along the curve $a_1= 0$. This line is plotted in Fig.~\ref{fig-S3} (the left panel). This plot shows that the transition temperature decreases as $\Omega$ increases.

%

\subsubsection{Phase diagram of the Yang-Mills theory on $S^3$ with an imaginary angular velocity}
\label{subsubsec-S3-Im}

Taking $\Omega_1= i \Omega_I >0$ and $\Omega_2=0$ in the effective potential \eqref{effective-action-free-YM}, we study the phase diagram of the system with an imaginary angular velocity. (Again the results do not change qualitatively so much even if both $\Omega_1$ and $\Omega_2$ are set to non-zero.)
Then, we obtain
\begin{align} 
    a_m=  \frac{1}{m \beta} \left(1- \frac{2x^{2m}(1-x^m+2 \cos m \beta \Omega_I)}{(1-x^m)(1+x^{2m}-2 x^m \cos m \beta \Omega_I )}\right).
    \label{an-YM-S3-Im}
\end{align} 
Now, $Z_m$ phases ($m=1,2,3$ and 4) appear. The transitions from the confinement phase to the $Z_m$ phases occur on the curves $a_m=0$. Beyond these curves, the transitions between the $Z_m$ phase and the $Z_n$ phase would occur on the curves $a_m=a_n (<0)$ as we discussed in Sec.~\ref{subsec-free-MQM-Im}. The results are shown in Fig.~\ref{fig-S3} (the right panel). We see that this phase diagram is qualitatively similar to that of the free matrix quantum mechanics and the bosonic BFSS matrix model at $(D,\tilde{D})=(3,1)$.

\subsection{Four-dimensional pure Yang-Mills theory at high temperature}
\label{subsec-high-T}

We study phases in the four-dimensional SU($N$) Yang-Mills theory on $S^1_{\beta}\times R^3$ at high temperatures, where the perturbative calculation works \cite{RevModPhys.53.43}.
At the leading order of the perturbative analysis, the effective potential density with an imaginary angular velocity $\Omega_I$ has been computed in Ref.~\cite{Chen:2022smf}\footnote{Here, we define $a_n$ as a coefficient in the effective Lagrangian density rather than the effective potential. Thus, for example, when we apply the formula \eqref{u_n-conf}, we have to replace $a_n$ with $V_{3} a_n $, where $V_{3}$ is the volume of the three-dimensional space. 
},
\begin{align} 
    {\mathcal V}(\beta,\{ u_n \})=&   
        \sum_{n=1}^\infty a_n   |u_n|^2, \quad
        a_n=  -\frac{2}{\pi^2 (n \beta )^4 } \cos \left(
            n \beta \Omega_I  \right).
            \label{action-4dYM-eff}
\end{align}
This result is for the rotation center. An analysis away from the center was also studied in Ref.~\cite{Chen:2024tkr} but we do not consider it here for simplicity. Since the effective potential has the symmetries $\beta \Omega_I \to  \beta \Omega_I + 2\pi$ and $\beta \Omega_I \to - \beta \Omega_I$, it is sufficient to evaluate the region $0 \le  \beta \Omega_I \le \pi$.

We again assume that a $Z_m$ phase is stable when the corresponding $a_m$ is the minimum. Then, the phase transition occurs at $a_m=a_n$, and the phase diagram is obtained as shown in Fig.~\ref{fig-4dYM}. $Z_m$ phases ($m=1,2,3$ and 4) appear similarly to the $S^3$ case.\\

In Appendix \ref{app-finite-N}, we explore the phase diagrams at finite $N$.
Interestingly, the authors of Ref.~\cite{Chen:2022smf} found stable high temperature confinement phases in the SU(2) and SU(3) cases near $\beta \Omega_I=\pi$. On the other hand, we did not observe a high temperature confinement phase at large $N$.
We will study larger $N$ ($N=4,5,6$ and 12) and argue the connection between the results at the $N=2$ and 3 cases and our large-$N$ results. We will see that the phase diagrams highly depend on $N$, and the large-$N$ approximation does not work for $N \le 5$ at least.

\begin{figure}
    \begin{center}
                    \includegraphics[scale=1]{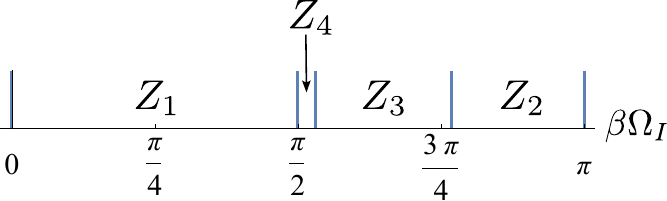}
        \caption{Phase diagram of the four-dimensional SU($N$) Yang-Mills theory on $S^1_{\beta}\times R^3$ at high temperatures \eqref{action-4dYM-eff} with an imaginary angular velocity $\Omega_I$ in the large-$N$ limit. This diagram is similar to the $S^3$ case shown in Fig.~\ref{fig-S3} at high temperatures. 
        }
        \label{fig-4dYM}
    \end{center}
\end{figure}
%

\section{Discussions}
\label{Sec_discussion}

In this paper, we have shown that the Polyakov loop effective potentials in finite temperature large-$N$ gauge theories obey the scaling relation \eqref{F-scaling}. In particular, we have seen that it holds even in the Monte-Carlo simulation of the bosonic BFSS matrix model, and it suggests that the scaling relation is valid in the non-perturbative regime. We have also studied the phase diagrams of the large-$N$ gauge theories in the presence of imaginary chemical potentials and imaginary angular velocities, and have observed that the $Z_m$-type deconfinement phases appear stably. The possible applications of our results are discussed in the following subsections.

\subsection{The gauge/gravity correspondence}
\label{subsec-discussion-GR}

We have found the stable $Z_m$ phases in the large-$N$ gauge theories. Through the gauge/gravity correspondence \cite{Itzhaki:1998dd, Witten:1998zw}, these predict the existence of the corresponding stable solutions in gravitational theories.

For example, the gravity duals of the deconfinement phases (the $Z_1$ phases) in the bosonic BFSS matrix model \cite{Aharony:2004ig} and the pure Yang-Mills theories \cite{Mandal:2011ws} are the single caged black hole solutions, where the Euclidean temporal $S^1_\beta$ in the gauge theories is mapped to a spatial $S^1$ in the gravitational theories. (Here the confinement/deconfinement transitions in the gauge theories correspond to the Gregory-Laflamme transitions in the gravitational theories \cite{Gregory:1994bj}.) Thus, the $Z_m$ phases would correspond to the multiple caged black hole solutions where $m$ identical black holes are symmetrically arranged on the spatial circle \cite{Harmark:2004ws, Azuma:2012uc}.

Usually, black holes are attracted to each other and such a multiple black hole configuration is unstable. Correspondingly, $Z_m$ solution in large-$N$ gauge theories is usually unstable \cite{Azuma:2012uc}. However, we have shown that the imaginary chemical potentials or the imaginary angular velocities make the $Z_m$ solutions stable. In the gauge/gravity correspondence, these potentials in the gauge theories are mapped to background supergravity fields in the gravitational theories. Therefore, our results predict that the multiple caged black holes are stabilized by these background fields. It would be interesting to understand the mechanism of the stabilization in gravity.
(Recently, a related study in an anti-de Sitter (AdS) soliton has been done in Ref.~\cite{Kumar:2024pcz}.)

Also, the deconfinement phases in the large-$N$ supersymmetric gauge theories are duals of the black hole solutions in the supergravity theories, where the Euclidean temporal $S^1_\beta$ in the gauge theories is identified with the temporal $S^1_\beta$ in the gravitational theories \cite{Witten:1998zw, Itzhaki:1998dd}. In Appendix \ref{app-SYM}, we find stable $Z_m$ phases in four-dimensional ${\mathcal N}=4 $ SYM theory on $S^3$ in the small volume limit, where the free field approximation is available \cite{Sundborg:1999ue, Aharony:2003sx}. Thus, we naively expect the existence of the dual solutions in supergravity. However, the corresponding gravitational solutions have not been known so far, and it is valuable to explore such solutions. One possibility is that the dual solutions are possible only in the higher spin theory corresponding to the ${\mathcal N}=4 $ SYM theory in the free limit \cite{Gopakumar:2003ns}.

In addition to these questions about the $Z_m$ solutions, it would be interesting to understand the implications of the scaling properties of the Polyakov loop effective potential in the gravitational theories.

\subsection{Scaling relation}
\label{subsec-discussion-scaling}

We have shown the scaling relation \eqref{F-scaling} for the Polyakov loop effective potentials in the large-$N$ gauge theories. However, we have considered the gauge theories coupled to only adjoint matter fields, and it would be important to study the contributions of other matter fields. It is also interesting to study the effects of the $\theta$-term \cite{Witten:1998uka} and the Chern-Simons couplings, which we have not considered.

If there are several compact circles in the theory, we can introduce the loop operators winding each circle. Then, the effective theory of these loop operators would obey a similar scaling relation with respect to these cycles \cite{Aharony:2005ew, Hanada:2007wn, Mandal:2011hb, Fujii:2024llh}. This should be investigated further.

Besides, if fermions are included, our scaling relation predicts the connection between the Polyakov loop effective potential and the Witten index \cite{Witten:1982df} (see Eq.~\eqref{F-scaling-fermion-even}). Since we can compute the indexes in supersymmetric gauge theories even at strong coupling \cite{Kinney:2005ej}, this connection may be useful to investigate the thermodynamics of supersymmetric gauge theories.

Another interesting avenue is to study systems coupled to $N_f$ fundamental matters, which break the $Z_N$ center symmetry. Since the $Z_N$ center symmetry is essential in deriving the scaling relation (see footnote \ref{ftnt-fundamental}), the fundamental matters would spoil the scaling relation. However, if $N_f/N_c$ is small, some predictions may be possible through a $N_f/N_c$ expansion. This question is particularly important in applying the scaling relation to QCD.

\subsection{$Z_m$ phases and phase diagrams}
\label{subsec-discussion-Zm}

In this study, we show that the phase diagrams of large-$N$ gauge theories with imaginary chemical potentials and imaginary angular velocities are very rich, and that stable $Z_m$ phases exist. We have studied several models: the gauged free matrix quantum mechanics, the bosonic BFSS matrix model, the four-dimensional YM theories and the $\mathcal{N}$=4 SYM theory on $S^3$. It would be valuable to further analyze other models. 

From the viewpoint of the gauge/gravity correspondence, the analysis of the phase diagrams of supersymmetric gauge theories 
\cite{Itzhaki:1998dd}
 and Chern-Simons theories \cite{Jain:2013py, Aharony:2008ug} is also an interesting problem. In particular, we can tune the imaginary chemical potentials such that the fermions become periodic along the temporal circle while keeping the periodicity of the bosons (e.g. $\beta \mu_I = 2\pi $). Then, supersymmetry would hold and calculations in this region may be simplified. This could be a clue to revealing the whole phase structures of the supersymmetric gauge theories.

\subsection{Real world from matrix?}
\label{subsec-discussion-phenomenology}

Many attempts have been made to describe our real world using matrix models such as the IKKT matrix model \cite{Nishimura:2001sx, Kim:2011cr, Brahma:2022ifx} (see Ref.~\cite{Anagnostopoulos:2022dak} for recent developments). In these studies, it is an important issue to clarify the mechanism by which the symmetries of the Standard Model in particle physics arise \cite{Aoki:2010gv, Aoki:2014cya}. In this paper, it has been demonstrated that even in simple (S)YM theories, the non-trivial $Z_m$ symmetries appear. It would be worthwhile to apply this result to gain a better understanding of the emergence of symmetries in large-$N$ gauge theories. A related topic is the Hosotani mechanism in particle phenomenology \cite{Hosotani:1983xw}, in which the gauge symmetries are broken by the Aharonov-Bohm effect on the compact circles, which are similar to imaginary chemical potentials on the Euclidean temporal circle discussed in this paper.

\subsection{Real angular velocities}
\label{subsec-discussion-angular}

Although the present study focused on the analysis of imaginary angular velocities, progress was also made for real angular velocities. 

As discussed in Sec.~\ref{subsubsec-S3-Re}, we found that, in the Yang-Mills theory on $S^3$, the deconfinement transition temperature decreases as the angular velocity increases. (Similar behaviors have been found in the $\mathcal{N}$=4 SYM theory on $S^3$ too \cite{Murata:2008bg}.) In Refs.~\cite{Morita:2010vi, Azuma:2023pzr}, a similar analysis was performed for the Yang-Mills theory on $T^D$ in a small volume limit, where the theory reduces to the bosonic BFSS matrix model \eqref{action-BFSS}, and the same behavior was observed. These results suggest that angular velocities lower the transition temperatures in pure Yang-Mills theories on finite volume spaces, and it may be so even in the four-dimensional Yang-Mills theory on $R^3$ with an appropriate IR-cut off. 

An important question is what happens in the SU(3) YM theory on $R^3$, which is related to QCD in our real world. However, the phase transitions discussed so far are the large-$N$ phase transitions, and any transitions do not occur in gauge theories on finite volume spaces with finite gauge groups. Therefore, we need to clarify whether the present results can be extended to infinite volume systems with finite gauge groups.

Another way to resolve this question is to use an analytic continuation from imaginary angular velocity to real one. We found that the phase diagram of the YM theory on $R^3$ with the imaginary angular velocity at high temperature (Fig.~\ref{fig-4dYM}) is very similar to that of $S^3$ (Fig.~\ref{fig-S3}, right panel). It may indicate that the phase diagram of the YM on $R^3$ at low temperature is also similar to that of $S^3$. If so, the imaginary angular velocity increases the transition temperature in the YM on $R^3$. The same behavior in the SU(3) YM on $R^3$ has also been proposed by the authors of Ref.~\cite{Chen:2022smf}\footnote{A lattice computation of SU(3) YM theory shows an opposite behavior that the imaginary angular velocity decreases the transition temperature \cite{Braguta:2021jgn}, and we should clarify this discrepancy.}.  Then, a naive analytic continuation predicts that the real angular velocity lowers the transition temperature, which is consistent with our results of the finite volume systems. However, it is quite subtle whether such an analytic continuation always works even for small $\Omega_I$ \cite{Chen:2022smf, Chernodub:2022qlz, Ambrus:2023bid}. Thus, although our results are suggestive, they are still not conclusive for understanding the phase diagram of SU(3) YM on $R^3$ with real angular velocity.

\paragraph*{Acknowledgment.---}
We thank H.~Aoki, M.~Chernodub, Y.~Hayashi, Y.~Hidaka, M.~Honda, K.~Murata, Y.~Tanizaki, and A.~Tsuchiya for valuable discussions and comments. We would also like to thank the members of Yukawa Institute for Theoretical Physics and Saga University for useful discussions at seminars, where parts of this work were presented. The work of T.~M. is supported in part by Grant-in-Aid for Scientific Research C (No. 20K03946) from Japan Society for the Promotion of Science. Numerical calculations were carried out using the computational resources, such as KEKCC and NTUA het clusters.

\appendix

\section{Observables in $Z_m$ solution}
\label{app-scaling-general}

In Sec.~\ref{subsec-free-MQM-Im}, we argued that the free energy of the $Z_m$ phase at $\beta \Omega_I=2 \pi k/m $ is related to that of the $Z_1$ phase at $\beta \Omega_I=0 $ through the scaling relation \eqref{F-scaling} as shown in Eq.~\eqref{F-Zm-free-MQM}. We extend this relation to several observables in general systems including higher dimensional gauge theories.

We assume that, in a $Z_m$ solution, the Polyakov loops satisfy
    \begin{align} 
         u_{mk}  \neq 0  \quad (k  \in  \mathbf{Z}), \quad  u_n  =0 \quad (n  \notin m \mathbf{Z}).
        \label{un-Zm-app}
     \end{align}
We will see that the values of the observables in the $Z_m$ solutions are related to each other. These relations hold even if the $Z_m$ solutions are unstable. Such relations would also be useful in Monte-Carlo calculations and gravitational calculations in holography, when we explore $Z_m$ solutions. Actually some relations shown in this appendix are observed in the Monte-Carlo computation in the bosonic BFSS matrix model as shown in Sec.~\ref{subsubsec-MC_abridge}.
     
In the following arguments, we assume that an imaginary chemical potential $\mu_I$ is introduced in the system. This assumption is not necessary for the derivation of the relations. But it would be useful, since various imaginary chemical potentials are used in this paper. 

The relations change depending on whether fermions are coupled to the system or not, and we discuss the bosonic case in Appendix \ref{app-scaling-no-fermion} and the case of systems coupled to fermions in Appendix \ref{app-scaling-fermion}.

\subsection{Bosonic systems}
\label{app-scaling-no-fermion}
We consider a system that is not coupled to fermions. We assume that the system has a periodicity with respect to the chemical potential $\mu_I = \mu_I + 2\pi/\beta$.

\subsubsection{Pseudo free energy}
\label{app-scaling-no-fermion-F}

By using the assumption \eqref{un-Zm-app} and the scaling relation \eqref{F-scaling}, the pseudo free energy of the $Z_m$ solution, which is $N^2 V$ at the solution via the saddle point approximation, is related to that of the $Z_1$ solution as
\begin{align} 
\left.  F(\beta, \mu_I)\right|_{Z_m} = \left.  F(m \beta, \mu_I) \right|_{Z_1}= \left.  F(m \beta, \mu_I+2\pi k/m \beta) \right|_{Z_1}, \quad k  \in  \mathbf{Z},
\label{F-Z_m}
\end{align}
where $|_{Z_m}$ represents the quantities of the $Z_m$ solution.

\subsubsection{Internal energy}
\label{app-scaling-no-fermion-E}
The internal energy $E$ is obtained from the free energy via
\begin{align} 
	E= \frac{\partial }{ \partial \beta} (\beta F).
\end{align}
Hence we obtain
\begin{align} 
    \left.  E(\beta, \mu_I)\right|_{Z_m} = \left.  E(m \beta, \mu_I) \right|_{Z_1}= \left.  E(m \beta, \mu_I+2\pi k/m \beta) \right|_{Z_1}, \quad k  \in  \mathbf{Z},
    \label{E-Z_m}
\end{align}
 from the relation \eqref{F-Z_m}.

\subsubsection{$u_m$ and $\rho(\alpha)$ in the $Z_m$ solution}
\label{app-scaling-no-fermion-um}

As discussed in Sec.~\ref{app-scaling-derivation}, the Polyakov loops $u_n$ obey the Schwinger-Dyson equation \eqref{SD-equation}, and thus the $Z_m$ solution \eqref{un-Zm-app} should satisfy
\begin{align} 
    \left.  u_{nm}(\beta, \mu_I) \right|_{Z_m}=  \left.  u_{n}(m \beta, \mu_I) \right|_{Z_1}= \left.  u_{n}(m \beta, \mu_I+2\pi k/m \beta) \right|_{Z_1}, \quad k  \in  \mathbf{Z}.
    \label{u_n-Z_m}
\end{align}
This is different from the relation \eqref{u_n-conf-2} for $\langle |u_n| \rangle$ in the confinement by a factor of $\sqrt{n}$. This is not a contradiction, since Eq.~\eqref{u_n-conf} is for the confinement phase while Eq.~\eqref{u_n-Z_m} is for the $Z_m$ solution.

Also, by applying Eq.~\eqref{u_n-Z_m} to Eq.\eqref{rho-un}, the eigenvalue density function $\rho(\alpha,\beta, \mu_I)$ is obtained as
\begin{align} 
    \left.  \rho(\alpha, \beta, \mu_I) \right|_{Z_m} = \frac{1}{2\pi} \sum_{n \in {\mathbf Z}}\left.  u_{n}(m \beta, \mu_I) \right|_{Z_1} e^{-imn \alpha} = \left.  \rho(m\alpha, m\beta, \mu_I+2\pi k/m \beta) \right|_{Z_1} .
    \label{rho-Zm}
\end{align}
Note that $ \left.  \rho(m\alpha) \right|_{Z_1} $ has a periodicity $\alpha = \alpha +2\pi/m$, since $\rho(\alpha)=\rho(\alpha+2\pi) $. Thus, if $ \left.  \rho(\alpha) \right|_{Z_1} $ has a single peak, $ \left.  \rho(\alpha) \right|_{Z_m} =  \left.  \rho(m\alpha) \right|_{Z_1}  $ has $m$ peaks as shown in Fig.~\ref{fig-rho-Z2} for $m=2$ and 3.

\subsubsection{$R^2$}
\label{app-scaling-no-fermion-R}

So far, we have studied the quantities in large-$N$ gauge theories in general dimensions. We consider here only matrix quantum mechanics.
There, the quantity $R^2$ defined by
\begin{align} 
	R^2(\beta,\mu_I):= \frac{1}{\beta N} \int_0^\beta dt \sum_{I=1}^D  \langle \Tr X^I X^I  \rangle,
    \label{R2}
\end{align}
is often evaluated in Monte-Carlo calculations.

By adding a source term $J \sum_{I=1}^D  \Tr X^I X^I  $ to the Lagrangian, this quantity can be evaluated by differentiating the obtained free energy by $J$. Since such a source term affects only the left-hand side of the Schwinger-Dyson equation \eqref{SD-equation}, the scaling relation \eqref{F-scaling} still holds. Hence we obtain the relation
\begin{align} 
    \left.  R^2(\beta, \mu_I)\right|_{Z_m} = \left.  R^2(m \beta, \mu_I) \right|_{Z_1}= \left.  R^2(m \beta, \mu_I+2\pi k/m \beta) \right|_{Z_1}, \quad k  \in  \mathbf{Z}.
	\label{R-Z_m}
\end{align}

\subsection{Systems coupled to fermions}
\label{app-scaling-fermion}
We study a system coupled to fermions. Here, we assume that the system has a periodicity $\mu_I = \mu_I + 4\pi/\beta$.
In this case, the effective potential obeys the scaling relations \eqref{F-scaling-fermion-odd} and \eqref{F-scaling-fermion-even} but not Eq.~\eqref{F-scaling-app}. Correspondingly, the relationships for the observables in the bosonic systems in Appendix \ref{app-scaling-no-fermion} should be modified depending on whether $m$ is odd or even. 

For pseudo free energy, we obtain the relations
\begin{align} 
    \left.  F(\beta, \mu_I)\right|_{Z_{2l+1}} =& \left.  F((2l+1) \beta, \mu_I) \right|_{Z_1}= \left.  F((2l+1) \beta, \mu_I+4\pi k/(2l+1) \beta) \right|_{Z_1}, \quad k  \in  \mathbf{Z}, \\
    \left.  F(\beta, \mu_I)\right|_{Z_{2l}} =& \left.  F(l \beta, \mu_I) \right|_{Z_2}= \left.  F(l \beta, \mu_I+4\pi k/l \beta) \right|_{Z_2}, 
\end{align}
from Eqs.~\eqref{F-scaling-fermion-odd} and \eqref{F-scaling-fermion-even}.
We obtain similar relations for the internal energy $E$ \eqref{E-Z_m} and $R^2$ \eqref{R2}
\begin{align} 
    \left.  E(\beta, \mu_I)\right|_{Z_{2l+1}} =& \left.  E((2l+1) \beta, \mu_I) \right|_{Z_1}= \left.  E((2l+1) \beta, \mu_I+4\pi k/(2l+1) \beta) \right|_{Z_1}, \\
    \left.  E(\beta, \mu_I)\right|_{Z_{2l}} =& \left.  E(l \beta, \mu_I) \right|_{Z_2}= \left.  E(l \beta, \mu_I+4\pi k/l \beta) \right|_{Z_2},\\
    \left.  R^2(\beta, \mu_I)\right|_{Z_{2l+1}} =& \left.  R^2((2l+1) \beta, \mu_I) \right|_{Z_1}= \left.  R^2((2l+1) \beta, \mu_I+4\pi k/(2l+1) \beta) \right|_{Z_1}, \\
    \left.  R^2(\beta, \mu_I)\right|_{Z_{2l}} =& \left.  R^2(l \beta, \mu_I) \right|_{Z_2}= \left.  R^2(l \beta, \mu_I+4\pi k/l \beta) \right|_{Z_2}. 
\end{align}
For the Polyakov loops $u_n$ of the $Z_m$ solution, we obtain
\begin{align} 
    \left.  u_{(2l+1)n}(\beta, \mu_I)\right|_{Z_{2l+1}} =& \left.  u_n((2l+1) \beta, \mu_I) \right|_{Z_1}= \left.  u_n((2l+1) \beta, \mu_I+4\pi k/(2l+1) \beta) \right|_{Z_1}, \\
    \left.  u_{2ln}(\beta, \mu_I)\right|_{Z_{2l}} =& \left.  u_{2n}(l \beta, \mu_I) \right|_{Z_2}= \left.  u_{2n}(l \beta, \mu_I+4\pi k/l \beta) \right|_{Z_2}, 
\end{align}
where other Polyakov loops $u_n$ ($n \notin m Z_m$) are zero in the $Z_m$ phase $(m \ge 2)$.
Correspondingly, we obtain the eigenvalue densities 
\begin{align} 
    \left.  \rho(\alpha, \beta, \mu_I) \right|_{Z_{2l+1}} = \left.  \rho((2l+1)\alpha, (2l+1)\beta, \mu_I) \right|_{Z_1} ,\quad     \left.  \rho(\alpha, \beta, \mu_I) \right|_{Z_{2l}} = \left.  \rho(l\alpha, l\beta, \mu_I) \right|_{Z_2}.
\end{align}

\section{Details of the numerical simulation based on the Fourier expansion regularization}
\label{subsubsec-MC_appendix}

In this appendix, we present the details of the numerical simulation of the bosonic BFSS matrix model \eqref{action-BFSS-J} based on the Fourier expansion regularization. Similarly to Ref.~\cite{Azuma:2023pzr}, where the lattice regularization has been used,
we rewrite the action \eqref{action-BFSS-J} as 
\begin{eqnarray}
    S &=& N \int^{\beta}_{0} dt \textrm{Tr } \Biggl\{ \frac{1}{2} \sum_{I=1}^D (D_{t} {X}^I (t))^2 - \sum_{I,J=1}^D \frac{1}{4} [{X}^I (t),{X}^J (t)]^2 - \frac{{\Omega}^2}{2} \sum_{K=1}^{2{\tilde D}} ({X}^K (t))^2 \Biggr. \nonumber \\
    & & \ \ \Biggl. + \Omega i \sum_{K=1}^{{\tilde D}} \{ (D_{t} {X}^K (t)) {X}^{K+{\tilde D}} (t) - (D_{t} {X}^{K+{\tilde D}} (t)) {X}^{K} (t) \} \Biggr\}, \label{action-BFSS-J2}
\end{eqnarray}
where $\displaystyle D_{t} = \partial_{t} - i [A (t), ]$. $\beta = \frac{1}{T}$ is the inverse temperature. We take the U$(N)$ gauge group rather than SU$(N)$, since U$(N)$ is simpler in numerical calculations and the difference is irrelevant at large $N$. At $\Omega=0$, this action is invariant under the transformations
\begin{eqnarray}
    & & X^I (t) \to X^I(t) + x^I I_N, \label{x_inv} \\
    & & A(t) \to A(t) + \alpha(t) I_N, \label{a_inv}
\end{eqnarray}
where $I_N$ is an $N \times N$ unit matrix. $x^I$ and $\alpha(t)$ are c-numbers, and $x^I$ is independent of $t$. The $\Omega \neq 0$ case maintains the invariance under the transformation (\ref{a_inv}), but breaks the invariance under the transformation (\ref{x_inv}). In the following, we study the pure imaginary $\Omega$ case, which does {\it not} involve a sign problem.
While we focus on the bosonic model here, we adopt a Fourier expansion for regularization \cite{0706_1647,0707_4454} in light of future extensions to the case including fermion.
\begin{eqnarray}
 X^{I}_{ij} (t) = \sum_{n = - \Lambda}^{\Lambda} X^{I,n}_{ij} e^{i \theta_0 n t}, \textrm{ where } \theta_0 = \frac{2\pi}{\beta}. \label{fourier_tr_boson}
\end{eqnarray}
This satisfies $X^{I,n}_{ij} = (X^{I,-n}_{ji})^{*}$ when $X^I(t)$ is Hermitian. We adopt a static diagonal gauge \eqref{gauge-diagonal}. $\alpha_k$ are independent of time and not subject to the Fourier expansion. This gauge gives rise to the gauge fixing term
\begin{eqnarray}
 S_{\textrm{g.f.}} = - 2 \sum_{1 \leq i < j\leq N} \log \left| \sin \frac{\alpha_i - \alpha_j}{2} \right|. \label{gauge_fixing_term}
\end{eqnarray}
The action we put on a computer is rewritten as
{\small
\begin{eqnarray}
 S_{\textrm{eff}} &=& \frac{N\beta}{2} \sum_{I=1}^D \left\{  \sum_{i,j,k,\ell=1}^N \sum_{-\Lambda \leq n,p,q , n+p+q \leq \Lambda} \left( X^{I,n}_{ij} X^{I,p}_{jk} X^{J,q}_{k \ell} X^{J,-(n+p+q)}_{ \ell i} -  X^{I,n}_{ij} X^{J,p}_{jk} X^{I,q}_{k \ell } X^{J,-(n+p+q)}_{ \ell i} \right)  \right. \nonumber \\
 & & + \left. \sum_{i,j=1}^N \sum_{n = -\Lambda}^{\Lambda} \left( \theta_0 n - \frac{\alpha_i - \alpha_j}{\beta} \right)^2 X^{I,n}_{ij} X^{I,-n}_{ji} \right\}  - \frac{N\beta \Omega^2}{2} \sum_{K=1}^{2{\tilde D}} \sum_{i,j=1}^N \sum_{n = -\Lambda}^{\Lambda}  X_{ij}^{K,n} X_{ji}^{K,-n} \nonumber \\
 & &  - N \beta \Omega \sum_{K=1}^{\tilde D} \sum_{i,j=1}^N \sum_{n = -\Lambda}^{\Lambda} \left( \theta_0 n - \frac{\alpha_i - \alpha_j}{\beta} \right)  (X^{K,n}_{ij} X^{K+{\tilde D},-n}_{ji} - X^{K+{\tilde D},n}_{ij} X^{K,-n}_{ji})  + S_{\textrm{g.f.}}. \label{boson_BFSS_fourier} 
\end{eqnarray}
}
The (real) Langevin equation we solve is 
\begin{eqnarray}
 \frac{d X_{ij}^{I,n}}{d\sigma} = - \frac{\partial S_{\textrm{eff}}}{\partial X_{ji}^{I,-n}} + \eta_{ij}^{I,n}, \ \ \frac{d \alpha_i}{d \sigma} = - \frac{\partial S_{\textrm{eff}}}{\partial \alpha_i} + \eta^{(\alpha)}_{i}. \label{real_langevin_eq}
\end{eqnarray}
Here, $\sigma$ is the fictitious Langevin time, which should not be confused with the time $t$. $\eta_{ij}^{I,n}$ and $ \eta^{(\alpha)}_{i}$ are the white noises. 
$\eta^{I,n}_{ij}$ are complex numbers satisfying $\eta^{I,n}_{ij} = (\eta^{I,-n}_{ji})^{*}$ so that $\eta^I (t)$, whose $(i,j)$ elements are $\eta^{I}_{ij} (t) = \sum_{n = - \Lambda}^{\Lambda} \eta^{I,n}_{ij} e^{i \theta_0 n t}$, are Hermitian matrices. $\eta^{(\alpha)}_{i}$ are real numbers. They obey the distribution proportional to 
\begin{eqnarray}
  \exp \left( \frac{-1}{4}  \int d \sigma \sum_{I=1}^{D} \sum_{i,j=1}^{N} \sum_{n=-\Lambda}^{\Lambda} \eta^{I,n}_{ij} \eta^{I,-n}_{ji}  \right), \ \ \exp \left( \frac{-1}{4} \int d \sigma \sum_{i=1}^{N} (\eta^{(\alpha)}_{i})^2  \right). \label{white_noise}
\end{eqnarray}
When we solve the Langevin equations \eqref{real_langevin_eq} numerically, they are discretized similarly to Sec.~5.1 of Ref.~\cite{Azuma:2023pzr}.
Since the action $S_{\textrm{eff}}$ has no sign problem, the Hermiticity of $X^I(t)$ is maintained in the course of solving the equation \eqref{real_langevin_eq}. 
Using the Fourier expansion \eqref{fourier_tr_boson}, the observables we study can be rewritten as 
\begin{eqnarray}
	R^2 &=& \frac{1}{\beta N} \int_0^\beta dt \sum_{I=1}^D  \textrm{Tr} X^I (t) X^I (t)  =\frac{1}{N} \sum_{I=1}^D \sum_{i,j=1}^N \sum_{n=-\Lambda}^{\Lambda} X^{I,n}_{ij} X^{I,-n}_{ji}, \label{rsq_fourier} \\
	\frac{E}{N^2} &=& \frac{-3}{4N\beta} \int^{\beta}_0 dt [X^I,X^J]^2 \label{fsq_fourier} \\
	& & \hspace*{-8mm} = \frac{3}{2N} \sum_{I,J=1}^D \sum_{i,j,k,\ell=1}^N \sum_{-\Lambda \leq n,p,q , n+p+q \leq \Lambda} \left( X^{I,n}_{ij} X^{I,p}_{jk} X^{J,q}_{k \ell} X^{J,-(n+p+q)}_{ \ell i} -  X^{I,n}_{ij} X^{J,p}_{jk} X^{I,q}_{k \ell } X^{J,-(n+p+q)}_{ \ell i} \right). \nonumber
\end{eqnarray}
The Polyakov loop is written as \eqref{polyakov_static_diag}. In a rotating system, we remove the trace part and implement the constraints\cite{Azuma:2023pzr}
\begin{eqnarray}
 \frac{1}{N \beta} \int^{\beta}_0 \textrm{Tr} dt X^I (t) =0, \ \ \textrm{Tr} A = 0. \label{traceless}
\end{eqnarray}

\begin{figure} [htbp]
    \centering
    \includegraphics[width=0.49\textwidth]{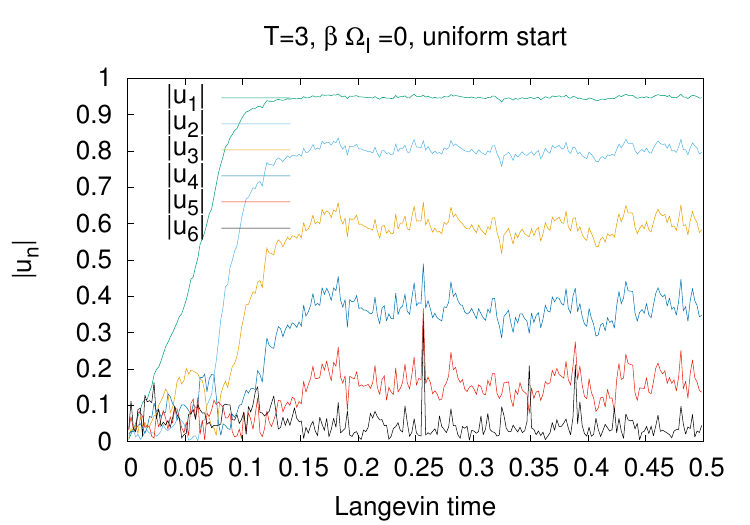}
    \includegraphics[width=0.49\textwidth]{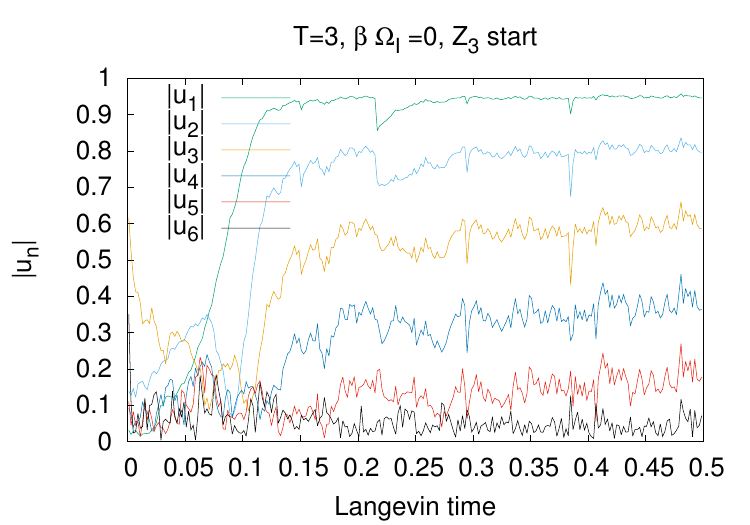}
    \includegraphics[width=0.49\textwidth]{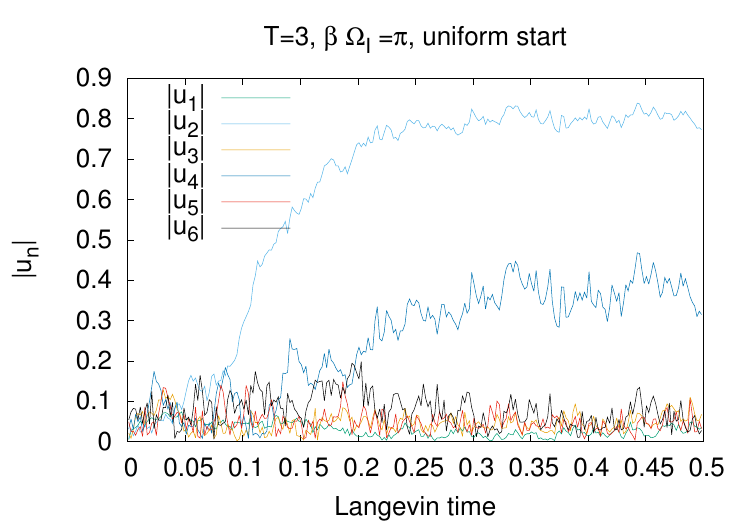}
    \includegraphics[width=0.49\textwidth]{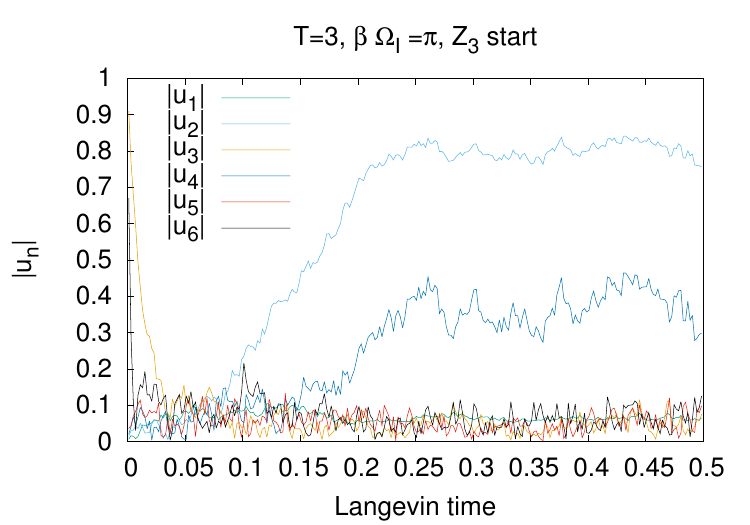}
    \includegraphics[width=0.49\textwidth]{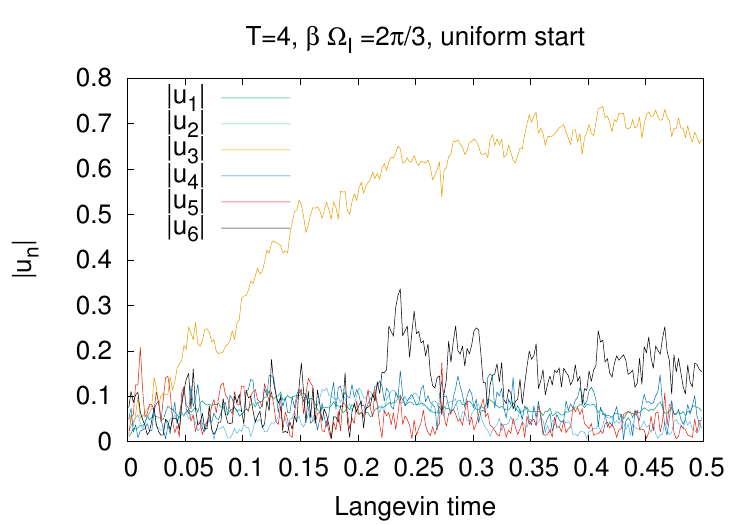}
    \includegraphics[width=0.49\textwidth]{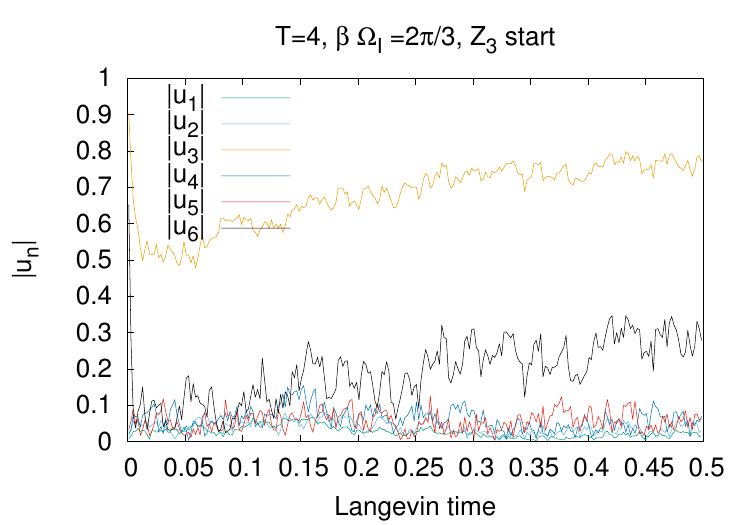}
    \caption{The history of $|u_k|$ $(k=1,2,3,4,5,6$) is plotted against the Langevin time. The initial configurations are the uniform  (left) and $Z_3$ configuations (right), respectively. The results for $T=3, \beta \Omega_I=0$ (top), $T=3$, $\beta \Omega_I=\pi$ (middle) and $T=4$, $\beta \Omega_I = \frac{2\pi}{3}$ (bottom) are presented.}\label{cascade_decay1}
\end{figure}

We solve the Langevin equation \eqref{real_langevin_eq} without imposing the constraints \eqref{traceless}, and instead calculate the observables \eqref{polyakov_static_diag}, \eqref{rsq_fourier} and \eqref{fsq_fourier} in the measurement routine in terms of the matrices
\begin{eqnarray}
 X^{I,n \textrm{(m)}}_{ij} = X^{I,n}_{ij}- \delta^{n,0} \frac{1}{N} \sum_{k=1}^N X^{I,n=0}_{kk}, \ \ \alpha_i^{\textrm{(m)}} = \alpha_i - \frac{1}{N} \sum_{k=1}^N \alpha_k. \label{traceless2}
\end{eqnarray}
To supplement the results in Sec. \ref{scale_MC} that verify the scaling relation, we present the stability of the $Z_m$ configurations of $\alpha_k$. In the following, we adopt the Fourier expansion regularization \eqref{fourier_tr_boson} and take the parameters to be
\begin{eqnarray}
 D=3, \ {\tilde D}=1, \ N=30, \ \Lambda=3. \label{parameters_MC}
\end{eqnarray}

We take the ``uniform configuration" (which is defined by the configuration of uniformly distributed $\alpha_k$) and the ``$Z_3$ configuration" as the initial configuration of the simulation. We present the history of $|u_k|$ ($k=1,2,3,4,5,6$) against the Langevin time in Fig.~\ref{cascade_decay1}. At $T=3$, $\beta \Omega_I=0$, we see that all of $u_k$ grow, which suggests the $Z_1$ solution in the deconfinement phase. On the other hand, at $T=3, \beta \Omega_I=\pi$ and $T=4$, $\beta \Omega_I = \frac{2\pi}{3}$, only $|u_{2,4}|$ and $|u_{3}|$ grow, respectively (in both cases, the growth of $|u_6|$ is not tangible). This suggests the stability of the $Z_2$ and $Z_3$ configurations at $T=3$, $\beta \Omega_I=\pi$ and $T=4$, $\beta \Omega_I = \frac{2\pi}{3}$, respectively.

\begin{figure}[htbp]
    \centering
    \includegraphics[width=0.32\textwidth]{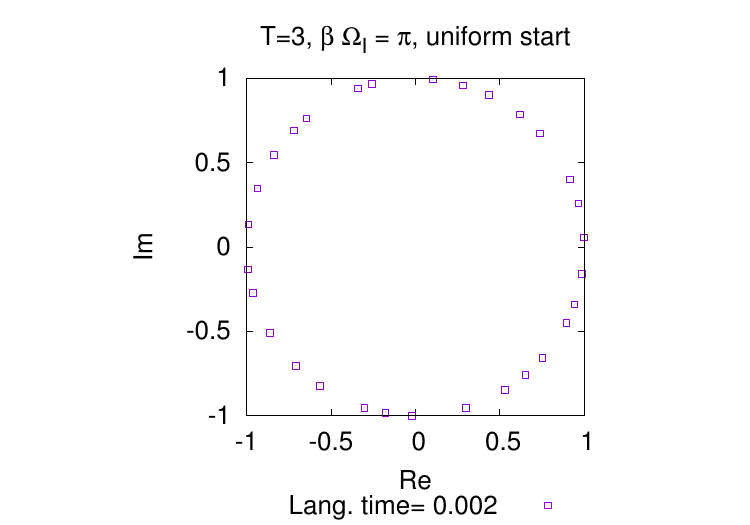} 
    \includegraphics[width=0.32\textwidth]{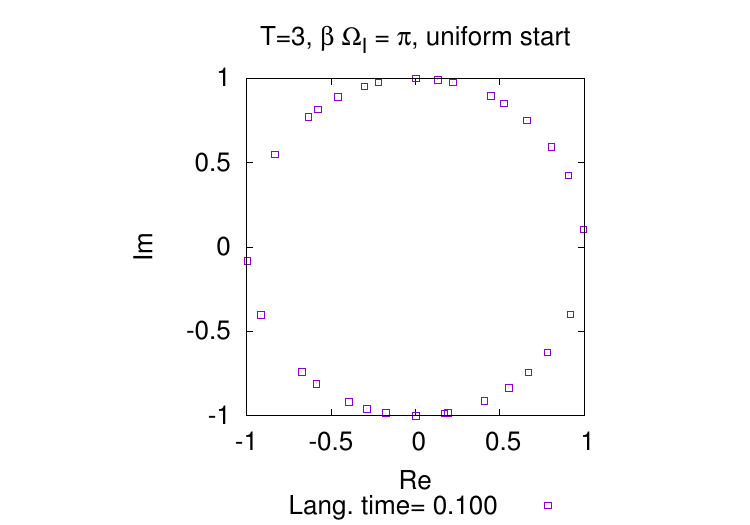}
    \includegraphics[width=0.32\textwidth]{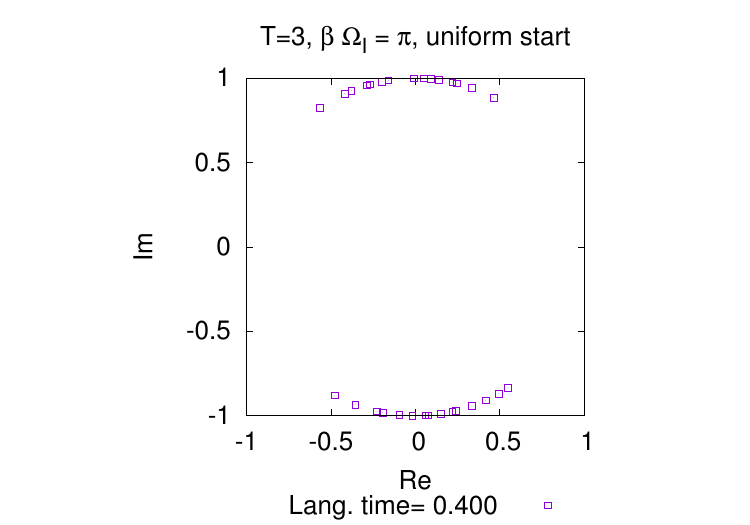}
    \includegraphics[width=0.32\textwidth]{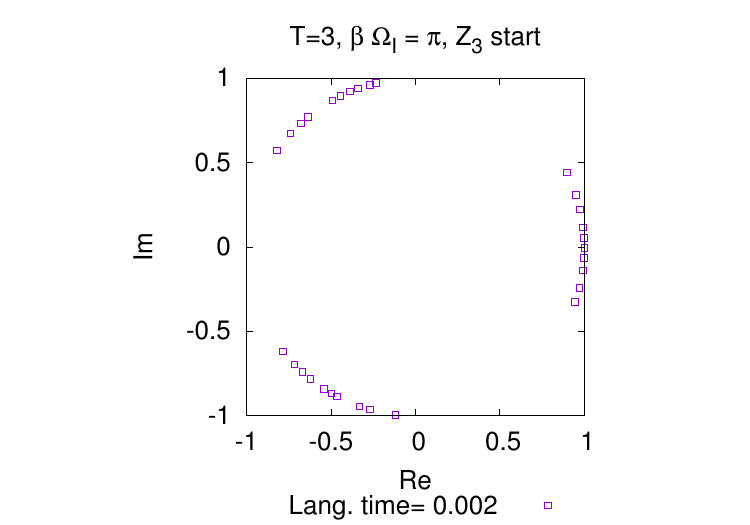}
    \includegraphics[width=0.32\textwidth]{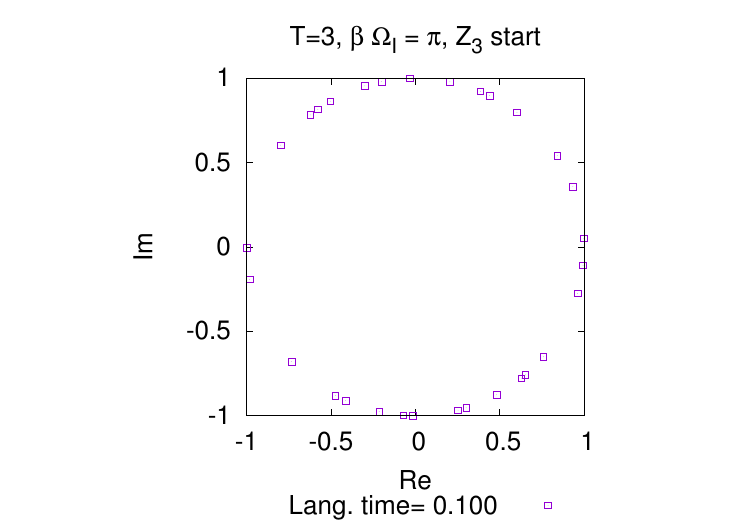}
    \includegraphics[width=0.32\textwidth]{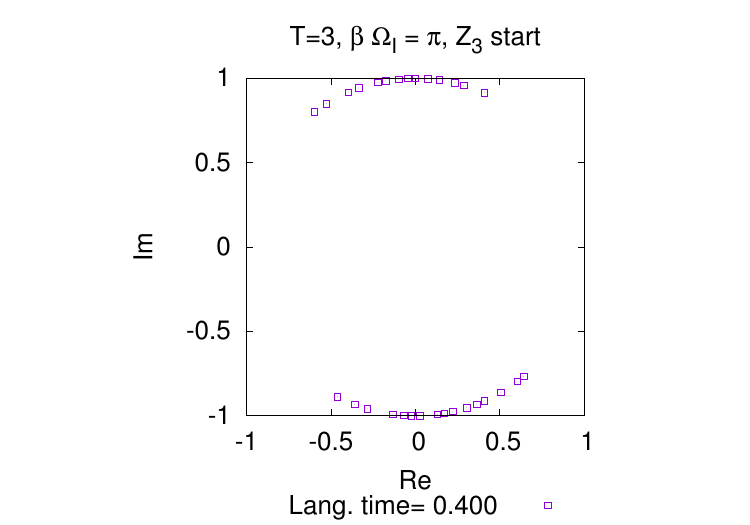}
    \caption{$e^{i \alpha_k}$, which move on the unit circle on the complex plane, are plotted for the $T=3$, $\beta \Omega_I=\pi$ case, which corresponds to Fig.~\ref{cascade_decay1} (middle). 
    We plot the cases of uniform (top) and $Z_3$ (bottom) initial configurations at the Langevin time 0.002 (left), 0.1 (middle) and 0.4 (right).}\label{cascade_decay2}
\end{figure}

In Fig.~\ref{cascade_decay2}, we plot the evolution of the eigenvalues distribution of $\alpha_k$ \cite{Azuma:2012uc} for the $T=3$, $\beta \Omega_I=\pi$ case, which corresponds to Fig.~\ref{cascade_decay1} (middle), as examples. We plot $e^{i \alpha_k}$ which moves on the unit circle on the complex plane. This illustrates the way the uniform and $Z_3$ solutions culminate in the $Z_2$ solution as the system is thermalized. \footnote{The movies of $e^{i \alpha_k}$ on the complex plane in the course of solving the Langevin equation \eqref{real_langevin_eq} for Fig.~\ref{cascade_decay2}, as well as the other cases in Fig.~\ref{cascade_decay1}, are available on the following website: \\ \noindent \url{http://www2.yukawa.kyoto-u.ac.jp/~takehiro.azuma/multi_cut/index.html} }

\section{$Z_m$ phases in the ${\mathcal N}=4 $ SYM theory on $S^3$}
\label{app-SYM}

In this appendix, we study the phase diagrams of the ${\mathcal N}=4$ supersymmetric Yang-Mills (SYM) theory on $S^3$. This model has two commutating angular momenta \cite{Hawking:1999dp} and three U$(1)_R$ changes $Q_i$ ($i=1,2,3$) \cite{Basu:2005pj, Yamada:2006rx}. Correspondingly, we introduce five (real) chemical potentials (two angular velocities $\Omega_i$ ($i=1,2$) and three U$(1)_R$ chemical potentials $\mu_i$ ($i=1,2,3$) ). The field contents of this theory is the gauge field, the six real scalars and the fermions that are the superpartners of the bosonic fields. Their U$(1)_R$ changes are summarized in Table \ref{tab-SYM} \cite{Yamada:2006rx, Murata:2008bg}. 

We can analytically investigate this model by taking the small volume limit, similarly to the YM theory on $S^3$ discussed in Sec.~\ref{subsec-S3}. The Polyakov loop effective potential with the five chemical potentials has been computed in Ref.~\cite{Murata:2008bg}, and the result is given by 
\begin{align} 
  &  V(\beta,\{ u_n \})=   
        \sum_{n=1}^\infty a_n   |u_n|^2, \nonumber \\
        &a_n:= \frac{1}{n \beta} \left( 1-z_V(x^n)
        -z_S(x^n)-(-1)^{n+1}z_F(x^n) 
        \right) .
        \label{effective-action-free-SYM}
\end{align}
Here  $x:=e^{-\beta}$ and we have taken the radius of the sphere 1. $z_V$ is the single-particle partition function for the spatial gauge field defined in Eq.~\eqref{z_V}. $z_S$ and $z_F$ are the single-particle partition functions for the scalar fields and the fermion fields, respectively, and they are defined as 
\begin{align} 
z_S(x)=&\frac{x(1-x^2)(x^{\mu_1}+x^{-\mu_1}+x^{\mu_2}+x^{-\mu_2}+x^{\mu_3}+x^{-\mu_3})}{(1-x^{1+\Omega_1})(1-x^{1+\Omega_2})(1-x^{1-\Omega_1})(1-x^{1-\Omega_2})}, \\
z_F(x)=&\frac{x^{\frac{3}{2}}\left[x^{\frac{\Omega_1+\Omega_2}{2}}+x^{-\frac{\Omega_1+\Omega_2}{2}}-x\left( x^{\frac{\Omega_1-\Omega_2}{2}}+x^{-\frac{\Omega_1-\Omega_2}{2}} \right)\right]}{(1-x^{1+\Omega_1})(1-x^{1+\Omega_2})(1-x^{1-\Omega_1})(1-x^{1-\Omega_2})} \nonumber \\
&\times \left( x^{\frac{\mu_1-\mu_2-\mu_3}{2}}+ x^{\frac{-\mu_1+\mu_2-\mu_3}{2}} + x^{\frac{-\mu_1-\mu_2+\mu_3}{2}} + x^{\frac{\mu_1+\mu_2+\mu_3}{2}} \right) \nonumber \\
&+ \left[ (\mu_1,\mu_2,\mu_3,\Omega_2) \to - (\mu_1,\mu_2,\mu_3,\Omega_2) \right] .
\end{align} 
We can easily check that the effective potential \eqref{effective-action-free-SYM} is consistent with the scaling relation \eqref{a-scaling-fermion}.

The phase diagrams of the effective potential \eqref{effective-action-free-SYM} with real chemical potentials have been investigated in Ref.~\cite{Murata:2008bg}, and here we focus on the analysis of the phase diagrams with imaginary chemical potentials.

\begin{table}
\centering
\begin{tabular}{|c|c|}
\hline
Field & U$(1)_R$ charges $(Q_1, Q_2, Q_3)$ \\
\hline\hline
6 scalars & $(\pm 1, 0, 0)$, $(0, \pm 1, 0)$, $(0, 0, \pm 1)$ \\
\hline
 Fermion  $(\mathbf{2}, \mathbf{\bar{4}})$ & $(\frac{1}{2}, \frac{1}{2}, -\frac{1}{2})$, $(\frac{1}{2}, -\frac{1}{2}, \frac{1}{2})$,  $(-\frac{1}{2}, \frac{1}{2}, \frac{1}{2})$, $(-\frac{1}{2}, -\frac{1}{2}, -\frac{1}{2})$ \\
\hline
 Fermion  $(\mathbf{\bar{2}}, \mathbf{4})$ & $(\frac{1}{2}, \frac{1}{2}, \frac{1}{2})$, $(\frac{1}{2}, -\frac{1}{2}, -\frac{1}{2})$,  $(-\frac{1}{2}, \frac{1}{2}, -\frac{1}{2})$, $(-\frac{1}{2}, -\frac{1}{2}, \frac{1}{2})$ \\
\hline
Vector & $(0, 0, 0)$ \\
\hline
\end{tabular}
\caption{U$(1)_R$ charges of the fields in the ${\mathcal N}=4$ SYM theory on $S^3$}
\label{tab-SYM}
\end{table}

\subsection{Phase diagrams of ${\mathcal N}=4 $ SYM theory with imaginary chemical potentials}
\label{subsec-SYM-Im}

We set the five chemical potentials pure imaginary
\begin{align} 
     \Omega_{1}=i\Omega_{I1}, \quad \Omega_{2}=i\Omega_{I2}, \quad \mu_{1}=i \mu_{I1} , \quad \mu_{2}=i\mu_{I2},\quad \mu_{3}=i\mu_{I3},
\end{align}
and explore the phase diagrams. 
Since the fermions take half-integer values for the angular momenta and the $R$-charges, while the scalar and the gauge field take integer values as shown in Table \ref{tab-SYM}, the system has the five periodicities: $\beta \Omega_{Ii}=\beta \Omega_{Ii}+4\pi$ ($i=1,2$) and $\beta \mu_{Ii}=\beta \mu_{Ii}+4\pi$ ($i=1,2,3$). 

In this system, we have the five different chemical potentials, which would make the phase diagrams very complicated. Thus, we analyzed the phase diagrams by setting some chemical potentials to zero and the remaining chemical potentials to a common value, say $\Omega_I$. Due to this prescription, the periodicity of the system becomes $\beta \Omega_{I}=\beta \Omega_{I}+4\pi$ when the number of non-zero chemical potentials is odd, while it becomes $\beta \Omega_{I}=\beta \Omega_{I}+2\pi$ when the number of non-zero chemical potentials is even.

\begin{figure}
    \begin{center}
        \begin{tabular}{cc}
            \begin{minipage}{0.5\hsize}
                \begin{center}
                    \includegraphics[scale=0.5]{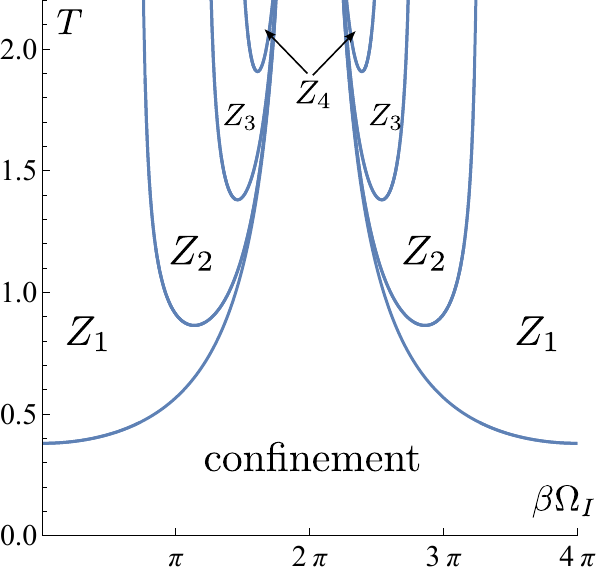}\\
                     $ \Omega_{I}=\Omega_{I1}$, $\Omega_{I2}=\mu_{I1}=\mu_{I2}=\mu_{I3}=0$ \vspace{12 pt}  \\
                \end{center}
            \end{minipage}
            \begin{minipage}{0.5\hsize}
                \begin{center}
                    \includegraphics[scale=0.5]{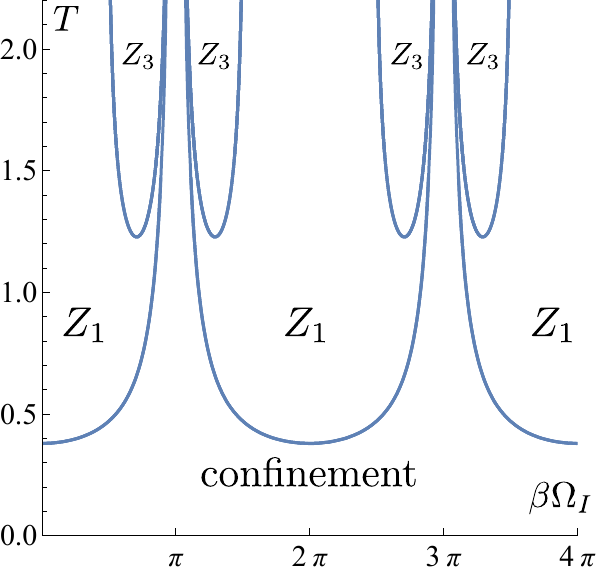}\\
                    $\Omega_{I}= \Omega_{I1} = \Omega_{I2} $, $\mu_{I1}=\mu_{I2}=\mu_{I3}=0$ \vspace{12 pt} \\
                \end{center}
            \end{minipage} \\

            \begin{minipage}{0.5\hsize}
                \begin{center}
                    \includegraphics[scale=0.5]{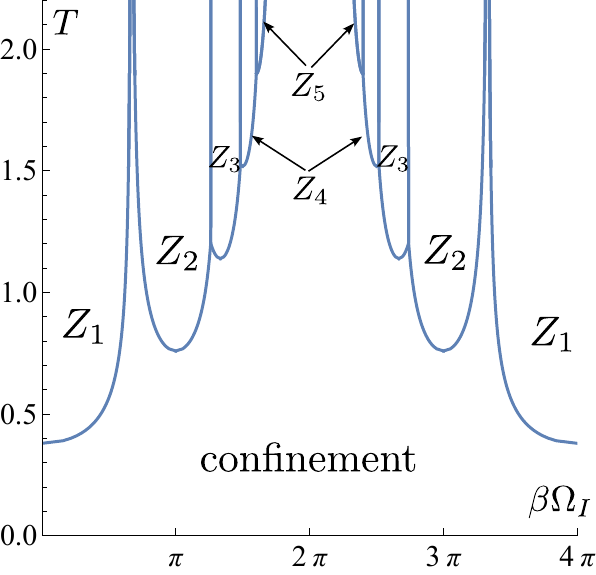}\\
                    $ \Omega_{I}=\mu_{I1} =\mu_{I2}=\mu_{I3}$, $\Omega_{I1}=\Omega_{I2}=0$ \vspace{12 pt}  \\
                \end{center}
            \end{minipage}
            \begin{minipage}{0.5\hsize}
                \begin{center}
                    \includegraphics[scale=0.5]{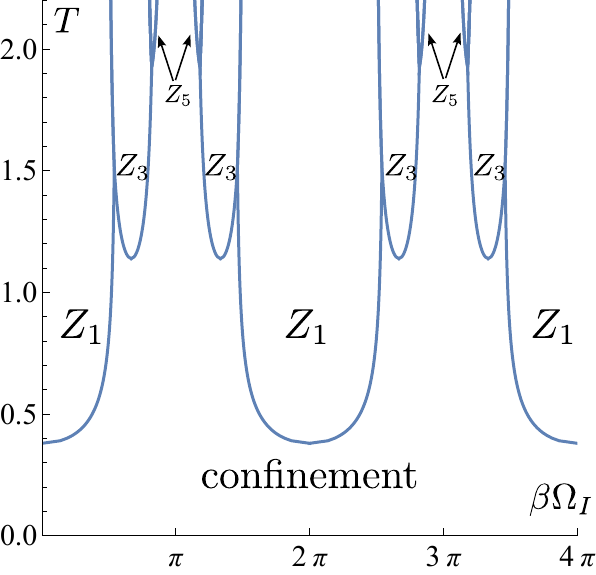}\\
                    $\Omega_{I}= \Omega_{I1}=\Omega_{I2}=\mu_{I1} =\mu_{I2} $, $\mu_{I3}=0$ \vspace{12 pt}  \\
                \end{center}
            \end{minipage} \\
            \begin{minipage}{0.5\hsize}
                \begin{center}
                    \includegraphics[scale=0.5]{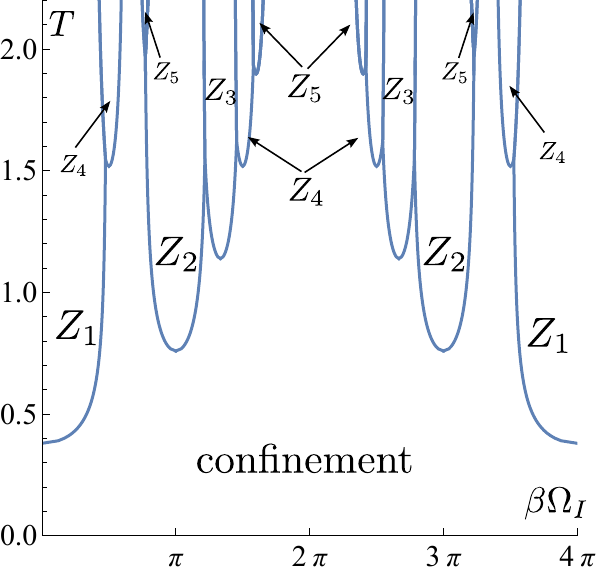}\\
                    $\Omega_{I}= \Omega_{I1}=\Omega_{I2}=\mu_{I1} =\mu_{I2}=\mu_{I3} $  \\
                \end{center}
            \end{minipage}
            \begin{minipage}{0.5\hsize}
                \begin{center}
                    \includegraphics[scale=0.5]{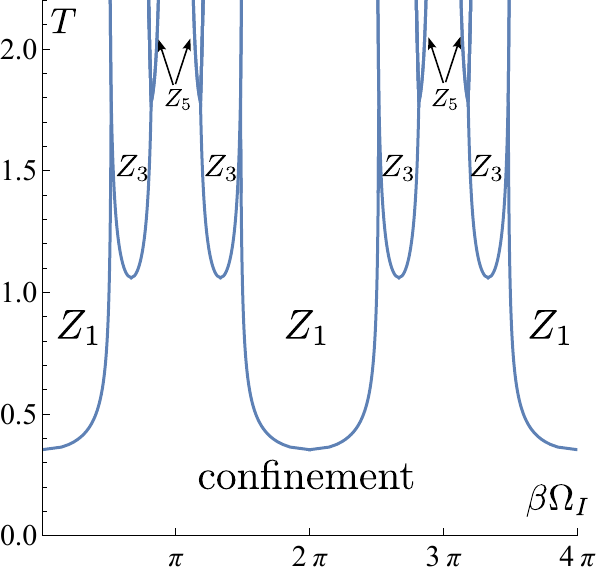}\\
                    $ \Omega_{I}= \Omega_{I1}=\Omega_{I2}=\mu_{I1} =\mu_{I2}$, $\mu_{3}=1/2$  \\
                \end{center}
            \end{minipage}
        \end{tabular}
\caption{
Phase diagrams of the free ${\mathcal N}=4 $ SYM theory on $S^3$.
There are five imaginary chemical potentials ($ \Omega_{I1}$, $\Omega_{I2}$, $\mu_{I1} $, $\mu_{I2}$ and $\mu_{I3}$) and we set some of them zero and set the remaining chemical potentials to a common value $\Omega_I$. 
When one or two chemical potentials are non-zero, the phase transitions from the confinement phase to the $Z_m$ phase ($m \ge 2$) do not occur (the top panels). Only when three or more chemical potentials are non-zero, they occur.
When the number of non-zero imaginary chemical potentials is even, the periodicity is $\beta \Omega_I=\beta \Omega_I +2\pi$ and the $Z_{2m}$ phases do not appear (the right panels), while when it is odd, the periodicity is $\beta \Omega_I=\beta \Omega_I +4\pi$ and the $Z_{2m}$ phases appear (the left panels).
In the bottom right panel, we set the real chemical potential $\mu_3=1/2$. But the phase diagram is qualitatively similar to the $\mu_3=0$ case (the middle right panel).  
}
        \label{fig-SYM1}
    \end{center}
\end{figure}

Similarly to the free matrix quantum mechanics studied in Sec.~\ref{subsec-free-MQM-Im}, phase transitions to $Z_m$ phases from a confinement phase occur on the curves $a_m=0$. Also, the transitions between the $Z_m$ phase and the $Z_n$ phase would occur on the curves $a_m=a_n$. We plot the obtained phase diagrams in Fig.~\ref{fig-SYM1}. There, we find that the phase diagram qualitatively changes significantly depending on how many chemical potentials are set to non-zero. Here, it is not qualitatively important which chemical potentials ($\Omega_{Ii}$ or $\mu_{Ii}$) are set to non-zero, and we exhibit only typical diagrams in Fig.~\ref{fig-SYM1}.

If an even number of chemical potentials are non-zero, no $Z_{2n}$ phases appear (the right column in Fig.~\ref{fig-SYM1}).  Besides, when one or two chemical potentials are set to non-zero, the direct transitions from the confinement phase to the $Z_m$ phase ($m\ge 2$) does not occur, and only the transitions to the $Z_1$ phase occur. The $Z_m$ phases appear only inside other $Z_n$ phases. See the top two panels in Fig.~\ref{fig-SYM1}. 
When three or more chemical potentials are set to non-zero, the transitions from the confinement phase to the $Z_m$ phases occur. Also, when all the five chemical potentials are set to non-zero, $a_n$ become complex numbers\footnote{
Since the expression of $a_n$ \eqref{effective-action-free-SYM} is complicated, it is not that obvious that $a_n$ becomes a complex number when all the imaginary chemical potentials are non-zero. For example, when we take $T \to \infty$, $a_1$ becomes 
\begin{align} 
a_1 \to 2iT  \frac{\sin \left( \beta \mu_{I1} \right) \sin \left( \beta \mu_{I2} \right)\sin \left( \beta \mu_{I3} \right) }{\sin \left( \beta \Omega_{I1} \right)\sin \left( \beta \Omega_{I2} \right)}.
\end{align}
Thus, we explicitly see that it becomes a complex number. This is due to the contributions of the fermions. }. In this case, the configuration that minimizes the real part of the effective potential would dominate the partition function.  Therefore, the phase transitions would occur on the curves Re$(a_m)=0$ or Re$(a_m)=$Re$(a_n)$ (See the bottom left panel in Fig.~\ref{fig-SYM1}).

So far, the chemical potentials are taken to be pure imaginary.  We can consider a mixture of real and imaginary chemical potentials too. The result for the $\mu_3=1/2$ case is plotted in the bottom right panel in Fig.~\ref{fig-SYM1}. We see that the real chemical potential does not change the phase diagram qualitatively.\\

We have found the stable $Z_m$ phases in the ${\mathcal N}=4 $ SYM theory. Since this theory has a dual gravity description in the AdS/CFT correspondence \cite{Witten:1998zw, Aharony:2003sx, Sundborg:1999ue}, our result suggests the existence of the corresponding $Z_m$ solutions in gravity. It seems that such solutions are not known in gravity so far, and exploring them is an interesting problem. Since the free SYM theory is directly related to the tensionless string theory \cite{Gopakumar:2003ns}, the $Z_m$ solutions should exist at least there.

\section{Phases of four-dimensional SU($N$) Yang-Mills theory on $R^3$ at high temperatures at finite $N$}
\label{app-finite-N}

In Sec.~\ref{subsec-high-T}, we have shown that, when the imaginary angular velocity is introduced, the $Z_m$ phases ($m=1,\cdots, 4$) appear in the four-dimensional SU($N$) Yang-Mills theory on $R^3$ at high temperatures in the large-$N$ limit. On the other hand, the phases of the SU(2) and SU(3) Yang-Mills theories have been investigated in Ref.~\cite{Chen:2022smf}, and it has been shown that the systems can be confined when the imaginary angular velocity is turned on. Thus, it is interesting to ask how the phase diagrams change as $N$ increases, and understand the connection to the SU(2) and SU(3) cases.

When $N$ is finite, it is convenient to compute the $\sum_n$ in the effective potential density \eqref{action-4dYM-eff} and use the following form \cite{Chen:2022smf},
\begin{align} 
    {\mathcal V}(\beta,\{ u_n \})=&   
        \frac{\pi^2 T^4}{3} \sum_{i,j=1}^N \sum_{s= \pm 1} B_4 \left(
        \left( \frac{\alpha_i-\alpha_j+s \beta \Omega_I }{2\pi}  \right)_{\text{mod 1}}
        \right) .
            \label{action-4dYM-eff-finite-N}
\end{align}
Here $B_4(x):=x^4-2x^3-x^2-\frac{1}{30}$ is the 4th Bernoulli polynomial, and $\{\alpha_i \}$ obey the constraint $\sum_{i=1}^N \alpha_i =0 $ mod $2\pi$ in SU($N$) gauge theories.

We numerically evaluate the effective potential \eqref{action-4dYM-eff-finite-N} and find the configuration $\{ \alpha_i \}$ that minimizes the potential. If the solution is $Z_m$ symmetric, we call it a $Z_m$ phase. To see whether the obtained solution has the $Z_m$ symmetry, we evaluate the Polyakov loops $u_n$. If the solution is $Z_m$ symmetric, $u_n$ exhibits
\begin{align} 
    \left\langle u_{mk} \right\rangle \neq 0  \quad (k  \in  \mathbf{Z}), \quad \left\langle u_n \right\rangle =0 \quad (n  \notin m \mathbf{Z}).
    \label{un-Zm}
 \end{align}
Thus, these quantities are useful to investigate the phase diagrams.

\begin{figure}[htbp]
    \begin{center}
        \begin{tabular}{rcl|rcl}
            \includegraphics[scale=0.45]{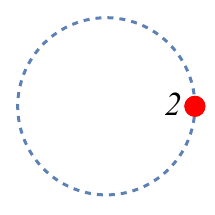} &
            \includegraphics[scale=0.45]{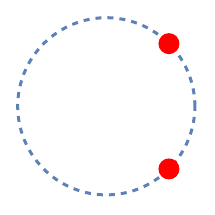} &
            \includegraphics[scale=0.45]{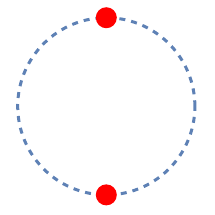} &
            \includegraphics[scale=0.45]{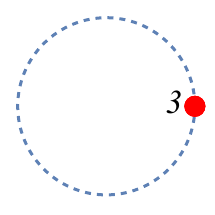} &
            \includegraphics[scale=0.45]{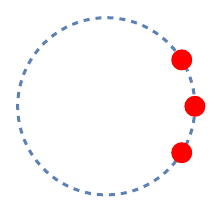} &
            \includegraphics[scale=0.45]{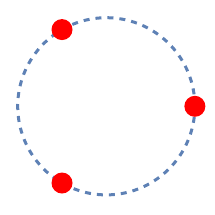}  \\
            $\beta \Omega_I=0$ &  $\beta \Omega_I= \pi /2$ &  $\beta \Omega_I= \pi $ & 
            $\beta \Omega_I=0$ &  $\beta \Omega_I=9 \pi /20$ &  $\beta \Omega_I= \pi $ \\
            \multicolumn{3}{c|}{
                    \includegraphics[scale=0.55]{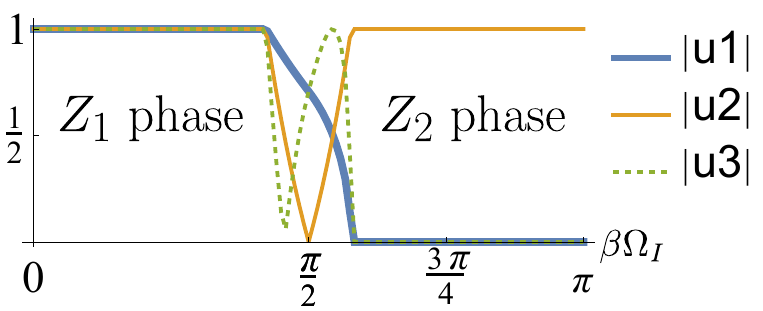}
             } &
             \multicolumn{3}{c}{
                    \includegraphics[scale=0.55]{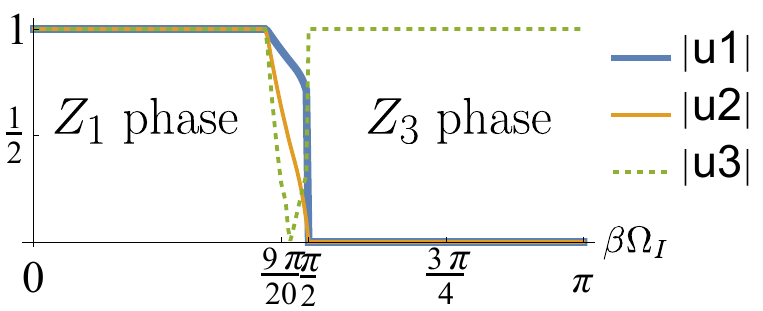}
             } \\
             \multicolumn{3}{c|}{
                SU(2)
         } &
         \multicolumn{3}{c}{
            SU(3)
     }  \\
                    \hline        
    \includegraphics[scale=0.4]{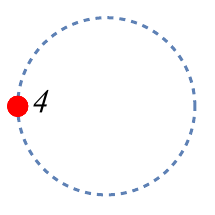} &
            \includegraphics[scale=0.45]{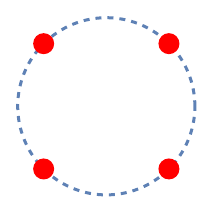} &
            \includegraphics[scale=0.45]{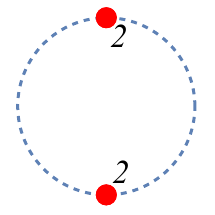} &
            \includegraphics[scale=0.45]{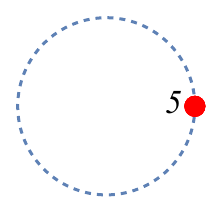} &
            \includegraphics[scale=0.45]{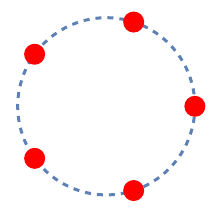} &
            \includegraphics[scale=0.45]{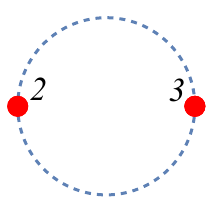}  \\
            $\beta \Omega_I=0$ &  $\beta \Omega_I=2 \pi /3$ &  $\beta \Omega_I= \pi $ & 
            $\beta \Omega_I=0$ &  $\beta \Omega_I=7 \pi /12$ &  $\beta \Omega_I= \pi $ \\
            \multicolumn{3}{c|}{
                    \includegraphics[scale=0.55]{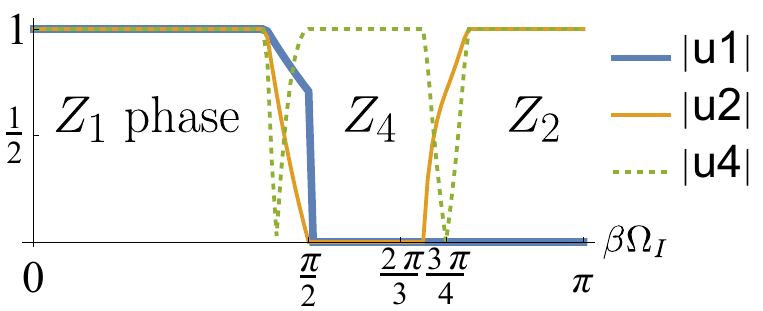}
             } &
             \multicolumn{3}{c}{
                    \includegraphics[scale=0.55]{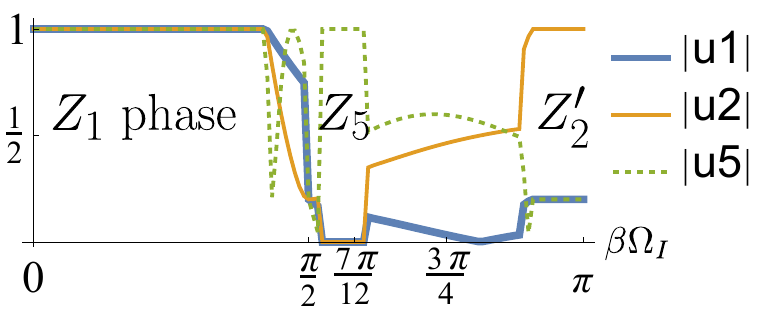}
             } \\
             \multicolumn{3}{c|}{
                SU(4)
         } &
         \multicolumn{3}{c}{
            SU(5)
     }  \\
                    \hline
                    \includegraphics[scale=0.45]{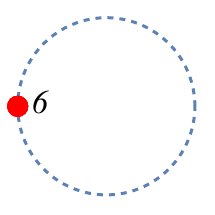} &
                    \includegraphics[scale=0.45]{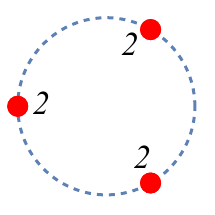} &
                    \includegraphics[scale=0.45]{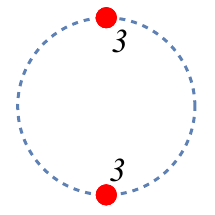} &
                    \includegraphics[scale=0.45]{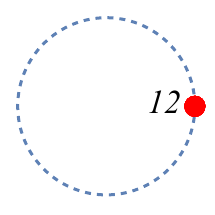} &
                    \includegraphics[scale=0.45]{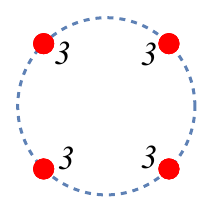} &
                    \includegraphics[scale=0.45]{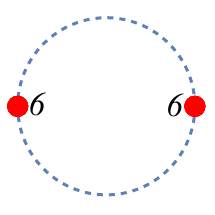}  \\
                    $\beta \Omega_I=0$ &  $\beta \Omega_I=2 \pi /3$ &  $\beta \Omega_I= \pi $ & 
                    $\beta \Omega_I=0$ &  $\beta \Omega_I= \pi /2$ &  $\beta \Omega_I= \pi $ \\
                    \multicolumn{3}{c|}{
                            \includegraphics[scale=0.55]{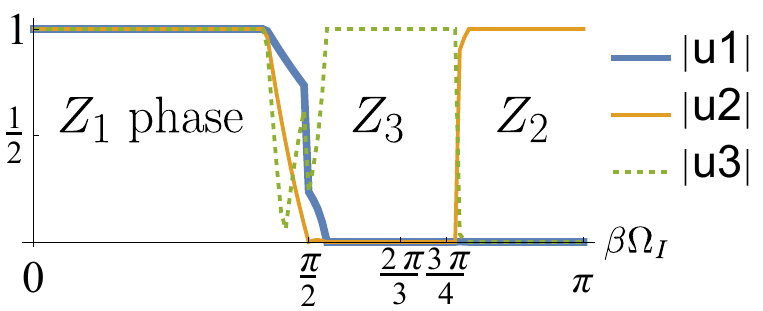}
                     } &
                     \multicolumn{3}{c}{
                            \includegraphics[scale=0.55]{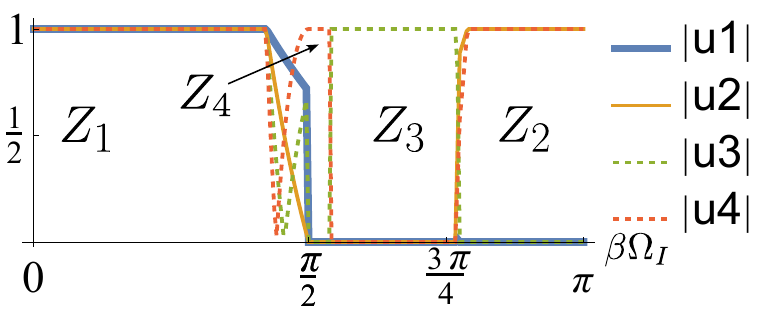}
                     } \\
                     \multicolumn{3}{c|}{
                        SU(6)
                 } &
                 \multicolumn{3}{c}{
                    SU(12)
             }  \\
                            \hline        \end{tabular}
        \caption{
The Polyakov loops $ |u_n|$ vs. the imaginary angular velocity $\beta \Omega_I$, and the eigenvalue distribution $\{ e^{i \alpha_i} \}$ ($i=1,\cdots,N $) in four-dimensional SU($N$) Yang-Mills theory at high temperatures. The results for SU($N$) ($N=2,\cdots,6$ and $12$) are shown. In the plots of the eigenvalue distributions, if there are several eigenvalues at one point, the number indicating the number of the eigenvalues is written. The values of the Polyakov loops indicate the phases such that, if $u_{mk}=1$ ($k \in \mathbf{Z}$) and $u_{n}=0$ ($n \notin m \mathbf{Z}$), it is the $Z_m$ phase as shown in Eq.~\eqref{un-Zm}. The $Z_N$ phase in SU($N$) is a confinement phase, since the $Z_N$ center symmetry is preserved, and this phase may continue to the low temperature conventional confinement phase. These plots show how the phase diagrams in the SU($N$) theories approach that of the large-$N$ limit in Fig.~\ref{fig-4dYM}. Although the SU(12) case is similar to the large-$N$ case, the diagrams up to $N=5$ are very different.
        }
        \label{fig-finite-N}
    \end{center}
\end{figure}

Our numerical results for SU($N$) $(N=2,\cdots,6$ and 12) are summarized in Fig.~\ref{fig-finite-N}. These plots show that the $Z_1$ phase is stable from $\beta \Omega_I=0$ to the vicinity of $\beta \Omega_I=\pi/2$ for all $N$. This is consistent with the large-$N$ case shown in Fig.~\ref{fig-4dYM}. Besides, in the SU(2) and SU(3) cases, the eigenvalues distribute ``uniformly" beyond $\beta \Omega_I=\pi/2$ and $u_1$ becomes 0. Thus, the systems are confined as the authors of Ref.~\cite{Chen:2022smf} pointed out. Note that these uniform distributions in the SU(2) and SU(3) cases can be regarded as the $Z_2$ phase and the $Z_3$ phase, respectively. This is because the $Z_N$ phase in the SU$(N)$ theory preserves the $Z_N$ center symmetry, and it indicates a confinement. In Fig.~\ref{fig-finite-N}, we observe the $Z_4$ phase in SU(4) and the $Z_5$ phase in SU(5), which are also confinement phases. These high temperature confinement phases may continue to the low temperature conventional confinement phases. However, we do not observe such high temperature confinement phases for $N \ge 6$.

In the SU(4), SU(6) and SU(12) cases, the $Z_2$ phases are stable around $\beta \Omega_I=\pi$. In the SU(5) case, since 5 is odd, the $Z_2$ phase is impossible. However, the eigenvalue distribution around $\beta \Omega_I=\pi$ resembles that of the $Z_2$ phase, and we call it the $Z'_2$ phase in Fig.~\ref{fig-finite-N}. These results are  consistent with the large-$N$ case, where the $Z_2$ phase appears around $\beta \Omega_I=\pi$.

In the SU(2) case, the stable phases are only the $Z_1$ and $Z_2$ phases, and intermediate configurations around $\beta \Omega_I=\pi/2$ connect these two phases. Thus, no other stable phases appear. The SU(3) case is also similar. Only the two stable phases ($Z_1$ and $Z_3$) exist. However, in the SU(4), SU(5) and SU(6) cases, stable intermediate $Z_4$, $Z_5$ and $Z_3$ phases appear between the $Z_1$ phase and the $Z_2$ (or the $Z'_2$) phase, respectively. In the SU(12) case, intermediate stable $Z_4$ and $Z_3$ phases appear, and the whole phase diagram is very similar to the large-$N$ case. 

In this way, the phase diagrams in the SU($N$) cases approach the large-$N$ result as $N$ increases. One interesting observation is that the finite $N$ results up to $N=5$ are very different form the large-$N$ limit. It is a folklore that the qualitative properties of SU($N$) Yang-Mills theories are similar to those of large-$N$ in many aspects, even at $N=3$ \cite{Lucini:2001ej}. This seems correct if imaginary angular velocity is not introduced. However, once it is introduced,  we have found that it is not true. This is a good lesson that we should not blindly trust the large-$N$ approximation.

\bibliographystyle{utphys}
\bibliography{bBFSS} 

\end{document}